\documentclass{aa}
\usepackage{graphics}
\newcommand{\plotatlas}[1]{\center\resizebox{0.95\hsize}{!}{\includegraphics{#1}}}

\newcommand{\plotsmall}[1]{\resizebox{0.1\hsize}{!}{\includegraphics{#1}}}
\newcommand{\plotx}[1]{\resizebox{0.75\hsize}{!}{\includegraphics{#1}}}
\newcommand{\plotfull}[1]{\resizebox{\hsize}{!}{\includegraphics{#1}}}

\newcommand{\plotthree}[3]{\center {\resizebox{0.95\hsize}{!}{\includegraphics{#1}\includegraphics{#2}\includegraphics{#3}}}}

\newcommand{\rmn}[1] {{\rm #1}}

\newcounter{dummyyoyo}
\newcounter{dummyfofo}

\begin{document}

\title{Neutral Hydrogen and Optical Observations of Edge-on Galaxies:
  Hunting for Warps}

\author{I. Garc\'{\i}a-Ruiz$^1$, R. Sancisi$^{2,1}$, K. Kuijken$^1$}
\institute{$^1$Kapteyn Astronomical Institute, Postbus 800, 9700 AV, Groningen, The Netherlands\\
$^2$Osservatorio Astronomico di Bologna, Via Ranzani 1, I-40127
Bologna, Italy \\
}
\authorrunning{Garc\'\i a-Ruiz et al.}
\date{Received ??? / Accepted ???}
\thesaurus{11.11.1; 11.19.6; 11.09.2}
\maketitle

\begin{abstract}
We present 21-cm HI line and optical R-band observations for a sample
of 26 edge-on galaxies. The HI observations were obtained with the
Westerbork Synthesis Radio Telescope, and are part of the WHISP
database (Westerbork HI Survey of Spiral and Irregular Galaxies). We
present HI maps, optical images, and radial HI density profiles. We have also
derived the rotation curves and studied the warping and lopsidedness
of the HI disks. 20 out of the 26 galaxies of our sample are warped, confirming that
warping of the HI disks is a very common phenomenon in disk
galaxies. Indeed, we find that all galaxies that have an extended HI disk with
respect to the optical are warped. The warping usually starts around the
edge of the optical disk. The degree of warping varies considerably
from galaxy to galaxy. Furthermore, many warps are asymmetric, as they 
show up in only one side of the disk or exhibit large
differences in amplitude in the approaching and receding sides of the
galaxy. These asymmetries are more pronounced in rich environments, which may
indicate that tidal interactions are a source of warp asymmetry. A
rich environment tends to produce larger warps as well. 
The presence of lopsidedness seems to be related to the
presence of nearby companions. 

\end{abstract}


\section{Introduction}

It has long been known that many galaxies have warped disks. The
phenomenon occurs at large 
galactocentric radii, starting close to the edge of the optical
disk. This makes 21-cm line observations essential for the study of
warps: there are galaxies that have no optical warp and yet 
exhibit extraordinary warps in their outer HI layers.

The first indication that the HI distribution of disk galaxies was
sometimes bent or warped came in 1957 from observations of our own
Galaxy (Burke, 1957; Kerr, 1957). Early calculations showed that the
tidal field due to the Magellanic
Clouds was unable to account for the warp of the Galaxy
(Burke, 1957; Kerr, 1957; Hunter \& Toomre, 1969), and so it was regarded as a rare, uncommon
phenome\-non. Later on, Sancisi (1976) studied the HI layer
of 5 edge-on galaxies and discovered 4 of them to have warps. This
result indicated that warps are very common among galaxies. This discovery 
was not expected, because three of Sancisi's galaxies were quite
isolated and thus 
presented a problem with the tidal-warp scenario. The high incidence
of warps among spiral galaxies has later been confirmed with larger
samples both in the optical \cite{ss,res} and the HI \cite{bosma}.
The ubiquity of
warps means that either warps are long-lasting
phenomena or they are transient but easily and often excited. Neither of these has been
proven and no satisfactory explanation has been found yet. A study by
Briggs (1990) of a sample of 12 galaxies showed that: (1) warps start around
R$_{25}$ - R$_{\rm H_\circ}$; (2) the line of nodes of the warp is straight
inside R$_{25}$, and is a leading spiral outside. 

The absence of a mechanism that would regularly generate warps and the
fact that the line of nodes of the observed warps does not show a
severe winding effect \cite{briggs} suggest that warps are long-lived.
The main issue in trying to explain a long lasting warp is
differential precession, which in a few rotation periods will 
destroy the coherent warp pattern (this is similar to the winding
problem of material spiral arms). The differential precession arises from the non-sphericity
of the potential \cite{avner}. However, the
survival time of a coherent warp pattern can be larger if the halo in which
the disk is embedded deviates only slightly from spherical symmetry
\cite{sanders}. Likewise, a halo with a flattening that decreases outwards in
the right way would be able to maintain a coherent warped disk
for a Hubble time \cite{petrou}.

Extending the work of Hunter \& Toomre (1969), Sparke \&
Casertano (1988) studied long-lived warping modes of a
galactic disk inside an oblate halo potential, where 
no winding occurs. They found that the combined torque from a halo and
a self-gravitating disk allows a configuration in which the
precession frequency of the warp is the same at all radii, thus
allowing the warp to maintain its shape unchanged for many rotation 
periods. One of their assumptions was that of a fixed halo
potential that does not react to the potential of the disk. 

The back-reaction of the precessing disk onto the halo, analogous to dynamical
friction, was investigated by Dubinski \& Kuijken (1995), Nelson \&
Tremaine (1995), and Binney et al. (1998) after an earlier investigation
in terms of WKB density waves by Bertin and Mark (1980). It turns out
that it has a very strong influence on the evolution of the warp,
causing the disk and halo to reorient to a common plane of symmetry in
a few orbital times, and making the warp decay. 

\begin{center}
\begin{table*}
\begin{center}
\begin{tabular}{cclcccrccrr}
\hline
UGC & NGC & \multicolumn{1}{c}{type} & R$_{25}$ & R$_{\rm opt}$ & {\it i} & 
\multicolumn{1}{c}{PA} & B$_c$ & $\Delta$B & \multicolumn{1}{c}{d$_{\rm V}$} & 
\multicolumn{1}{c}{d$_{\rm TF}$} \\
 &  &  & $'$ & $'$ & $^\circ$ & $^\circ$ & & & Mpc  & Mpc \\
\hline
\hline
1281 &    & Sc  &  2.23 &  2.59 &  90.0 &  38 &  10.95 &  0.11 &  5.4 &  5.1 \\
2459 &    & Scd &  1.24 &  2.02 &  90.0 &  62 &  ----- &  -----&  36.3& 36.3 \\
3137 &    & Sbc &  1.78 &  2.66 &  90.0 &  74 &  13.61 &  1.00 &  18.3& 33.8 \\
3909 &    & SBc &  1.32 &  1.39 &  90.0 &  82 &  13.57 &  0.77 &  17.7& 24.5 \\
4278 &    & SBc &  2.27 &  2.69 &  90.0 &  172&  11.09 &  0.23 &  10.4 & 8.1 \\
4806 &2770& Sc  &  1.83 &  2.41 &  77.0 &  148&  11.62 &  0.07 &  29.6 &21.0 \\
5452 &3118& Sc  &  1.26 &  1.41 &  90.0 &  41 &  12.88 &  1.00 &  22.1 &21.7 \\
5459 &   &  SBc &  2.38 &  2.67 &  90.0 &  132&  11.58 &  0.11 &  19.6 &15.9 \\
5986 &3432& SBd &  3.44 &  3.52 &  81.2 &  38 &  10.36 &  0.08 &  8.9 &  8.5 \\
6126 &3510& SBcd&  2.03 &  2.13 &  90.0 &  163&  11.17 &  0.33 &  8.8 &  8.8 \\
6283 &3600& Sab &  2.03 &  2.15 &  87.0 &  3 &  11.51 &  0.59 &  11.4 & 11.3 \\
6964 &4010& SBcd&  2.10 &  2.13 &  85.5 &  66&  11.79 &  0.10 &  16.6 & 16.9 \\
7089 &   &  Sc  &  1.61 &  2.73 &  87.2 &  36&  12.29 &  0.14 &  13.2 & 11.6 \\
7090 &   4096  &  SBc    	 &  3.30 &  3.96 &  80.9 &  20 &  10.16 &  0.19 &  9.4 &  10.2 \\
7125 &   &  SBd    	 &  2.35 &  2.26 &  90.0 &  85 &  12.44 &  0.92 &  19.6 &  12.6 \\
7151 &   4144  &  SBc    	 &  3.06 &  3.34 &  81.6 &  104 &  10.72 &  0.16 &  4.3 &  6.0 \\
7321 &   &  Sc     	 &  2.78 &  2.91 &  90.0 &  82 &  11.99 &  0.16 &  4.0 &  14.9 \\
7483 &   4359  &  SBc    	 &  1.75 &  2.49 &  80.8 &  108 &  12.31 &  0.80 &  22.4 &  17.6 \\
7774 &   &  Sc     	 &  1.75 &  1.90 &  90.0 &  102 &  12.98 &  0.86 &  7.3 &  20.6 \\
8246 &   &  SBc    	 &  1.71 &  1.86 &  90.0 &  83 &  13.62 &  1.00 &  11.7 &  19.4 \\
8286 &   5023  &  Sc     	 &  3.03 &  3.58 &  90.0 &  28 &  11.09 &  0.13 &  6.3 &  8.0 \\
8396 &   5107  &  SBc    	 &  0.84 &  0.90 &  77.3 &  128 &  13.89 &  1.45 &  17.4 &  27.5 \\
8550 &   5229  &  SBc    	 &  1.67 &  2.10 &  90.0 &  167 &  12.97 &  1.00 &  6.3 &  13.2 \\
8709 &   5297  &  SBbc   	 &  2.72 &  3.02 &  83.6 &  148 &  11.13 &  0.21 &  37.3 &  19.8 \\
8711 &   5301  &  SBc    	 &  2.02 &  2.49 &  87.2 &  151 &  11.98 &  0.22 &  25.8 &  22.5 \\
9242 &   &  Sc     	 &  2.51 &  3.04 &  90.0 &  71 &  11.86 &  0.07 &  24.7 &  12.6 \\
\hline
\end{tabular}
\caption{Optical properties of the galaxies in the sample: Galaxy name
  (UGC and NGC numbers), Hubble type, R$_{25}$, R$_{\rm opt}$ (see
  Sect.~\ref{ropt} for definition), inclination angle, position angle of
  the major axis, apparent b magnitude, corrected for inclination and
  galactic extinction, error in the magnitude, distance derived with a
  Virgocentric inflow model, and Tully-Fisher distance. All the data
  have been extracted from the LEDA database, except for the distances
  (see section \ref{dist}). We have not used the corrected b magnitude
  of UGC 2459 because it is not reliable due to its low galactic
  latitude.}
\label{tab:opt0}
\end{center}
\end{table*}
\end{center}


More recently, several other mechanisms have been investigated. Debattista
\& Sellwood (1999) studied the warps generated by a disk
embedded in a halo whose angular momentum vector is misaligned with
that of the disk, and Ing-Guey \& Binney (1999) built a
model of a warp caused by a disk embedded in a halo which is
constantly accreting material of angular momentum misaligned
with that of the disk. Magnetic fields have also been proposed as a
cause for warps. This hypothesis is mainly  based on the alignment of
the warps of different galaxies that are close in space
\cite{bat}. Such alignment would be hard to explain in a warping model
that is based only on the interaction of a disk and a halo of an individual
galaxy without involving the surroundings.

In spite of the several models proposed, none is fully
convincing, and many questions remain. Observationally the main
problem lies in the fact that the current sample of warped galaxies is
very small and inhomogeneous, and often 
the resolution of the data is poor. Furthermore, in
many cases the warps are inferred by modeling the HI
kinematics, using assumptions like axisymmetry and
circularity of the orbits, which may be not completely correct.

Although most of the warps start outside the optical disk,
optical warps do exist. In fact, Sanchez-Saavedra et al. (1990) claimed
that the majority of galaxies have optical warps. A  study by
Reshetnikov \& Combes (1998) on optical warps showed the influence of the 
environment, in the sense that there are more 
optically warped galaxies in dense environments than in low-density
regions. The relation between the optical and HI
warps is not clear yet. There have been claims based on near infrared
data (DIRBE) that the Galactic warp is more pronounced in 
the HI than in the old stellar population (Porcel, Battaner \&
Jim\'enez-Vicente 1996), but an early 
cutoff of the old stellar disk \cite{robin} can also explain the data. If
magnetic fields have 
some influence in exciting and/or maintaining warps, the old stellar
disk is expected to have a milder warp than the HI disk at every
radius \cite{jorge}.

Our goal in this paper is to present an analysis of a sample of late-type spiral galaxies to
determine the occurrence rate of warps, and to study their properties
(symmetry, dependence on environment, etc).
We have selected a sample of edge-on galaxies that have been observed
with the Westerbork Synthesis Radio Telescope (WSRT) for the WHISP
project \cite{whisp}. We have
used optical R-band data as well \cite{sm}, to explore possible links
between the optical properties and the warps of
our galaxies. These images also serve to study the environment of
our sample galaxies and to identify possible 
satellite dwarf galaxies that may be orbiting them.

\section{The Sample}

The WHISP sample \cite {whisp} was selected from the Uppsala
General Catalogue of Galaxies \cite{ugc}. Only the galaxies north of
20\degr declination and with blue diameters larger than 1.5$'$
were selected. The galaxies are generally larger 
in HI than in the optical, and with the typical beam size of the WSRT 
(12$''$ x 12$''$/$\sin\delta$) they are well resolved. 

In order to be sure of a sufficiently high signal-to-noise ratio, only
galaxies that had a flux density larger than 100 mJy were included in the
WHISP sample. 

From this sample of 409 galaxies, we selected those that were highly
inclined, to be able to observe the warping of the HI layer directly
instead of having to infer it from the kinematics. For this purpose,
galaxies with optical inclination angles larger than 75\degr were
selected (we assumed an intrinsic axis ratio of 0.2 to calculate the 
inclination angles). The WHISP database was not yet complete when this project
started (August 1999), thus our sample was formed from the galaxies
with data available at that time. These are listed in
Table~\ref{tab:opt0}. A key property of the sample is
that it was unbiased towards the existence of warps in the selected galaxies.

Note that nearby large galaxies, which have already
been studied in the past have not been re-observed in the WHISP
project yet, and are therefore absent from our sample. Including 
galaxies from the literature might bias the sample towards more
``interesting'' galaxies.

\section{Optical Observations and Reduction}

For most of the galaxies in our sample there are R-band optical data
obtained with the Isaac Newton Telescope at La Palma \cite{sm}. These
data serve to study the morphology of the galaxies and
to search for possible satellites which might have an influence in the
bending of the HI disks. The observing procedure and data processing
(bias, flat field, calibration) are described in Swaters
(1999). There are three galaxies for which
data from the INT could not be obtained: for UGC 2459 and UGC 8396 we used
data from the Digital Sky Survey, and data for UGC 6964 were kindly
provided by M. Verheijen (1997).  

The apparent B magnitudes for the galaxies in the sample have been obtained
from the LEDA catalogue ({\em http://leda.univ-lyon1.fr}), after
correction for inclination and Galactic extinction.

\label{ropt}

We have used two quantities to measure the extent of the optical
disk. The first one is R$_{25}$, the radius where the R band surface
brightness falls to 25 $mag/arcsec^2$, which we have taken from the
LEDA database.  

Photometric radii, and especially the not very deep ones like R$_{25}$,
are not always good indicators for the size of a disk. To
circumvent this problem, we have estimated the size of the galaxy
(R$_{\rm opt}$) by visually inspecting the images of each galaxy in the
sample and determining where the disk ends. These radii are listed in
Table~\ref{tab:opt0}. They are larger than R$_{25}$, in some cases by more
than 50\%. They are somewhat subjective but do provide a useful lower
limit to the optical size of the galaxy. 

\section{HI Observations and Analysis}

The HI observations were obtained with the WSRT between 1995 and
1998. Most galaxies were observed for 12 hours. The typical beam size
is 12$''$ x 12$''$/$\sin\delta$, the 
channel separation 4.14 km/s for most of the galaxies, and 16.7 km/s for
the most distant ones. The observational parameters, exposure time,
resulting beam size and other relevant data are given in
Table~\ref{tab:hidata}. The primary beam is about 37$'$.

The reduction was done using the WHISP reduction pipeline. The
detailed information about the reduction steps can be found in the
WHISP web pages, http://www.astro.rug.nl/$\sim$whisp. This results in
3 sets of channel maps for each galaxy: at full resolution,
at 30$''$ and at 60$''$.          

The HI data presented in this article in Fig. \ref{fig:plotwl2}
(rotation curves, HI radial surface density profiles and warp curves)
and in Fig. \ref{atlas} (6 panel figures for each galaxy) is also
available in electronic form in the WHISP pages
(http://www.astro.rug.nl/$\sim$whisp), together with some other
diagrams that supply extra velocity information.

\subsection{Global HI line profiles, HI masses and total HI maps}

The global HI line profiles were determined by adding the intensities of the
CLEAN components from the 60$''$ Hanning smoothed datacube, corrected for
primary beam attenuation. 

The total HI masses were derived according to the formula
\begin{equation}
M_{\rm{HI}}=2.36\cdot 10^5 D^2 \int{SdV}
\end{equation}

where $M_{\rm{HI}}$ is the total mass in $M_{\odot}$, $D$ is the distance to
the galaxy in Mpc, $S$ is the flux density in Jy and the integral is
made over all the velocities in km s$^{-1}$.

\label{totalhi}

To construct the total (integrated over velocity) HI maps, we used the
same masks already used in the CLEANing to define the areas of
emission. Outside these areas the maps were set to zero to avoid adding
unnecessary noise to the total HI map. Then all the masked channels
were added, creating the column density map. 

\begin{table*}
\begin{tabular}{ccccclrcccrcc}
\hline
Galaxy & T$_{obs}$ & \multicolumn{1}{c}{Date} & R.A. & Dec & \multicolumn{1}{c}{f$_{cen}$} & V$_{hel}$ & S. beam & Bw & N$_{ch}$ & res$_{vel}$ & \multicolumn{1}{c}{rms} & T$_B$/S \\
UGC & hours & & (1950) & (1950) & \multicolumn{1}{c}{MHz} & km/s & $\alpha$ x $\delta$ (") & MHz & & km/s & \multicolumn{1}{c}{mJy/beam} & K/mJy \\
(1) & (2) & \multicolumn{1}{c}{(3)} & (4) & (5) & \multicolumn{1}{c}{(6)} & (7) & (8) & (9) & (10) & (11) & (12) & (13)    \\
\hline\hline
1281 & 12 & 08 Nov 95 & 01 46 39 & 32 20 39 & 1419.79 &      157 & 11.14 x 20.96 & 2.48 & 127 & 5.0 & 3.2 & 2.6 \\
2459 & 12 & 04 Jun 97 & 02 57 06 & 48 49 59 & 1408.75 &     2464 & 13.08 x 17.17 & 4.92 &  63 & 20.1 & 2.1 & 2.7 \\
3137 & 12 & 24 Jan 96 & 04 39 24 & 76 19 59 & 1415.64 &     1020 & 11.70 x 11.88 & 2.48 & 127 & 5.0 & 2.9 & 4.4 \\
3909 & 12 & 11 Jun 97 & 07 30 52 & 73 49 27 & 1415.86 &      945 & 11.38 x 11.83 & 2.48 & 127 & 5.0 & 3.3 & 4.5 \\
4278 & 12 & 01 Apr 96 & 08 10 27 & 45 53 49 & 1417.68 &      581 & 12.05 x 17.23 & 2.48 & 127 & 5.0 & 2.8 & 2.9 \\
4806 & 12 & 16 Jun 97 & 09 06 29 & 33 19 37 & 1411.09 &     1951 & 12.57 x 23.85 & 4.92 &  63 & 20.0 & 1.8 & 2.0 \\
5452 & 24 & 23 Jun 97 & 10 04 17 & 33 16 27 & 1413.95 &     1342 & 12.06 x 22.43 & 2.48 & 127 & 5.0 & 2.5 & 2.3 \\
5459 & 12 & 24 Nov 97 & 10 04 54 & 53 19 36 & 1415.15 &     1111 & 14.39 x 17.89 & 2.48 & 127 & 5.0 & 4.1 & 2.4 \\
5986 & 12 & 09 Feb 96 & 10 49 42 & 36 53 04 & 1417.46 &      616 & 11.34 x 19.29 & 2.48 & 127 & 5.0 & 3.3 & 2.8 \\
6126 & 36 & 17 Nov 97 & 11 01 01 & 29 09 19 & 1416.99 &      705 & 12.47 x 27.34 & 2.48 & 127 & 5.0 & 2.6 & 1.8 \\
6283 & 12 & 19 Sep 97 & 11 13 06 & 41 51 48 & 1416.90 &      719 & 12.50 x 18.73 & 2.48 & 127 & 5.0 & 3.0 & 2.6 \\
6964 & 12 & 12 Mar 98 & 11 56 03 & 47 32 20 & 1416.11 &      907 & 10.24 x 13.29 & 2.48 & 127 & 5.0 & 1.8 & 4.5 \\
7089 & 12 & 27 Oct 97 & 12 03 26 & 43 25 37 & 1416.67 &      776 & 12.87 x 18.76 & 2.48 & 127 & 5.0 & 3.3 & 2.5 \\
7090 & 12 & 03 Nov 97 & 12 03 28 & 47 45 12 & 1417.64 &      566 & 12.48 x 17.15 & 4.92 &  63 & 19.9 & 1.7 & 2.8 \\
7125 & 12 & 29 Nov 96 & 12 06 10 & 37 04 51 & 1415.32 &     1071 & 12.32 x 20.54 & 2.48 & 127 & 5.0 & 3.3 & 2.4 \\
7151 & 12 & 02 May 97 & 12 07 28 & 46 44 07 & 1419.13 &      267 & 12.24 x 17.03 & 2.48 & 127 & 5.0 & 3.2 & 2.9 \\
7321 & 13 & 10 Nov 97 & 12 15 02 & 22 49 00 & 1418.39 &      409 & 12.14 x 32.76 & 2.48 & 127 & 5.0 & 3.5 & 1.5 \\
7483 & 12 & 12 Nov 97 & 12 21 42 & 31 47 57 & 1414.39 &     1253 & 11.88 x 24.21 & 2.48 & 127 & 5.0 & 3.4 & 2.1 \\
7774 & 24 & 10 Dec 97 & 12 33 57 & 40 16 49 & 1417.83 &      526 & 10.67 x 16.93 & 2.48 & 127 & 5.0 & 2.4 & 3.4 \\
8246 & 12 & 22 Dec 97 & 13 07 44 & 34 26 48 & 1416.49 &      813 & 12.05 x 22.71 & 2.48 & 127 & 5.0 & 3.1 & 2.2 \\
8286 & 24 & 20 Dec 96 & 13 09 58 & 44 18 13 & 1418.40 &      407 & 12.30 x 18.11 & 2.48 & 127 & 5.0 & 2.3 & 2.7 \\
8396 & 12 & 05 Jan 98 & 13 19 09 & 38 47 58 & 1415.86 &      946 & 12.15 x 20.27 & 2.48 & 127 & 5.0 & 3.5 & 2.5 \\
8550 & 11 & 12 Dec 96 & 13 31 58 & 48 10 16 & 1418.61 &      364 & 12.88 x 16.75 & 2.48 & 127 & 5.0 & 3.6 & 2.8 \\
8709 & 12 & 09 Jan 98 & 13 44 19 & 44 07 23 & 1409.03 &     2407 & 12.56 x 18.56 & 4.92 &  63 & 20.1 & 1.7 & 2.6 \\
8711 & 12 & 12 Jan 98 & 13 44 21 & 46 21 28 & 1413.26 &     1503 & 10.87 x 15.04 & 4.92 &  63 & 20.0 & 1.7 & 3.7 \\
9242 & 12 & 26 Nov 98 & 14 23 59 & 39 45 00 & 1413.40 &     1470 & 9.83 x 15.28 & 2.48 & 127 & 5.0 & 3.1 & 4.1 \\
\hline
\end{tabular}
\caption{Observation parameters for the HI synthesis data: Galaxy UGC
  number (1), Length of the observation in hours (2), Date of
  observation (3), RA (4) and DEC (5) of the field center, central
  observing frequency (6), corresponding heliocentric velocity (7),
  synthesized beam in both RA and
  DEC directions (8), total bandwidth (9), number of channels (10),
  velocity resolution (11), rms noise of the resulting datacube (12),
  conversion factor from flux density to brightness temperature (13). }
\label{tab:hidata}
\end{table*}

This is a standard procedure which allows a higher S/N ratio to be
obtained, but it has the disadvantage 
that the noise level is not the same everywhere across the map. This
is caused by the fact that at each position of the map a different
number of channels are added. We can calculate the noise at
each position as  

\begin{equation}
\sigma_{\rm tot}=n_l^{0.5}\sigma_{\rm cs}\sqrt{(n_{\rm c}+n_{\rm l})
    /(n_{\rm c}-1)}
\end{equation}

where $\sigma_{\rm tot}$ is the noise in the total HI map and $\sigma_{\rm cs}$
is the noise in the fit to the line-free channels used to
form the continuum. The number of channels contributing to the 
total HI map is given by $n_{\rm l}$ and the number of continuum channels by
$n_{\rm c}$ (see Verheijen \& Sancisi (2001) for the derivation of this
formula). This expression is used to calculate the noise at each position  
in the total HI map, and to construct the map with the S/N
ratios. We then calculated the average noise value of all 
the points in this map which had 2.75 $< ({S\over{N}} ) <$ 3.25, and
adopted this as the '3$\sigma$' contour level. This threshold is
indicated by a thick line in the total HI maps of the Atlas (see
Fig. \ref{atlas}).  

Finally, we corrected the map for primary beam attenuation.

The global HI profiles, total HI maps, and position-velocity (XV)
diagrams along the major axis of each galaxy are shown in
Fig.~\ref{atlas}.

\subsection{Distance determination}
\label{dist}
In order to determine the distance to each galaxy in the sample we
first tried a simple Virgocentric inflow correction, as described in
Kraan-Korteweg (1986) with a 
Hubble constant of $H_{\circ} =$ 75 km s$^{-1}$ Mpc$^{-1}$. 
\begin{figure}
\resizebox{\hsize}{!}{\includegraphics{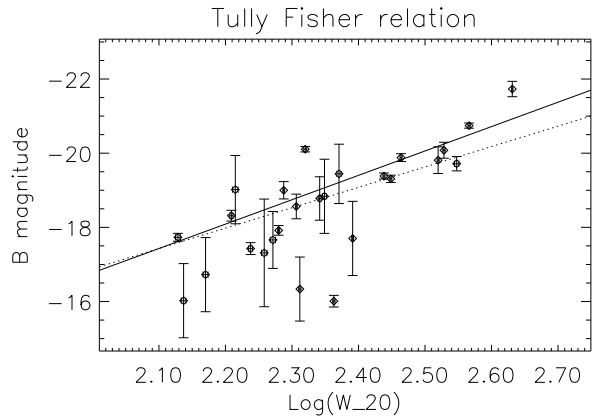}}
\caption{B band Tully-Fisher relation for the galaxies in the
  sample. We have taken apparent B magnitudes from the LEDA database,
  and converted them to absolute magnitudes with the distance
  determined as explained in section \ref{dist}. The dotted line is the TF
  from Verheijen (1997), the solid line indicates the
  fit to the data.}
\label{fig:tf}
\end{figure}

Fig.~\ref{fig:tf} shows the Tully-Fisher (TF) relation for our
sample obtained using the derived distances. We took the corrected apparent
magnitudes from the LEDA catalogue, and determined the absolute
magnitudes with the ``virgocentric'' distances. The widths of the
profiles were corrected for instrumental 
broadening (following Bottinelli et al. 1990) as well as for
inclination. The dotted line in Fig.~\ref{fig:tf} is the TF relation
derived by 
Verheijen (1997) for the B band. Some points do clearly
follow the relation, but there are many outliers. Some of the
deviations are greater than 1.5 mag. (a factor of two in
distance). There are various indications that the
Virgocentric inflow model may not give the correct distances: the
first is that when one looks at the distribution of galaxies in
supergalactic coordinates,
the galaxies with large discrepancies are not randomly distributed on
the sky, but grouped together. Another indication that there could be
problems with such simple, spherical virgocentric inflow model is the fact that the
Virgo Cluster may be far from spherical
\cite{virgo}. This will probably affect quite strongly the distance
determination for nearby galaxies.

We have, therefore, decided to calculate the distances using the TF
relation. The estimated distance uncertainty introduced by this is
expected to be 30\% (0.6 mag). Of course, galaxies for
which the apparent magnitude already has a large error will have a
more uncertain distance.

\subsection{Determination of the center of the galaxies}

It is necessary to know the position of the center of a galaxy to be
able to determine its rotation curve, its lopsidedness and the warp
asymmetry. In an axisymmetric galaxy with
gas in circular orbits this location is at the same time the center of the
potential, the center of the gas orbits, and the point of maximum
optical emission (if dust absorption is not very important). In the
case of an asymmetric galaxy, the center can be determined in several
ways:

If the galaxy has a nuclear radio continuum source, this marks
the position of the center with great accuracy. But this is rare. 

The center may also be derived from ellipse
fitting to the optical iso\-photes. This method has the problem that it
is very sensitive to dust, and can lead to large errors. Similarly,
ellipses can be fitted to the total HI maps, but since the HI
often has a lot of substructure such fits are unreliable. 
 
In both HI and optical ellipse fitting, one gets a set of ellipses (thus,
a set of centers) and has to extract
a single center from it. Sometimes this is easy and the weighted mean
of all those centers is a good measure of the 'true' center, but when
the galaxy is lopsided this is no longer the case and a
linear fit to the centers derived from the fits is necessary to find the
center of the galaxy. 


Another method is based on the position-velocity (XV)
diagram along the major axis of the galaxy and consists in finding the
center that minimizes the asymmetries of the XV diagram. In this way
one can be sure that the derived asymmetries are a lower limit to
the true asymmetries of the galaxy. Density and kinematic
information are thus combined in this method, which uses the whole XV
diagram and not only the rotation curve or only the density profile.
The center is determined in the following way: 
\begin{enumerate}
\item
Symmetrization of the XV diagram: The XV diagram is mirrored with
respect to an initially adopted center, and 
added to the original XV diagram. In this way we create a symmetric image that
will resemble the original one if the center is correct and the
galaxy is symmetric.

\item
We cross-correlate the symmetrized XV with the original XV and find
where the peak of the resulting image lies. This gives us the distance
and direction we need to move our center from the previously adopted one.

\item
Steps (i) and (ii) are repeated until convergence is achieved. The meth\-od
has proved to be very robust and converges in very few iterations.
\end{enumerate}

The centers used in this article were derived using a mixed approach:
first, the center was determined using the ellipse-fitting method to
the optical images. To determine the position angle we used both
the HI and optical ellipse fits. If the optical position angles gave a
consistent result, we chose that one. If, instead, the ellipses were
strongly influenced by dust or bright star wings, then we chose the
position angle determined from the HI ellipse fits. 

With a position angle and a center, we extracted the XV
diagram along the major axis defined by these quantities. At this
point we applied the ``XV-method'' to improve the determination of the
position of the center. Table~\ref{tab:koencen} lists the positions of
the centers
obtained with this method as well as those listed in LEDA. The
distances between both centers are usually not larger than 4$''$ and
there is only one case exceeding 10$''$ found for UGC 5986, a
very distorted galaxy both kinematically and morphologically.

\begin{center}
\begin{table*}
\begin{center}
\begin{tabular}{cccccc}
\hline
UGC & RA$_{2000}$ LEDA & DEC$_{2000}$ LEDA & RA$_{2000}$ XV &DEC$_{2000}$ XV & dist ($''$) \\
\hline
\hline
1281 & 01 49 32.0 & 32 35 21.5 & 01 49 31.7 & 32 35 16.2 & 6.2  \\
2459 & 03 00 37.1 & 49 02 41.3 & 03 00 36.5 & 49 02 34.2 & 9.0  \\
3137 & 04 46 16.0 & 76 25 07.2 & 04 46 15.5 & 76 25 06.4 & 2.1  \\
3909 & 07 36 59.1 & 73 42 50.1 & 07 36 58.8 & 73 42 48.9 & 1.7  \\
4278 & 08 13 58.8 & 45 44 35.2 & 08 13 59.0 & 45 44 38.0 & 3.7  \\
4806 & 09 09 33.5 & 33 07 29.4 & 09 09 33.8 & 33 07 25.1 & 5.8  \\
5452 & 10 07 11.5 & 33 01 38.4 & 10 07 11.6 & 33 01 39.3 & 2.0  \\
5459 & 10 08 10.1 & 53 04 58.4 & 10 08 10.3 & 53 04 58.9 & 2.2  \\
5986 & 10 52 30.9 & 36 37 03.4 & 10 52 31.7 & 36 37 16.3 & 17.  \\
6126 & 11 03 43.6 & 28 53 05.6 & 11 03 43.6 & 28 53 07.3 & 1.7  \\
6283 & 11 15 52.1 & 41 35 32.2 & 11 15 52.1 & 41 35 27.6 & 4.7  \\
6964 & 11 58 37.1 & 47 15 37.1 & 11 58 37.1 & 47 15 35.1 & 2.1  \\
7089 & 12 05 57.7 & 43 08 35.3 & 12 05 58.0 & 43 08 37.2 & 3.6  \\
7090 & 12 06 01.5 & 47 28 47.6 & 12 06 01.2 & 47 28 40.4 & 7.9  \\
7125 & 12 08 42.2 & 36 48 07.8 & 12 08 42.3 & 36 48 08.5 & 1.1  \\
7151 & 12 09 58.7 & 46 27 26.5 & 12 09 58.1 & 46 27 27.8 & 6.6  \\
7321 & 12 17 33.9 & 22 32 25.0 & 12 17 34.0 & 22 32 25.1 & 1.5  \\
7483 & 12 24 11.4 & 31 31 17.5 & 12 24 11.8 & 31 31 14.9 & 5.1  \\
7774 & 12 36 23.1 & 40 00 17.8 & 12 36 23.0 & 40 00 18.4 & 0.8  \\
8246 & 13 10 04.2 & 34 10 52.5 & 13 10 04.4 & 34 10 49.8 & 3.3  \\
8286 & 13 12 11.8 & 44 02 14.8 & 13 12 11.7 & 44 02 13.2 & 1.8  \\
8396 & 13 21 24.7 & 38 32 15.7 & 13 21 24.9 & 38 32 16.9 & 2.6  \\
8550 & 13 34 03.1 & 47 54 45.2 & 13 34 03.0 & 47 54 47.3 & 2.6  \\
8709 & 13 46 23.8 & 43 52 20.3 & 13 46 23.8 & 43 52 18.9 & 1.4  \\
8711 & 13 46 24.7 & 46 06 28.2 & 13 46 24.6 & 46 06 24.6 & 3.8  \\
9242 & 14 25 20.8 & 39 32 21.1 & 14 25 20.6 & 39 32 19.4 & 2.8  \\
\hline
\end{tabular}
\caption{Comparison between the positions of the centers listed in
  LEDA (columns 2,3) and the ones adopted in this work (columns
  4,5). The distance between the two centers is listed in column 6.} 
\label{tab:koencen}
\end{center}
\end{table*}
\end{center}

\subsection{HI profile and radius}
\label{hiprof}

In highly inclined systems, each line of sight to the galaxy contains
information about different portions of that galaxy. At each position
(see Fig.~\ref{fig:xvedge}) we get the
integrated signal coming from different radii, and we have to 
disentangle that information to get the radial density
profile. 

This has been done following the procedure developed by Warmels
(1988). First, all the emission from the galaxy is
integrated in the direction perpendicular to the major axis, resulting in
an HI strip integral. This is then deconvolved using the Lucy
(1974) deconvolution scheme, assuming that the HI
distribution is axisymmetric, and taking into account the size of the
beam. This algorithm has been applied separately to the two sides of the
galaxy. This allows us to estimate the density lopsidedness in the HI
gas by comparing one side with the other.

The HI radius (R$_{\rm HI}$) has been calculated using the average of the
two deconvolved profiles, and is defined as the radius where the HI surface density
drops to $1 M_{\odot}/pc^2$.

\subsection{Rotation Curves}
\label{rc}
The determination of a rotation curve for a highly inclined galaxy is
not simple. The most satisfactory approach would be to model the
kinematics of the galaxy as a whole \cite{rob}, but in an edge-on galaxy there
are many parameters to be fitted: the radial HI profile, the warp
profile and its line of nodes, and the rotation curve. Furthermore we are
also trying to determine deviations from axisymmetry. 

\begin{figure}
\plotfull{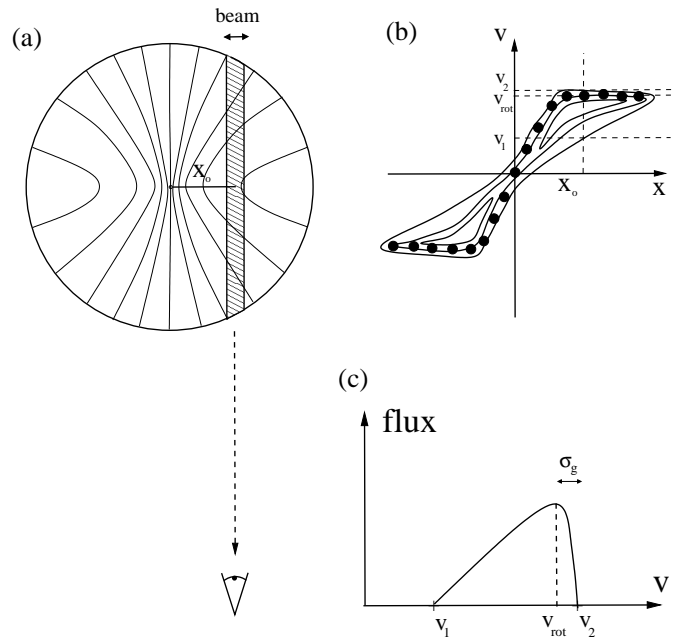}
\caption{Schematic representation of the HI detected
  at each position along the major axis in an edge-on galaxy. (a) At a
  certain distance $x_{\circ}$ 
  from the center of the galaxy, all the emission from gas in the dashed
  area is detected by the telescope. Isovelocity lines (with respect
  to the observer) have been drawn in the figure. The XV diagram along
  the major axis with the rotation curve (filled dots) is shown 
  in (b), and a velocity profile at a certain major axis distance ($x_{\circ}$)
  in (c). The rotation velocity at $x_{\circ}$ is indicated by
  $v_{\rm rot}$, where $\sigma_{\rm }g$ is the broadening due to instrumental
  and random motion effects.} 
\label{fig:xvedge}
\end{figure}

For these reasons, we have focused on the derivation of the rotation
curve from the position-velocity (XV) diagram along the major axis of
the galaxy. In 
galaxies with low inclinations, the beam samples the HI emission from a
small portion of the disk, and the profiles in the XV diagram are
basically Gaussian, unless beam-smearing plays an important role. In
a completely edge-on galaxy, however, the beam intercepts a large
portion of the disk, as illustrated in Fig.~\ref{fig:xvedge}. In this
case, the derivation of the rotation curve can be done as outlined in
Sancisi \& Allen (1979). At each position along the major
axis the maximum 
rotation velocity (relative to the systemic) is chosen, corrected for
the instrumental broadening and random motions of the gas. This method
assumes that there is gas everywhere along the line of sight, and may
give the wrong rotation velocity if this is not the case.

We have determined the velocity at each position along the major
axis by fitting half a Gaussian to the edge of the profile at
that position. The width of the Gaussian is held fixed with a value
of $\sigma_{\rm g}=\sqrt{\sigma_{\rm instr}^2+\sigma_{\rm rnd}^2}$, 
where $\sigma_{\rm instr}$
is the instrumental broadening and $\sigma_{\rm rnd}$ is the velocity
dispersion of the gas due to the random motions in the
disk. Measurements of velocity dispersion of galaxies in HI give
typical values that range from $\sim$ 10-12 km s$^{-1}$ in the
inner parts to $\sim$ 6 km s$^{-1}$ in the outer parts
\cite{kamphuis,di}. Here we have adopted a constant value of 8 km
s$^{-1}$ through the whole HI disk, which seems a reasonable assumption.

\begin{figure}
\plotfull{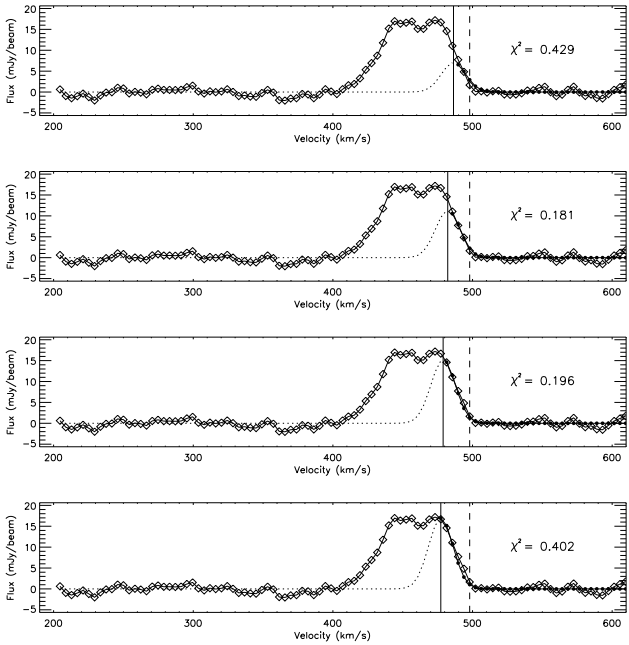}
\caption{Edge-fitting method. A profile of UGC 8286 is
  plotted, along with 4 different fits. The data are displayed as
  diamonds, and the fit as filled circles and dotted line. The
  dotted line shows the part of the Gaussian that is not used in the fit.
  From top to bottom, the fits
  make use of 2,3,4 and 5 'signal' points (points left of the vertical
  dashed line). Each panel also shows the reduced $\chi^2$ for each
  fit, and the resulting velocity (solid vertical line). 
  The final parameters are taken from
  the best fit (in this case, the second panel from the top).}
\label{fig:rotsample}
\end{figure}

Fig.~\ref{fig:rotsample} illustrates how the rotational velocity at
each radius is determined. We extract a velocity profile (see
Fig.~\ref{fig:xvedge}) at a position along the major axis, and determine
where the emission from the galaxy begins in velocity. This is done by
looking where in the profile $n_{\rm p}$ points lie above
3$\sigma$. After some experimentation we have adopted $n_{\rm }p=$\,3. At this
moment we make a series of Gaussian fits to the edge 
of the profile, varying the amount of signal points included in the fit (see
Fig.~\ref{fig:rotsample}). The rotation velocity will be the center
of the Gaussian with a smaller reduced $\chi^2$ of all
the fits. 

The error in the determined rotation velocity is calculated from
the amplitude of the fitted Gaussian. This error also depends on the
velocity resolution compared to the dispersion of the Gaussian, and on
the dispersion of the Gaussian itself. After extensive Monte-Carlo
simulations we derived the following formula to calculate the errors
for our edge-fitting method:

\begin{equation}
\sigma_{\rm v}={4.0\sigma_{\rm g}\over\sqrt{n_{\rm ps}}({S\over{N}})}
\end{equation}

where $\sigma_{\rm v}$ is the error in the rotation curve velocity,
$\sigma_{\rm g}$ is the dispersion of the Gaussian, $n_{\rm ps}$ is the number
of points every $\sigma_{\rm g}$ (sampling factor), and $S/N$ is the ratio
of the amplitude of the Gaussian to the noise level. This formula is valid
for $n_{\rm ps} \ga$ 2, and the error increases less than a 60\% for
$n_{\rm ps}\simeq$ 1. The typical velocity resolution of our observations
has a FWHM of 4 km\hspace{0.5mm}s$^{-1}$, so $n_{\rm ps}\simeq$ 2.

A rotation curve derived by making edge-fits in the velocity direction
as described above still suffers from the effects of beam
smearing. The
influence of this is to overestimate the velocities in the
inner rising parts of the rotation curve and in some galaxies the
effect is quite severe. To correct for this we
have performed edge-fits in the horizontal direction as well. In these
fits the width of the Gaussian depends on the slope of the
rotation curve, velocity resolution and the beam width. We determined
this width by fitting a full Gaussian around the center of the galaxy.

At this stage we have two determinations of the rotation curve, one
from the edge-fits in the velocity direction and the other from
those in the  spatial direction. The final rotation curve is the
minimum of the two determinations. 

To assess that the errors are realistic in the rotation curve for each
radius and correct artificial 'bumps' created by noise in the edge-fits, we determined
the mean velocity of the data within a beam size and its 
dispersion. The mean value is the finally adopted rotation velocity
and the error is the sum in quadrature of the formal error and the
dispersion.


The rotation velocities were derived separately for the approaching
and receding sides, to be able to compare them and to look for kinematic
asymmetries. 

We would also like to note that a number of galaxies (UGC 5986, 8246,
9242, see XV diagrams in Fig.~\ref{atlas}) have non-circular motions in the central
regions. This suggests either that orbits in these systems are
elliptical, or that there are radial motions of the gas, and caution
against the use of our rotation curves as tracers of the inner potential wells of these
galaxies. 

\subsection{Warping}
\label{warpcurve}

We have derived the curve of the warp from the total intensity maps,
by fitting Gaussians to the density profiles parallel to the minor axis 
of the galaxy. We thereby determine the position of the HI ridge at
each radius. This underestimates warps that have a line of nodes
significantly different from our line of sight. For example, the ridge
of the HI could warp very little, but there may be a low level envelope that
bends with greater amplitude, but that would not show up in a
Gaussian fit. Thus, the numbers we get from this procedure will have to be
treated as lower limits for the warp amplitude. 

To characterize this curve by a number, we have, first of all,
determined the warp starting radius ($r_{\rm warp}$). This is the radius
where the warp curve leaves significantly the plane defined 
by the inner parts of the disk. Many factors (as small corrugations and
noise in the warp curve) can hamper the determination of the warp
radius. To overcome these we have devised a complex procedure which
smoothes the data at different resolutions and takes the scatter in
the inner disk into account (see Appendix \ref{rwarp}  for
full details). The warp radius for one side of a galaxy (UGC 6126) was
strongly affected by the presence of spiral arms. We manually set the
warp radius of that side to that of the other side of the galaxy.
Once the warp radius has been determined, a straight
line is fitted to the points with $r > r_{\rm warp}$. With this we
calculate the warp angle according to the following definition (see
Fig.~\ref{fig:warpangle}): 

\begin{figure}
\plotfull{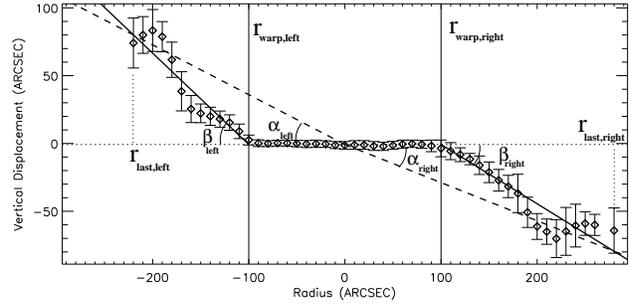}
\caption{Determination of the warp angles $\alpha_i$, for 
  $i$= left, right. $r_{{\rm warp},i}$ is where 
  the warp begins, $r_{{\rm last},i}$ the projected distance along the major
  axis to the last measured point, and $\beta_i$ the angle between the
  major axis and the straight line resulting from the fit to the
  points between those radii.}
\label{fig:warpangle}
\end{figure}

\begin{equation}
\tan(\alpha)=\tan(\beta) (r_{\rm last}-r_{\rm warp})/r_{\rm last}
\end{equation}

where $\beta$ is the angle resulting from the mentioned fit and
$r_{{\rm last}}$ is the maximum radius at which we can measure the centroid
of the HI ridge. Wiggles in the warp curve (caused by not completely
edge-on spiral arms, or by warps that bend first to one side and then to
the other) can cause important differences in the measured $r_{\rm warp}$,
but we have found that our measure of the warping angle ($\alpha$) is
quite robust to them. 

Even though for most galaxies the warps are well represented by this
angle, there are 
some that are not well fitted by a straight line from $r_{\rm warp}$ to
$r_{\rm last}$, either because they rise faster or because the gas layer
bends back towards the plane defined by the inner disk, as is the case
of the Galaxy \cite{bu}.

The warping angle is calculated separately for both sides of the galaxy, and in
this way we can estimate the asymmetry of the warp using the
difference between one side and the other. Most of the warps (all
except 2 systems that are strongly interacting) have S-shape, i.e.
the gas layer bends in opposite directions on either
side of the galaxy. Taking into account that the warp angles are
measured anticlockwise and that a symmetric S-shape warp would have
$\alpha_{\rm left}=-\alpha_{\rm right}$ we have defined the warp asymmetry as:

\begin{equation}
\alpha_{\rm asym}=|\alpha_{\rm right}+\alpha_{\rm left}|
\end{equation}

\begin{center}
\begin{table*}
\begin{center}
\begin{tabular}{crrcrrrrrrlrlc}
\hline
UGC   & \multicolumn{1}{c}{Lop$_{kin}$} & \multicolumn{1}{c}{Lop$_{\rho}$} & R$_{\rm HI}$ & \multicolumn{1}{c}{M$_{\rm HI}$} & \multicolumn{1}{c}{V$_{sys}$} & \multicolumn{1}{c}{W$_{20}$} & \multicolumn{1}{c}{W$_{50}$} & \multicolumn{1}{c}{PA} & \multicolumn{2}{c}{Warp$_1$} & \multicolumn{2}{c}{Warp$_2$}& Env \\
 & \multicolumn{1}{c}{\%} & \multicolumn{1}{c}{\%} & \multicolumn{1}{c}{$'$} & \multicolumn{1}{c}{$10^8$M$_\odot$} & \multicolumn{1}{c}{km/s} & \multicolumn{1}{c}{km/s} & \multicolumn{1}{c}{km/s} & \multicolumn{1}{c}{\degr} & \multicolumn{2}{c}{degree} & \multicolumn{2}{c}{degree} & \\
(1) & \multicolumn{1}{c}{(2)} & \multicolumn{1}{c}{(3)} & (4) & \multicolumn{1}{c}{(5)} & \multicolumn{1}{c}{(6)} & \multicolumn{1}{c}{(7)} & \multicolumn{1}{c}{(8)} & \multicolumn{1}{c}{(9)} & \multicolumn{2}{c}{(10)} & \multicolumn{2}{c}{(11)} & (12) \\
\hline
\hline
1281 & 3.1 & 0.0 & 3.40 & 1.2 & 157 & 132.4 & 123.3 & 39 & 5.3$\hspace{2.5mm}\pm$ &0.7 & -0.7$\hspace{2.5mm}\pm$ &0.7 & 0 \\
2459 & 2.2 & 7.4 & 3.87 & 134.1 & 2465 & 335.5 & 324.7 & 64 & -4.4$\hspace{2.5mm}\pm$ &0.6 & 1.4$\hspace{2.5mm}\pm$ &1.1 & 2 \\
3137 & 6.0 & 4.0 & 3.53 & 32.2 & 992 & 243.3 & 232.2 & 74 & -2.7$\hspace{2.5mm}\pm$ &0.5 & 3.3$\hspace{2.5mm}\pm$ &0.7 & 0 \\
3909 & 1.7 & 0.4 & 2.29 & 10.4 & 943 & 183.6 & 167.6 & 80 & -4.9$\hspace{2.5mm}\pm$ &2.6 & 8.8$\hspace{2.5mm}\pm$ &1.7 & 0 \\
4278 & 7.2 & 1.8 & 3.13 & 10.2 & 558 & 192.8 & 173.5 & 173 & \multicolumn{2}{c}{---------} & -2.9$\hspace{2.5mm}\pm$ &0.6 & 2 \\
4806 & 1.2 & 0.3 & 2.33 & 72.1 & 1950 & 351.6 & 328.1 & 148 & 3.2$\hspace{2.5mm}\pm$ &1.4 & \multicolumn{2}{c}{---------} & 2 \\
5452 & 3.1 & 9.3 & 1.94 & 23.3 & 1340 & 220.8 & 203.0 & 38 & -4.7$\hspace{2.5mm}\pm$ &2.1 & 17.4$\hspace{2.5mm}\pm$ &2.3 & 2 \\
5459 & 9.2 & 3.0 & 3.58 & 37.3 & 1108 & 287.6 & 269.0 & 131 & -8.1$\hspace{2.5mm}\pm$ &3.1 & 3.1$\hspace{2.5mm}\pm$ &0.9 & 1 \\
5986 & 5.4 & 6.9 & 5.29 & 27.1 & 624 & 265.6 & 243.1 & 38 & 4.6$\hspace{2.5mm}\pm$ &0.7 & 22.4$\hspace{2.5mm}\pm$ &3.0 & 2 \\
6126 & 3.7 & 11.6 & 3.07 & 10.3 & 701 & 198.7 & 184.8 & 166 & -18.6$\hspace{2.5mm}\pm$ &1.1 & 26.5$\hspace{2.5mm}\pm$ &1.1 & 2 \\
6283 & 2.9 & 0.4 & 3.48 & 15.1 & 715 & 214.9 & 201.8 & 8 & 6.1$\hspace{2.5mm}\pm$ &1.8 & -3.5$\hspace{2.5mm}\pm$ &1.3 & 0 \\
6964 & 5.4 & 3.2 & 2.83 & 19.0 & 906 & 274.4 & 262.7 & 65 & 20.2$\hspace{2.5mm}\pm$ &1.3 & -15.7$\hspace{2.5mm}\pm$ &0.8 & 1 \\
7089 & 13.4 & 15.1 & 2.13 & 5.8 & 782 & 158.3 & 141.7 & 36 & \multicolumn{2}{c}{---------} & \multicolumn{2}{c}{---------} & 2 \\
7090 & 9.8 & 10.2 & 3.82 & 12.9 & 575 & 341.3 & 312.8 & 20 & \multicolumn{2}{c}{---------} & \multicolumn{2}{c}{---------} & 2 \\
7125 & 5.6 & 9.5 & 3.70 & 34.2 & 1079 & 161.9 & 147.6 & 83 & \multicolumn{2}{c}{---------} & 5.9$\hspace{2.5mm}\pm$ &0.6 & 2 \\
7151 & 8.6 & 0.4 & 3.26 & 1.9 & 265 & 167.6 & 157.0 & 102 & \multicolumn{2}{c}{---------} & \multicolumn{2}{c}{---------} & 1 \\
7321 & 2.4 & 1.5 & 3.23 & 2.4 & 407 & 230.2 & 220.9 & 82 & \multicolumn{2}{c}{---------} & -2.6$\hspace{2.5mm}\pm$ &0.8 & 0 \\
7483 & 2.5 & 2.7 & 2.47 & 21.1 & 1247 & 227.1 & 198.8 & 108 & \multicolumn{2}{c}{---------} & \multicolumn{2}{c}{---------} & 2 \\
7774 & 5.2 & 1.9 & 2.72 & 3.4 & 522 & 203.3 & 190.3 & 102 & 15.2$\hspace{2.5mm}\pm$ &2.5 & -32.8$\hspace{2.5mm}\pm$ &2.7 & 1 \\
8246 & 3.9 & 1.8 & 2.19 & 4.8 & 809 & 145.4 & 134.6 & 83 & \multicolumn{2}{c}{---------} & -17.0$\hspace{2.5mm}\pm$ &2.5 & 2 \\
8286 & 2.1 & 0.6 & 4.25 & 5.4 & 406 & 188.9 & 179.0 & 28 & -2.3$\hspace{2.5mm}\pm$ &0.5 & 4.6$\hspace{2.5mm}\pm$ &1.0 & 0 \\
8396 & 7.4 & 4.9 & 1.78 & 9.8 & 951 & 173.3 & 142.2 & 128 & \multicolumn{2}{c}{---------} & 23.9$\hspace{2.5mm}\pm$ &6.7 & 2 \\
8550 & 1.9 & 3.6 & 2.85 & 2.0 & 359 & 135.4 & 124.4 & 167 & 2.5$\hspace{2.5mm}\pm$ &1.0 & \multicolumn{2}{c}{---------} & 1 \\
8709 & 4.4 & 9.5 & 3.46 & 174.2 & 2410 & 416.5 & 399.1 & 148 & \multicolumn{2}{c}{---------} & \multicolumn{2}{c}{---------} & 2 \\
8711 & 3.9 & 0.7 & 3.10 & 53.8 & 1506 & 330.5 & 311.2 & 151 & -16.4$\hspace{2.5mm}\pm$ &2.7 & 8.7$\hspace{2.5mm}\pm$ &1.7 & 0 \\
9242 & 6.8 & 2.9 & 3.13 & 31.0 & 1439 & 208.5 & 195.1 & 72 & \multicolumn{2}{c}{---------} & \multicolumn{2}{c}{---------} & 1 \\
\hline
\end{tabular}
\caption{Results from the HI analysis: Galaxy UGC number (1),
  kinematical lopsidedness (2), density lopsidedness (3), HI radius
  (at 1 $M_{\odot}/{\rm pc}^2$)
  (4), total HI mass (5), systemic velocity (6), width of the global
  profile at 20\% (7) and 50\% (8) levels, position angle of the major
  axis (9), warp angles on the East (10) and West (11) sides of
  the galaxy, environmental classification (12, for definitions see 
  Sect.~\ref{environment})}
\label{tab:results}
\end{center}
\end{table*}
\end{center}

\subsection{Lopsidedness}

There is ample evidence that a large fraction of galaxies depart from
axisymmetry. This can be inferred from the asymmetric shape of the
global HI profile. Such asymmetry can be caused by either a density
asymmetry or a 
kinematic one. Richter \& Sancisi (1994) estimated that
at least 50\% of the galaxies have asymmetric global profiles. Later
work by Haynes et al. (1998) confirmed this 
result. Swaters et al. (1999) discussed the lopsidedness in
the kinematics and concluded that probably at least half of all galaxies
are kinematically lopsided.

We therefore want to find out how lopsided our galaxies are and whether
there is any relationship between lopsidedness and warping. This will
be investigated both in the kinematics (using the rotation curves from the
approaching and receding sides) and in the density. 

To quantify the lopsidedness in the HI distribution we have calculated
the HI mass for each side of the galaxy ($M_1$,$M_2$) and defined the 
lopsidedness index as 

\begin{equation}
{\mbox{Lop}}_{\rho}= {|M_1-M_2|\over{M_1+M_2}}
\end{equation}

The kinematic lopsidedness has been derived from the difference in the
rotation curves between the receding and the approaching sides ($V_{\rm r}$
and $V_{\rm a}$ respectively, defined to be positive). The kinematic
lopsidedness index was calculated according to the formula 

\begin{equation}
{\rmn {Lop}}_{\rm kin}={\sum_x V_{\rm dif}(x)/{\sigma^2_{\rm dif}(x)} \over{\sum_x
    1/{\sigma^2_{\rm dif}(x)}}},  
\end{equation}
\begin{equation}
\mbox{where   } V_{\rm dif}(x)={|V_{\rm a}(x)-V_{\rm r}(x)| \over (V_{\rm a}(x)+V_{\rm r}(x))/2}
\end{equation}
and $\sigma_{\rm dif}$ are the uncertainties in $V_{\rm dif}$. 

\subsection{Environment}
\label{environment}

One of the issues that we want to address is the role of the
environment in warp phenomena. Reshetnikov \& Combes (1998) found that
the percentage of optically warped galaxies is higher in 
rich environments. We want
to find out if this also holds for the HI warps or not. For that
purpose, for each galaxy we have searched in the NASA/IPAC
Extragalactic Database (NED) looking for companions which are within 100 arc
minutes and have a radial 
velocity difference less than 150 km/s with respect 
to our galaxies. We have also looked for companions in our HI datacubes, at full
resolution, 30$''$ resolution, and 60$''$ resolution. The
primary beam at Westerbork has a FWHM of about 37$'$ at 21 cm., which
determines how far away from the galaxy 
we are able to detect other objects. We have listed the companions
from NED and those detected in HI in Table~\ref{tab:comp}. We have found two
new systems (companions to UGC 4806), and determined velocities for 4
systems that had no previous velocity determination. 

With all this information, we have assigned a number to each galaxy
depending on how close the companions are: 2 (companion within 50
arcmin), 1 (companion at distance between 50 and 100 arcmin), or 0 (no
companions in 100 arcmin). 

\section{Results}

Here we present (Table~\ref{tab:results}) the determined warp angles, the density and 
kinematical lopsidedness and the classification according to the
environment. We also present the rotation curves, the radial density
profiles, and the warp curves in the combined plots of
Figure~\ref{fig:plotwl2}.  

Finally, we study the influence of the environment on warping and
lopsidedness, and we compare the optical and HI warps of a number of galaxies.

\subsection{Warp statistics and shapes}
\label{stat}

\begin{center}
\begin{table}
\begin{center}
\begin{tabular}{l|c|ccc}
\hline
          & Total   & Poor    & Intermediate & Rich  \\ 
\hline                                                                  
Total     & 26      &  7      &    6      &  13      \\
\hline                                                           
No warp   &  6 (9)  &  0 (1)  &   2 (3)   &   4 (5)  \\
1-warps   &  7 (7)  &  1 (2)  &   1 (0)   &   5 (5)  \\
2-warps   & 13 (10) &  6 (4)  &   3 (3)   &   4 (3)  \\
\hline
All warps & 20 (17) &   7 (6) &   4 (3)   &  9 (8)   \\
\hline
\end{tabular}
\caption{Different types of warps depending on the environment. Each
  column represents a different environment (for details see
  Sect.~\ref{environment}). The first row is for galaxies with no warp, the
  second for galaxies with a warp on only one side, and the third is
  for galaxies that
  show a warp on both sides of their disks. The last row lists the
  total number of warped galaxies for each environment. Numbers in
  brackets count a side of a galaxy as warped only if it warps more
  than 2\degr within the errors.}
\label{tab:warp}
\end{center}
\end{table}
\end{center}

The first question to address
is how common warps are. Table~\ref{tab:results} already shows
that the outer parts of 
the majority of galaxies exhibit some departure from flat disks. We
detect warps in 20 of our galaxies, of which 
7 show a warp only on one 
side (see Table~\ref{tab:warp}). If we consider only warps larger
than 2\degr, the number of warps decreases to 17, of which 7 are
only on one side. Taking into account that we will probably miss
most of the warps with a line of nodes close to the direction
perpendicular to the line of sight from us to the galaxy, it means
that the vast majority of galactic disks are warped.

Warps normally start at the edge of the optical disk or further
away. This means that for the warp to be visible there must be gas
outside the optical disk. We looked at the galaxies in our sample with
no warp to see if any of them had extended HI disks, i.e. if there
was any galaxy with an extended HI disk (with respect to the optical emission)
and no warp. As a measure of the optical extent of the galaxy we used
R$_{\rm opt}$ (see section \ref{ropt}), which is more representative than
R$_{25}$, particularly in edge-on galaxies. We found that all the galaxies in our
sample that have HI disks more extended than the optical are
warped. If we consider each side of the galaxy independently, there is
only one exception to this rule: one side of UGC 7125 is warped, while the
other remains flat, and both sides have HI more extended than the
optical. 

One of the most striking characteristics of the warps observed here is their
asymmetry. To begin with, a considerable percentage of the warped
galaxies show a warp only on one side. This is not due to an absence
of HI on the opposite side: in all cases the gas disk on the unwarped
side does extend to beyond R$_{\rm warp}$. Sometimes (as in UGC 7321,
8246, 8396) there are hints of the beginning of a warp in that
side but the data do not allow a firm detection of it.

\begin{figure}
\plotfull{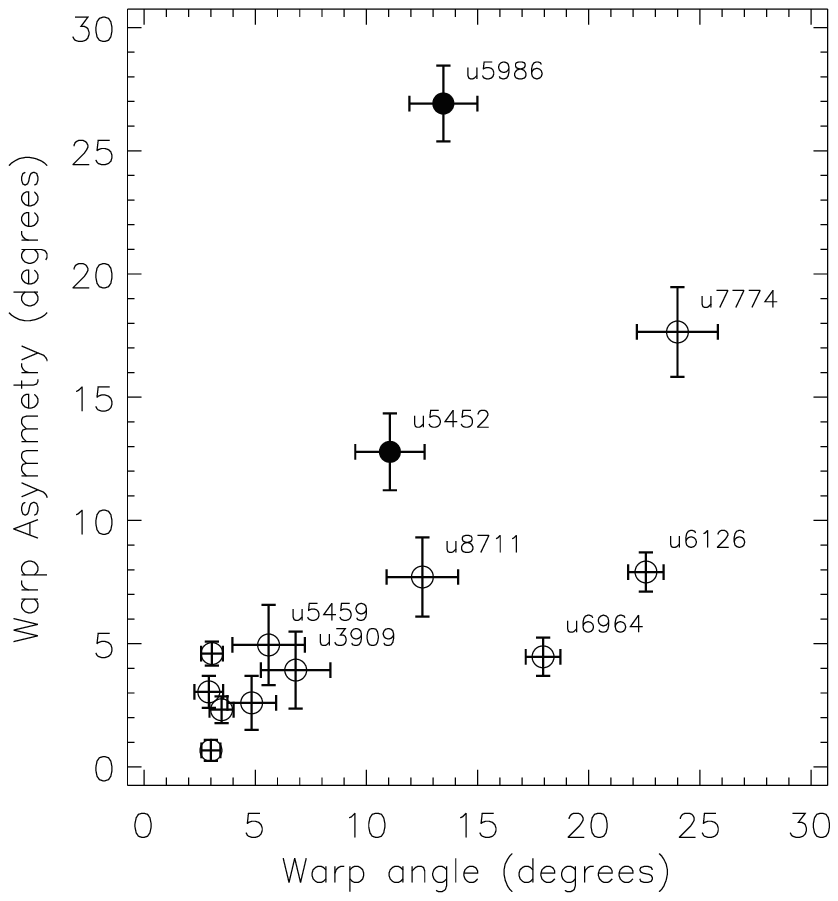}
\caption{Warp asymmetry vs. mean warp angle. Filled circles represent
  galaxies with obvious tidal features with nearby companions, thus
  indicating strong interaction. The most strongly warped galaxies are
identified by their UGC number.}
\label{fig:warpasym}
\end{figure}

However, the clearest sign of asymmetry in warps comes from
galaxies with both sides 
warped. Fig.~\ref{fig:warpasym} shows the warp asymmetry
($\alpha_{\rm asym}$) with
respect to the mean warp angle. Only galaxies with warps
detected on both sides of the disk are plotted. We inspected the total
HI maps of all 
our galaxies looking for tidal features that would clearly indicate
strong tidal interaction with a nearby companion. Such interaction
might produce a highly asymmetric warp. These galaxies are plotted as
filled circles in Fig.~\ref{fig:warpasym}, and indeed they appear to
possess more asymmetric warps than the rest of the galaxies. But the
remaining warped galaxies also have
large asymmetries. There are only 2 systems with large warps
and high symmetry: UGC 6964 and UGC 6126. The other galaxies
lie more or less on a straight line where the ratio of the asymmetry
to the mean warp is of about 0.7. A galaxy with this asymmetry and a
mean warp of 10\degr would have warps in each side of 6.5\degr and
13.5\degr, which is quite asymmetric. 

\begin{figure*}
\begin{center}
\plotx{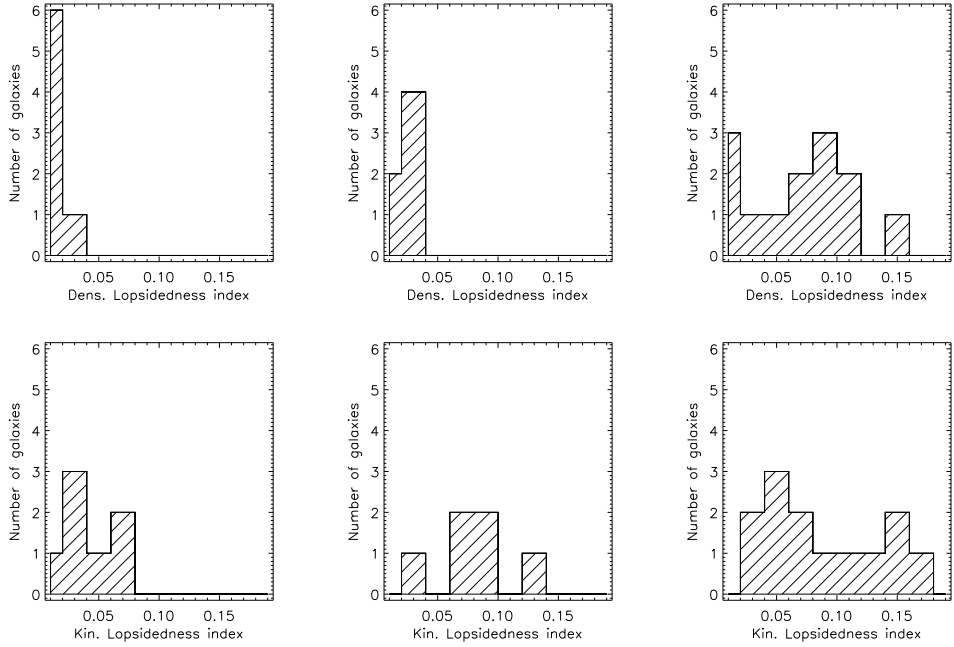}
\caption{Histograms showing the distribution of galaxies according to
  the density (upper panels) and kinematical
  (lower panels) lopsidedness index for different
  environments. Galaxies in poor environments are plotted in the left
  panels, galaxies in intermediate environments in the middle panels,
  and galaxies in rich environments in the right panels. For a
  detailed description of the environment classification, see
  Sect.~\ref{environment}.}
\label{fig:loplopk2}
\end{center}
\end{figure*}

Most of the galaxies with both sides warped are antisymmetric (S
shape warps). We only have two cases of U-type warps, and both these
galaxies are strongly interacting with nearby companions and are very
disturbed. 

The warp curves, together with the rotation curves and the radial
density profiles have been plotted in Fig. ~\ref{fig:plotwl2}. These plots
show the shapes and amplitudes of the warps, and allow a
comparison between features in the
radial density profiles, rotation curves and warp curves. It is quite
clear from Fig.~\ref{fig:plotwl2} that there are several distinct
warp shapes: some galaxies have monotonically increasing warps
(rising linearly or even faster), while others 
have warps that at some point rise slowly, and in some cases head back
to the plane defined by the inner regions, like the southern warp in
the Milky Way. Note that the rotation curves in
Fig.~\ref{fig:plotwl2} have been calculated along the major axis and
not along the warp line. This is the reason why the rotation curves
often stop before the HI density drops to zero, as in \mbox{UGC 6964}.

\subsection{Lopsidedness}

Lopsidedness may have some connection with warping. If, for instance, the lopsidedness
of a galaxy is caused by merging, and such merging process is also
causing the warping, we would expect more pronounced warps in lopsided
galaxies, and small or no warps in axisymmetric galaxies. 

The first indication that
lopsidedness may be the result of accretion or 
interactions with nearby companions comes from the fact that the lopsidedness
we measure seems to depend quite strongly on the
environment. In Fig.~\ref{fig:loplopk2} we have plotted histograms
for the galaxies according to their density lopsidedness index for
each type of environment: poor, intermediate and rich (see 
Sect.~\ref{environment}). Clearly, galaxies with no nearby companions
are quite symmetric in mass, galaxies with companions not closer
than 50$''$ are somewhat more lopsided, and finally, all galaxies with
a lopsidedness greater than 5\% have a companion closer than 50$''$. The
dependence of lopsidedness on environment found here is
remarkable if one takes into account the uncertainty of the 
classification of the environment. The kinematical lopsidedness seems also
to depend somewhat on the environment, as
isolated galaxies have lower kinematical lopsidedness than the rest of
the galaxies. The way both
lopsidedness indices are measured actually makes 
them lower limits due to the fact that the galaxy is edge-on: to
measure our lopsidedness indices we are comparing the approaching and receding
sides of the galaxy (integrated along the line of sight), thus if the
galaxy is asymmetric in the direction along the line of sight we would
not be able to detect it. We did not find a clear correlation between
kinematic and density lopsidedness in our galaxies.

\setcounter{dummyfofo}{\thefigure}

\begin{figure*}

\plotthree{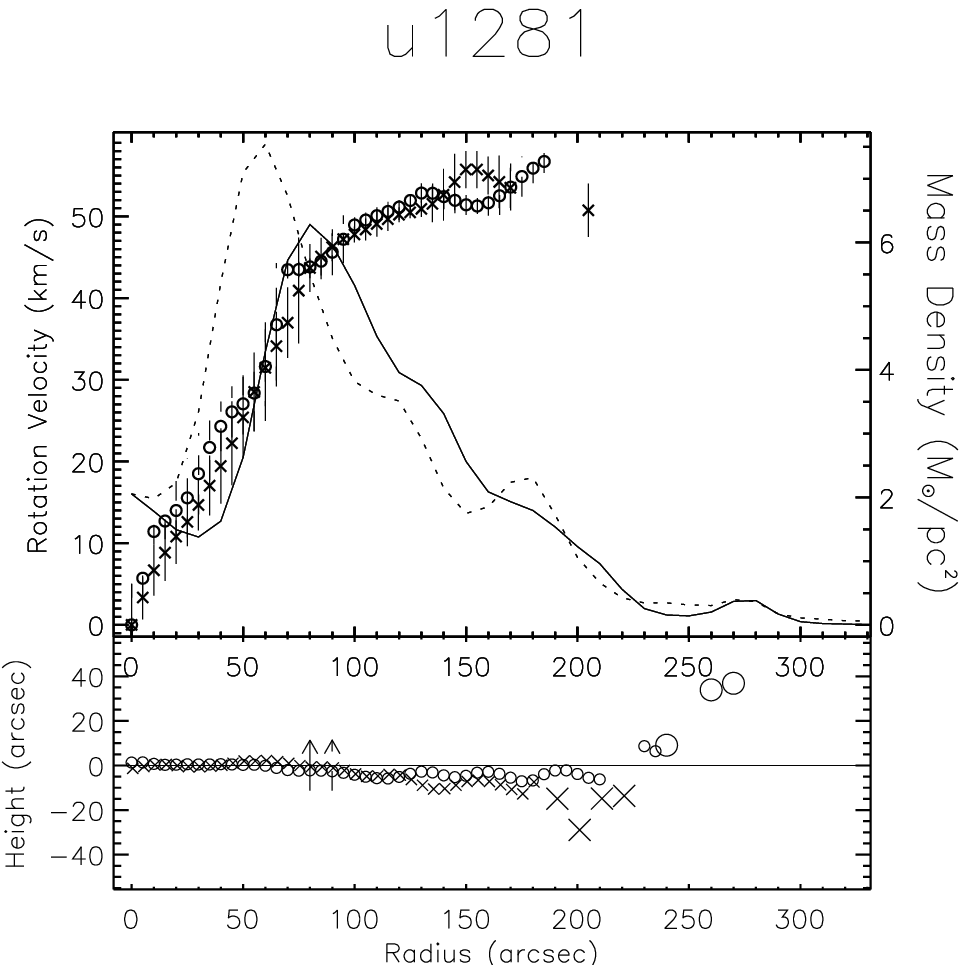}{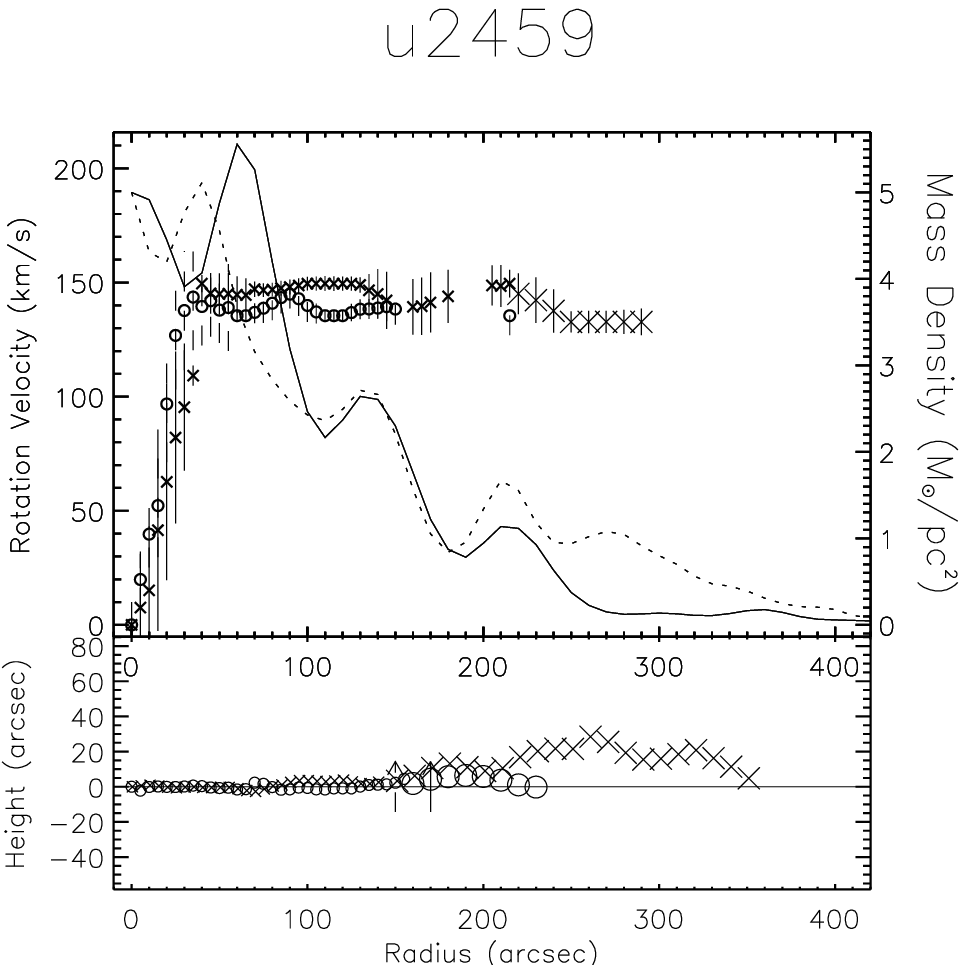}{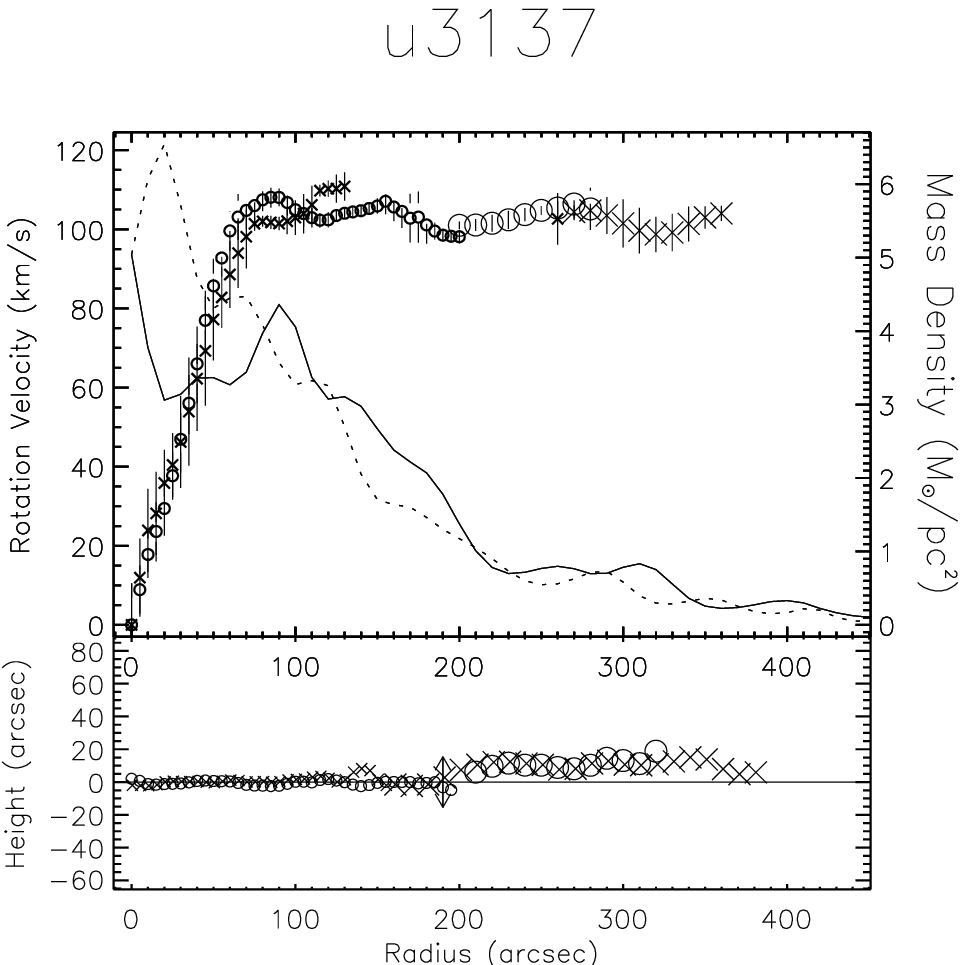}
\plotthree{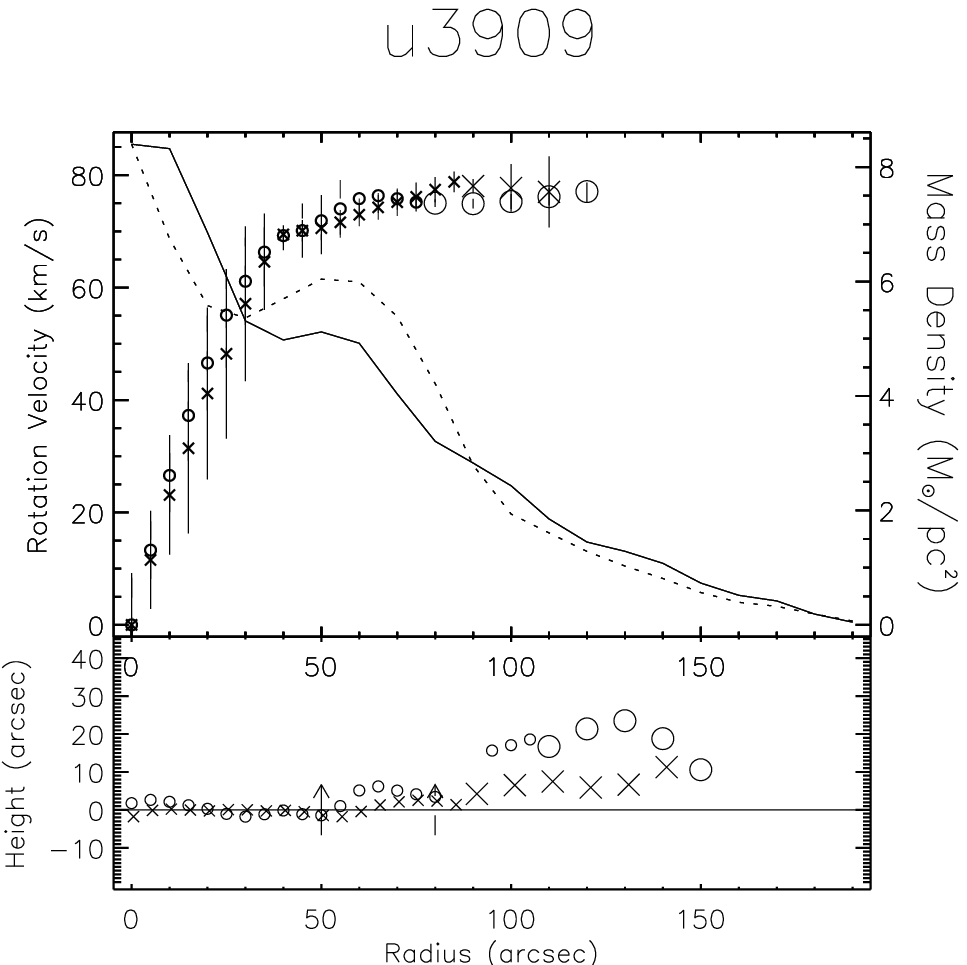}{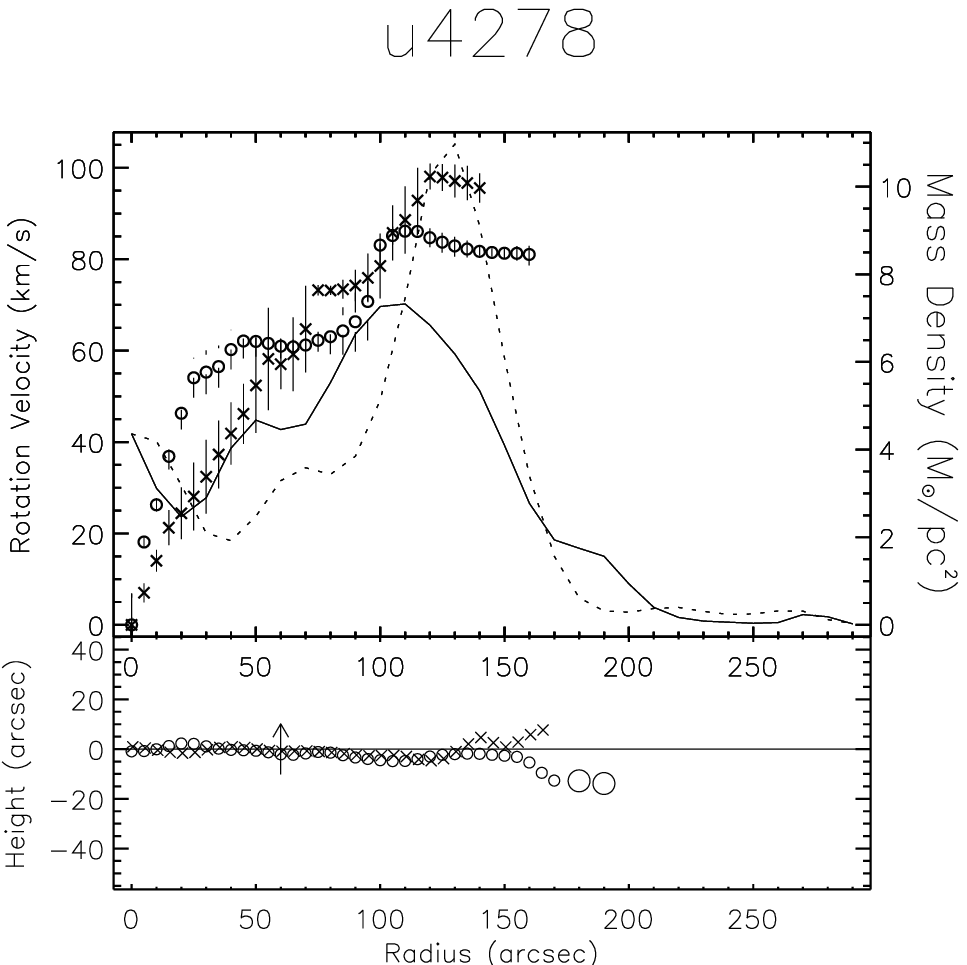}{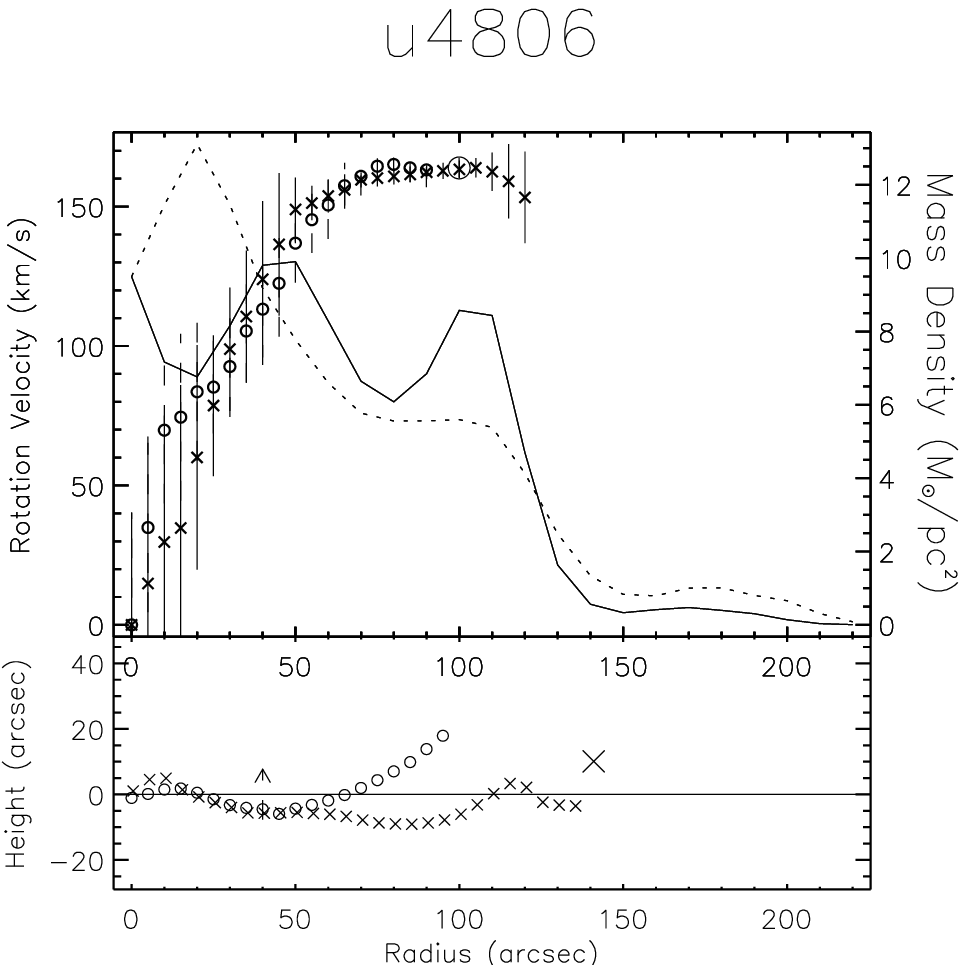}
\plotthree{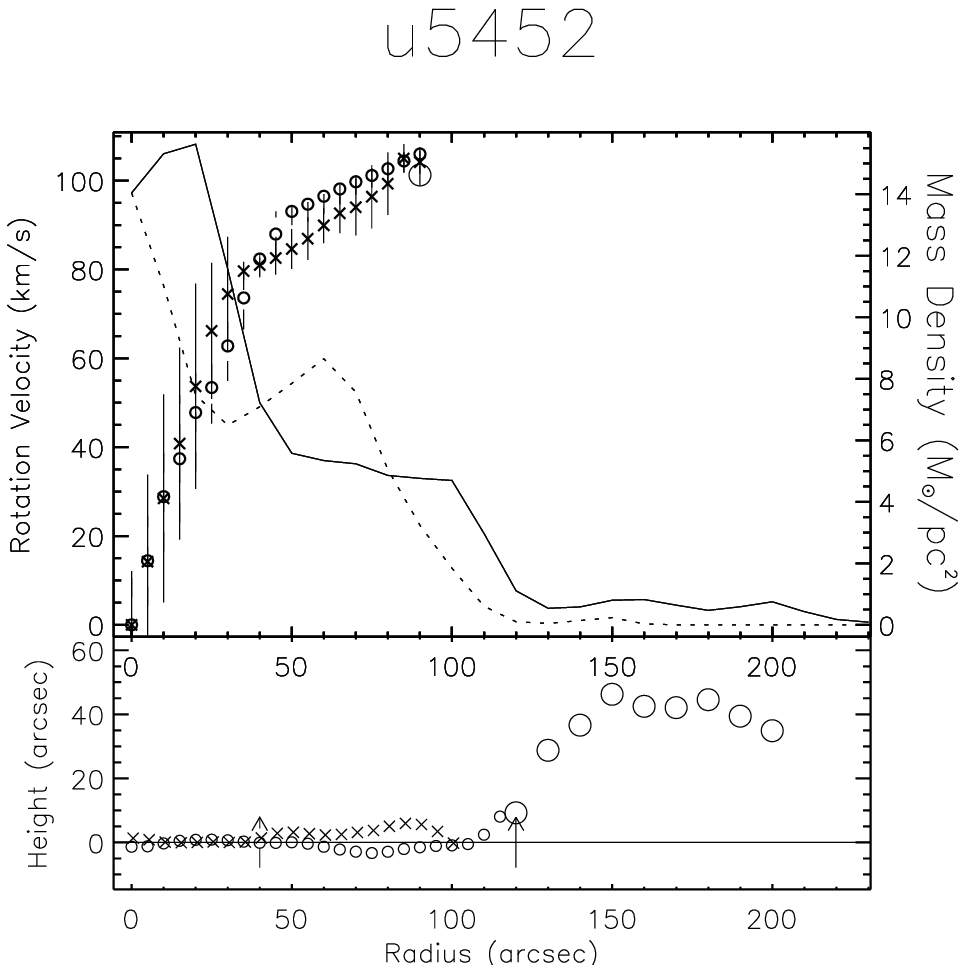}{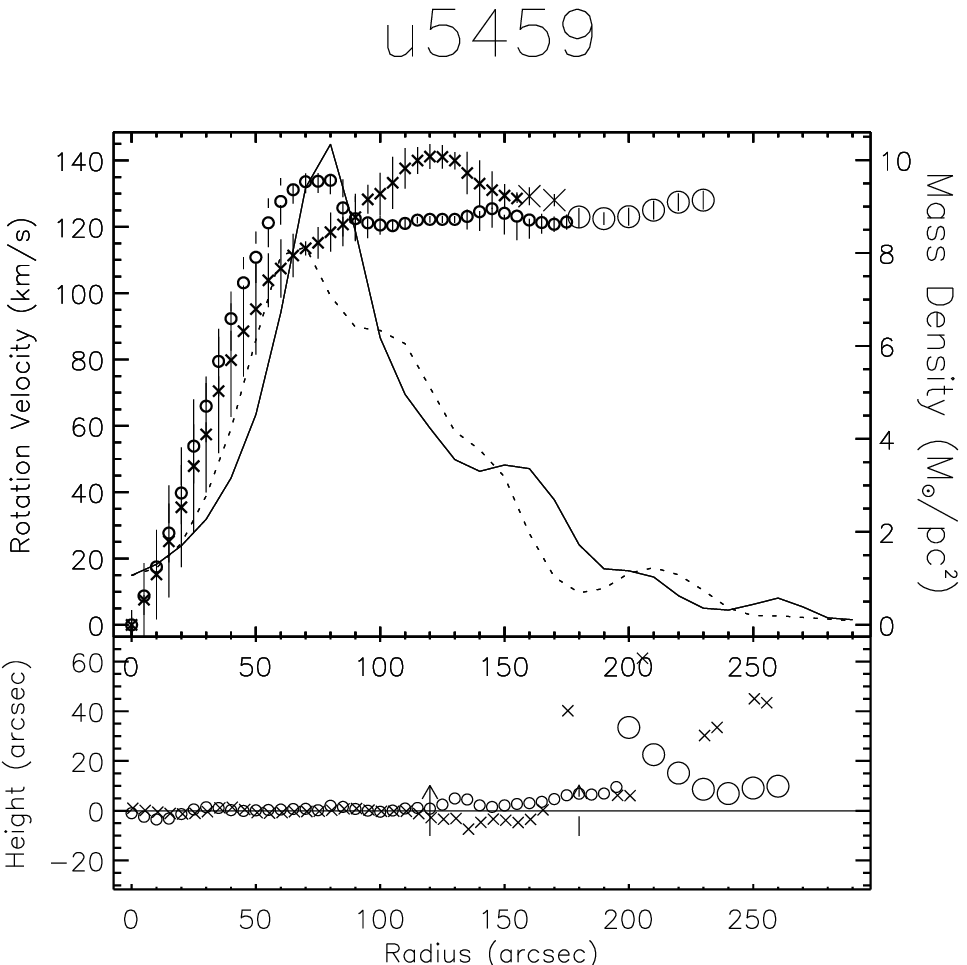}{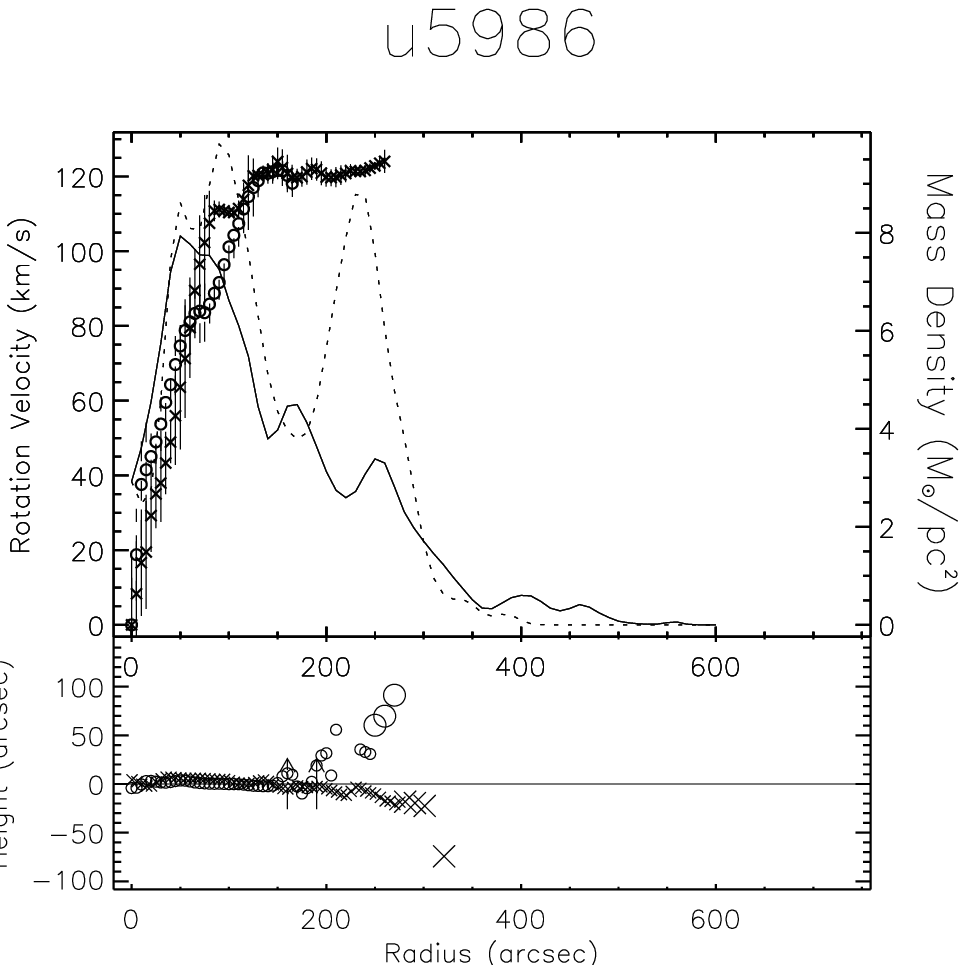}
\caption{Density - RC - warp plots. For each galaxy, the rotation
curve along the major axis, the radial HI density profile and the
curve of warp are plotted. We have measured these quantities
independently for both sides of the galaxies, and here we present them
together in the same plot. The top panel shows the rotation velocities
and radial density profiles for both sides of the galaxy. The left
axis shows the units for the rotation velocities, while the right axis
applies to the density profiles. The bottom panel shows the warp
curves for both sides. Because most of the warps have an S-shape, the
warp curve derived for the left side has been inverted, to highlight
symmetries or asymmetries between the two sides of the galaxy. With
this process, both warp curves of a galaxy with a perfect S-shape
symmetry (thus, antisymmetry) would fall on top of each other. One
side is plotted with crosses (rotation curve and warp curve) and
dotted line (density), and the other side is plotted with circles
(rotation curve and warp curve) and solid line (density). The rotation
curve and warp curves in the figure are shown at 2 different
resolutions. We derived these curves using the full resolution data
(small symbols), and then complemented it in the outer parts with the
curves obtained at 30'' resolution (bigger symbols). Errorbars have
not been plotted in the warp curves for sake of clarity, they are
shown in Figure 13.  }  \label{fig:plotwl2}

\end{figure*}

\setcounter{figure}{\thedummyfofo}

\begin{figure*}

\plotthree{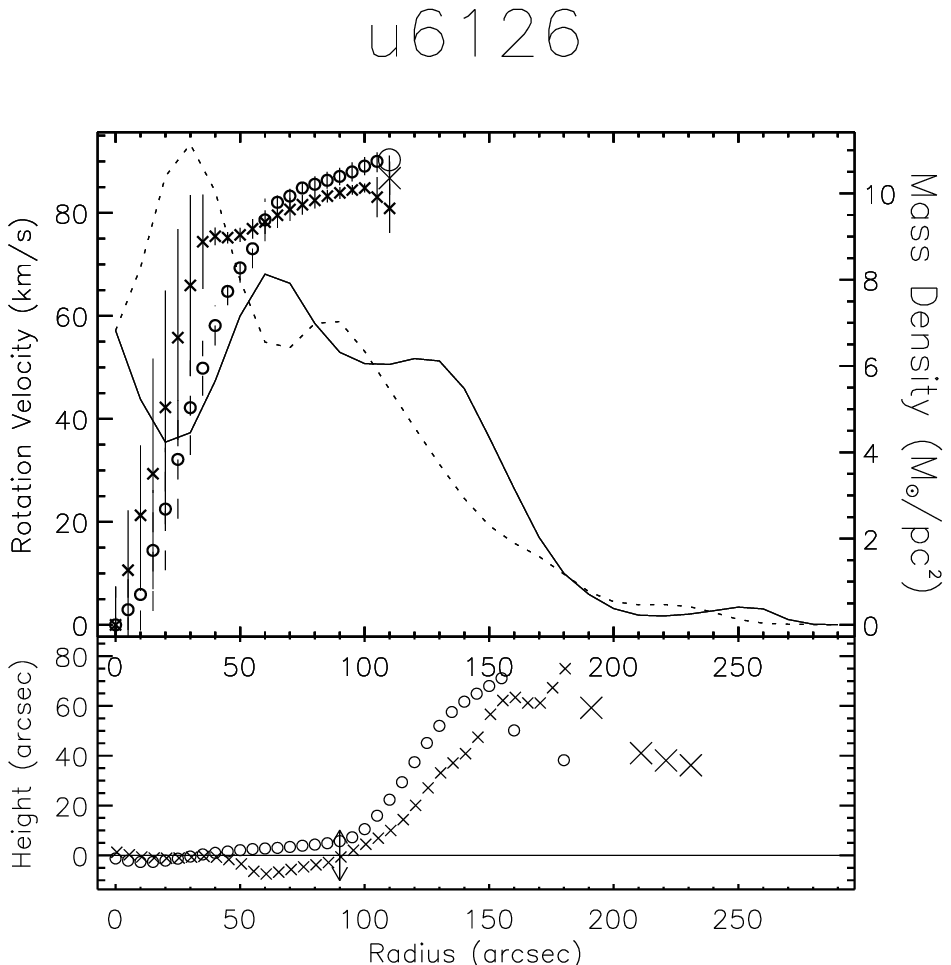}{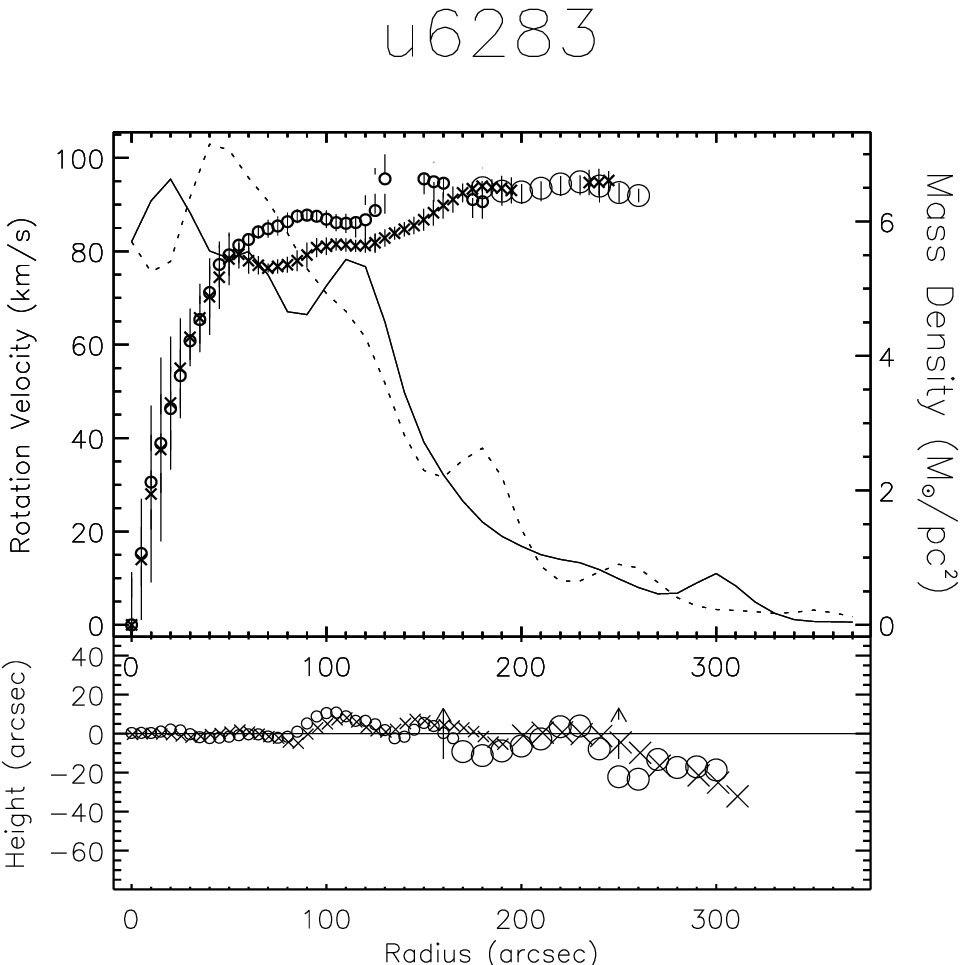}{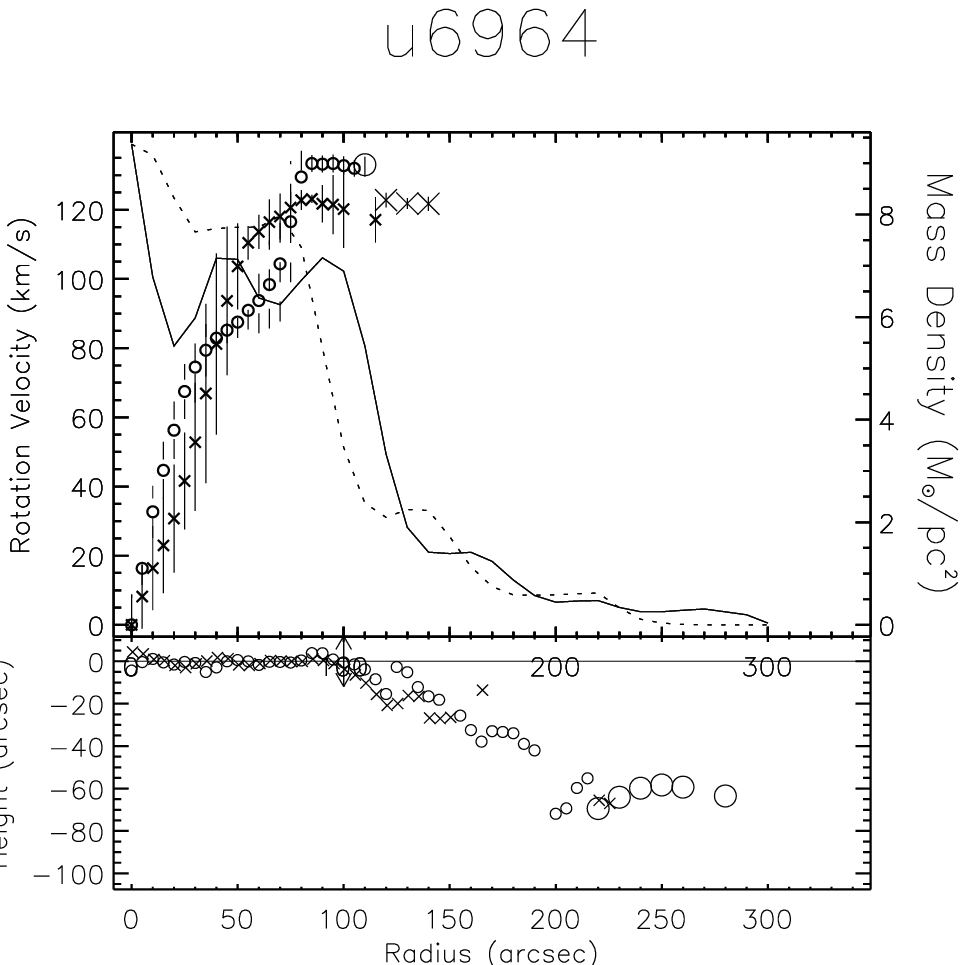}
\plotthree{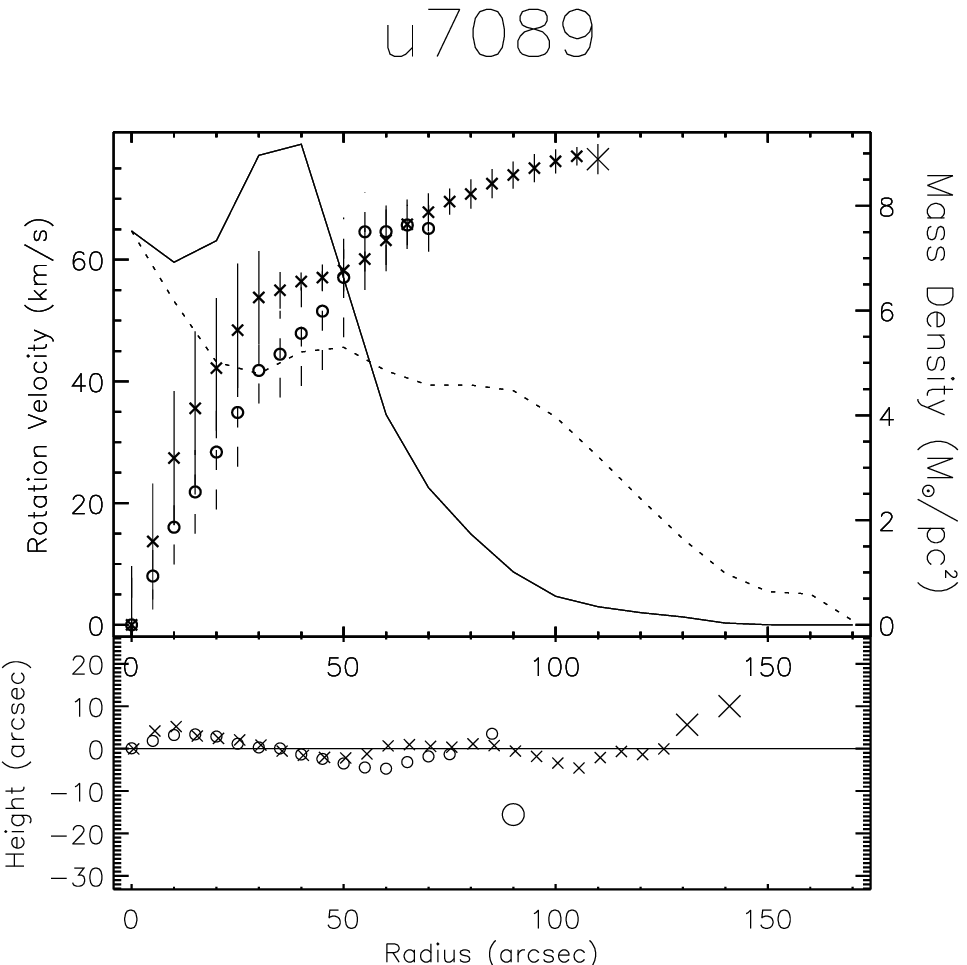}{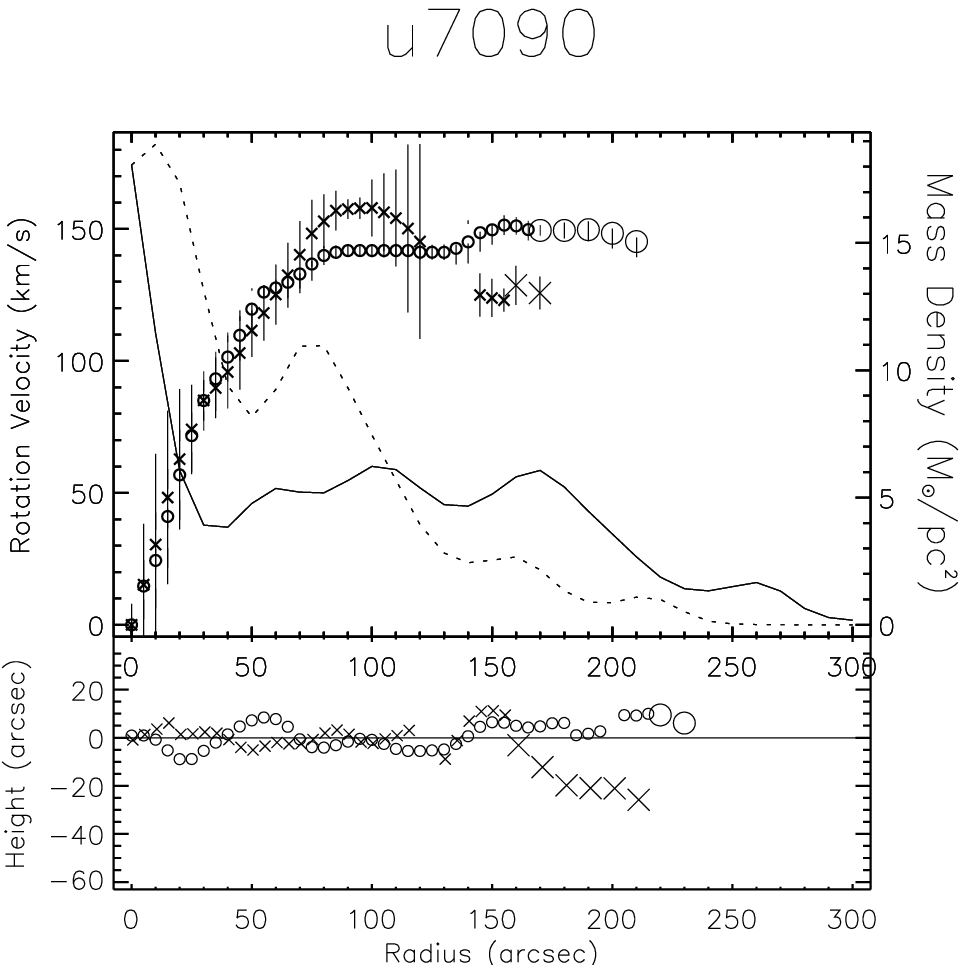}{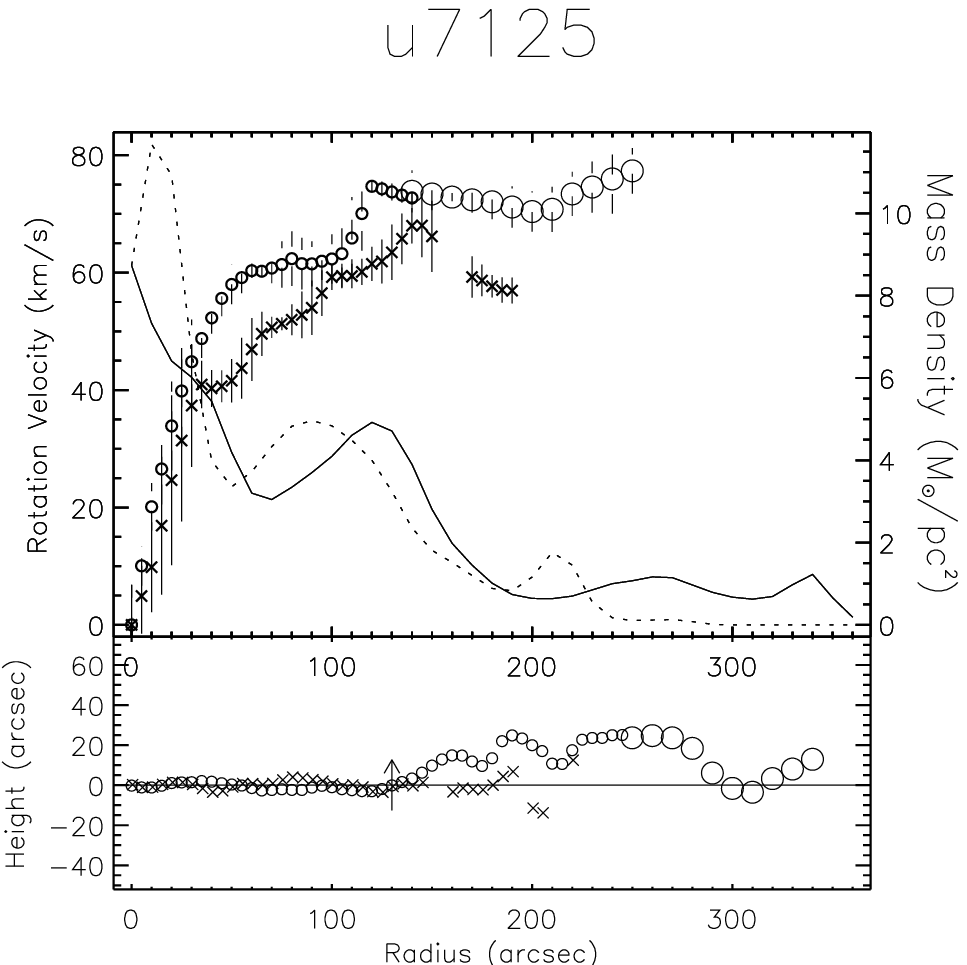}
\plotthree{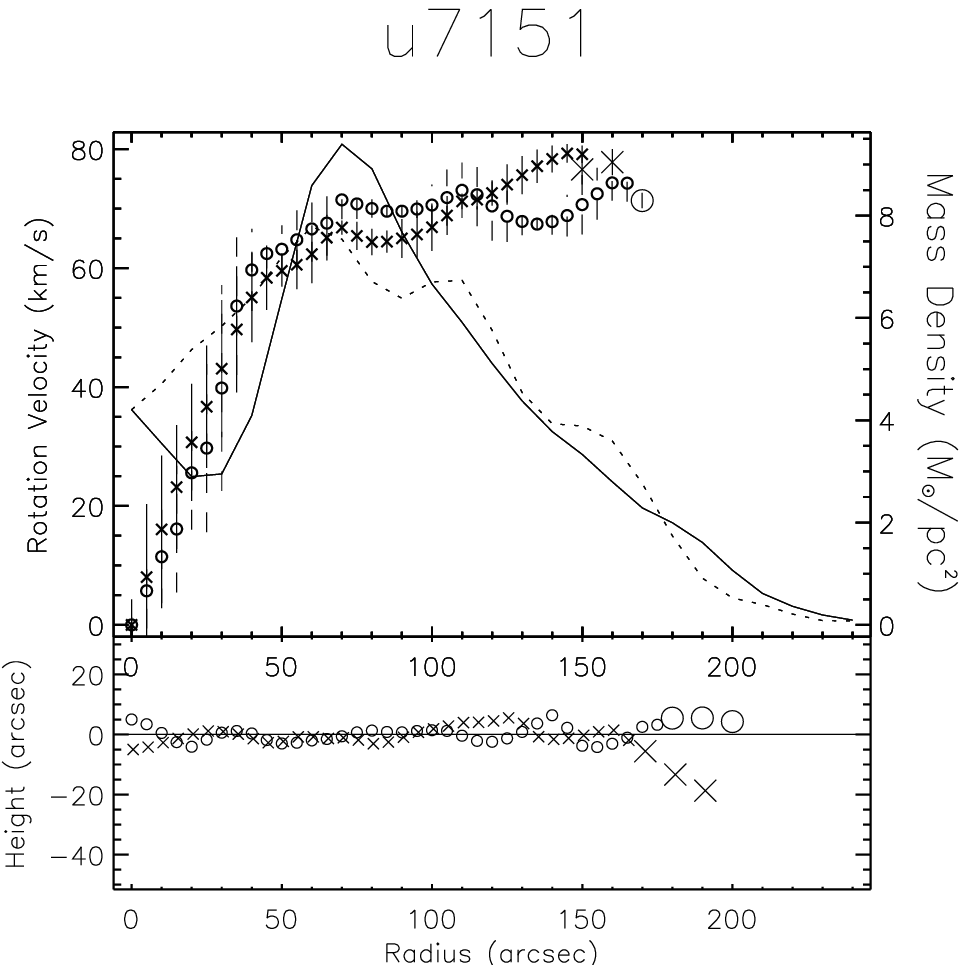}{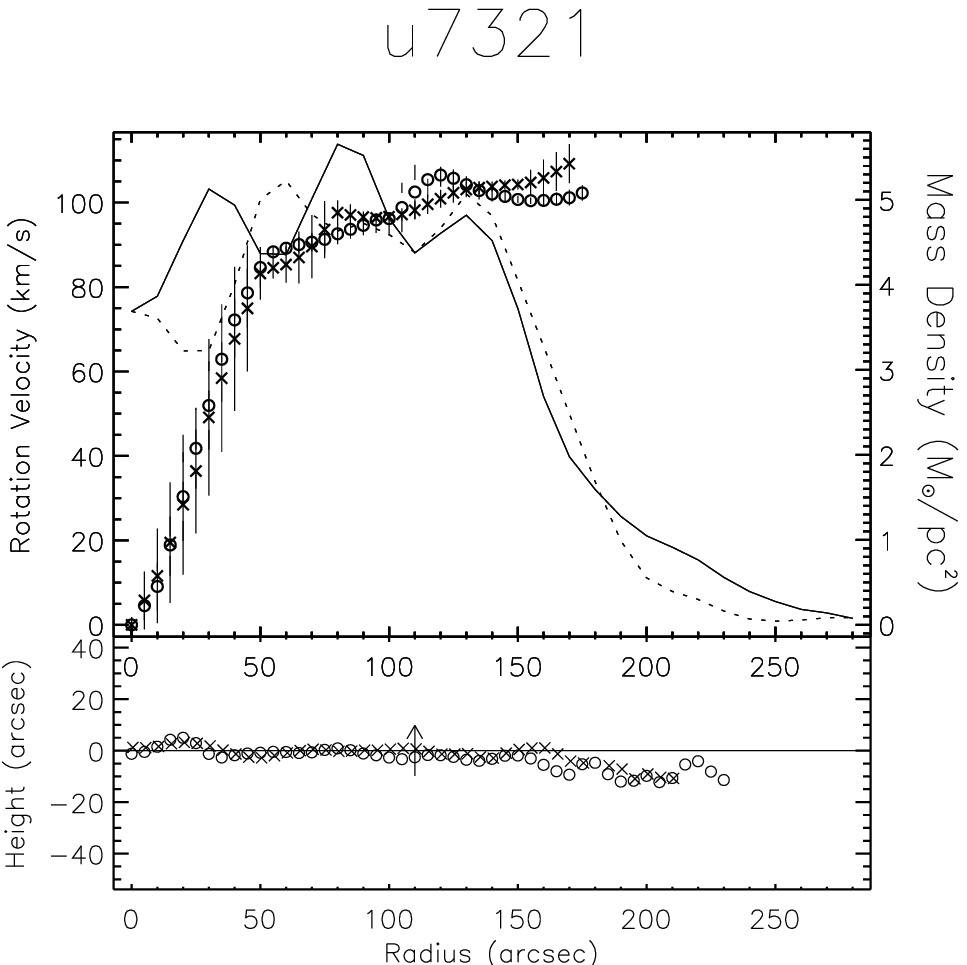}{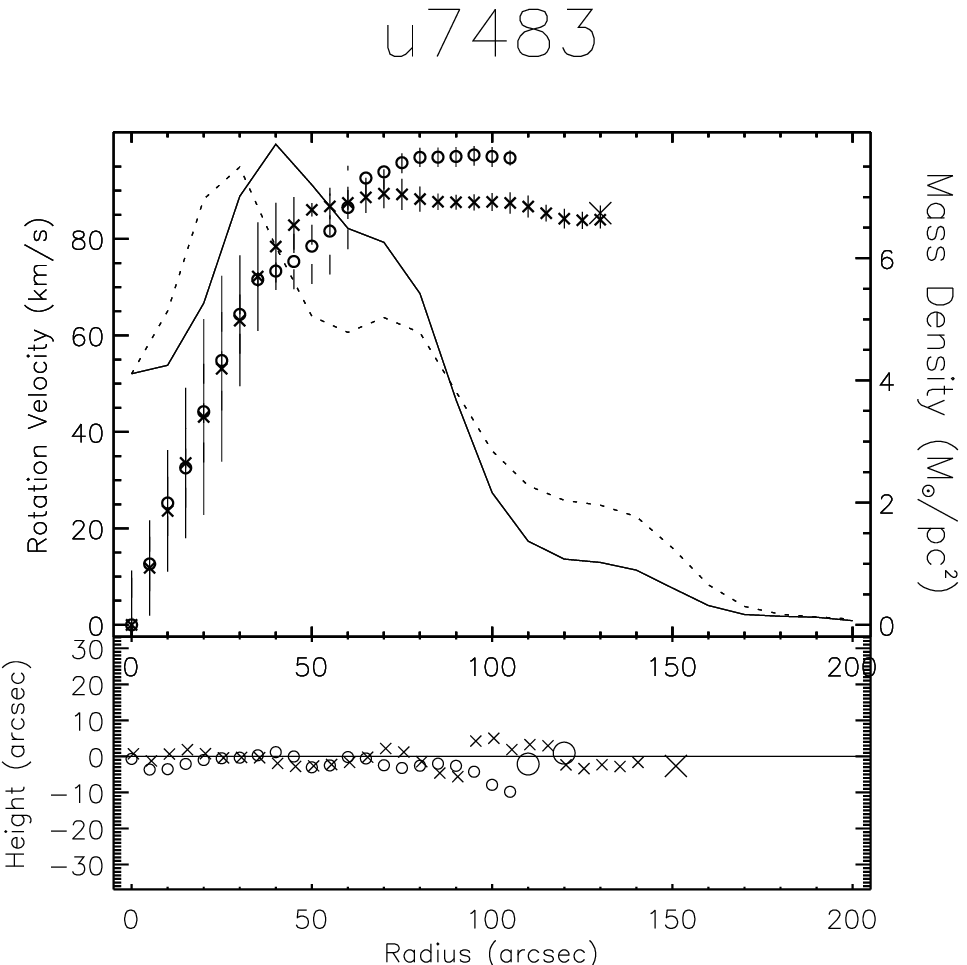}
\caption{-- {\it continued}} 

\end{figure*}

\setcounter{figure}{\thedummyfofo}

\begin{figure*}

\plotthree{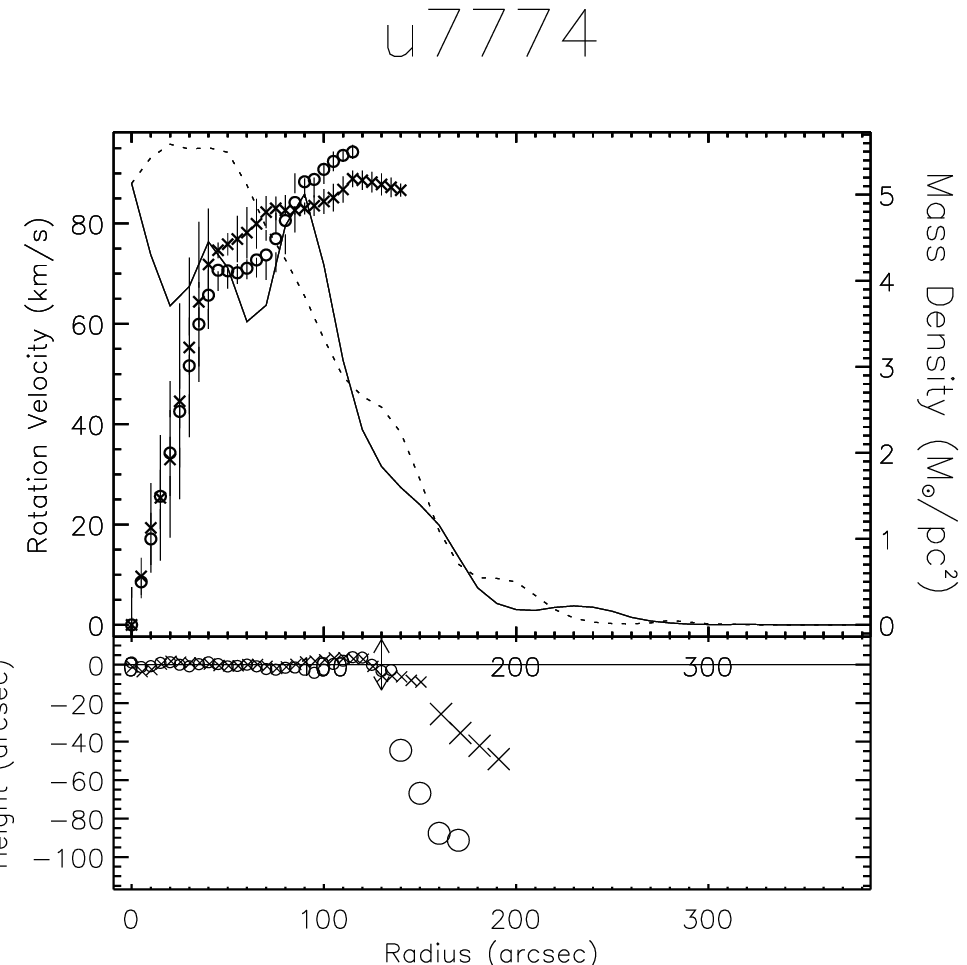}{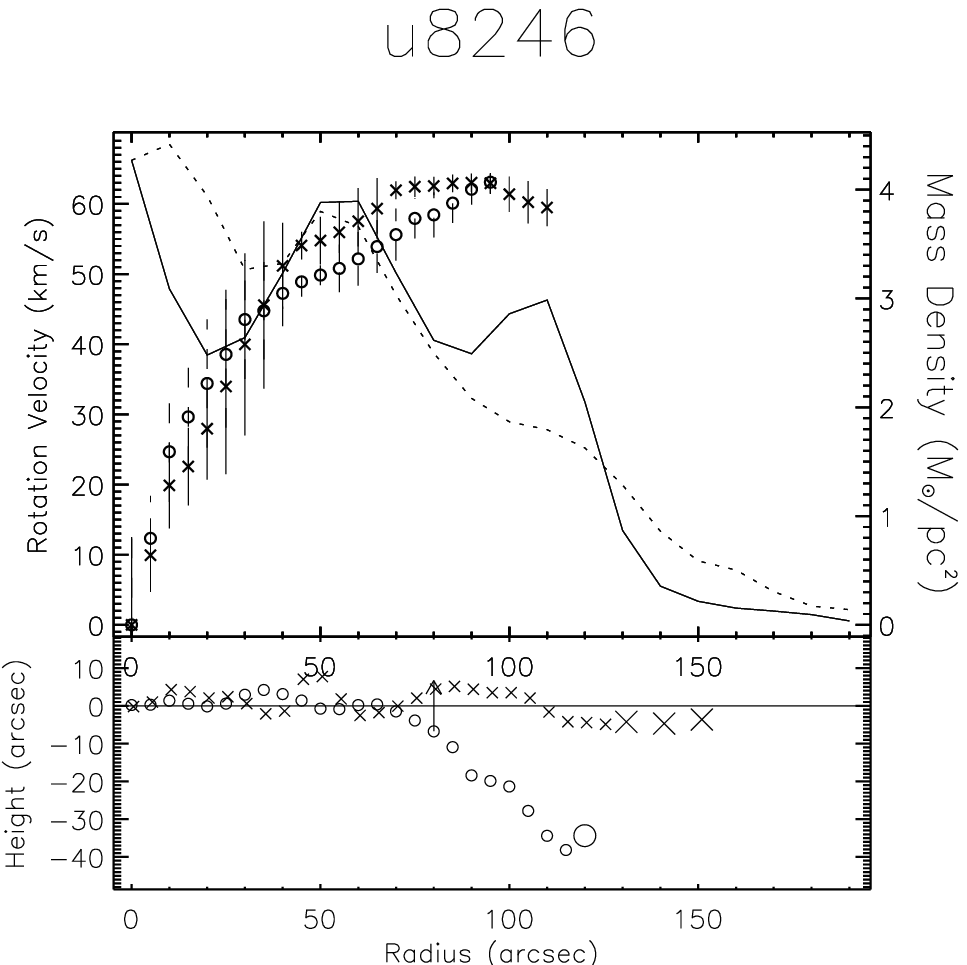}{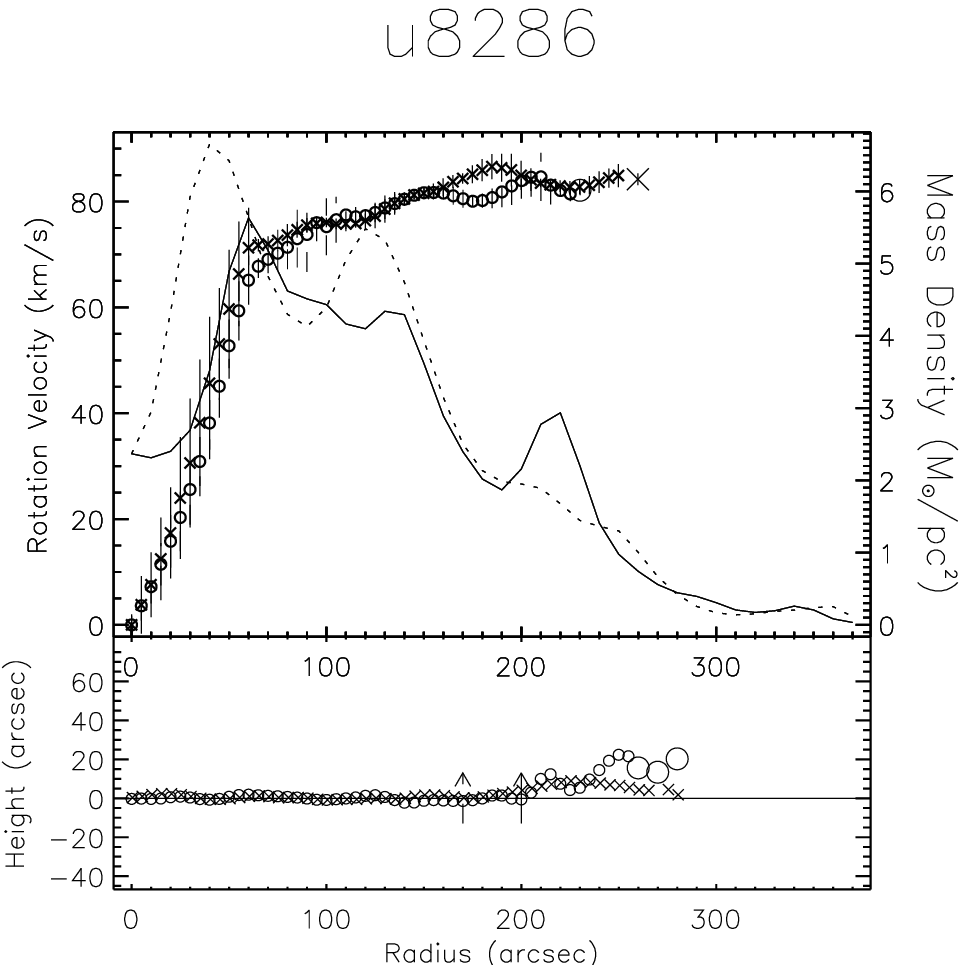}
\plotthree{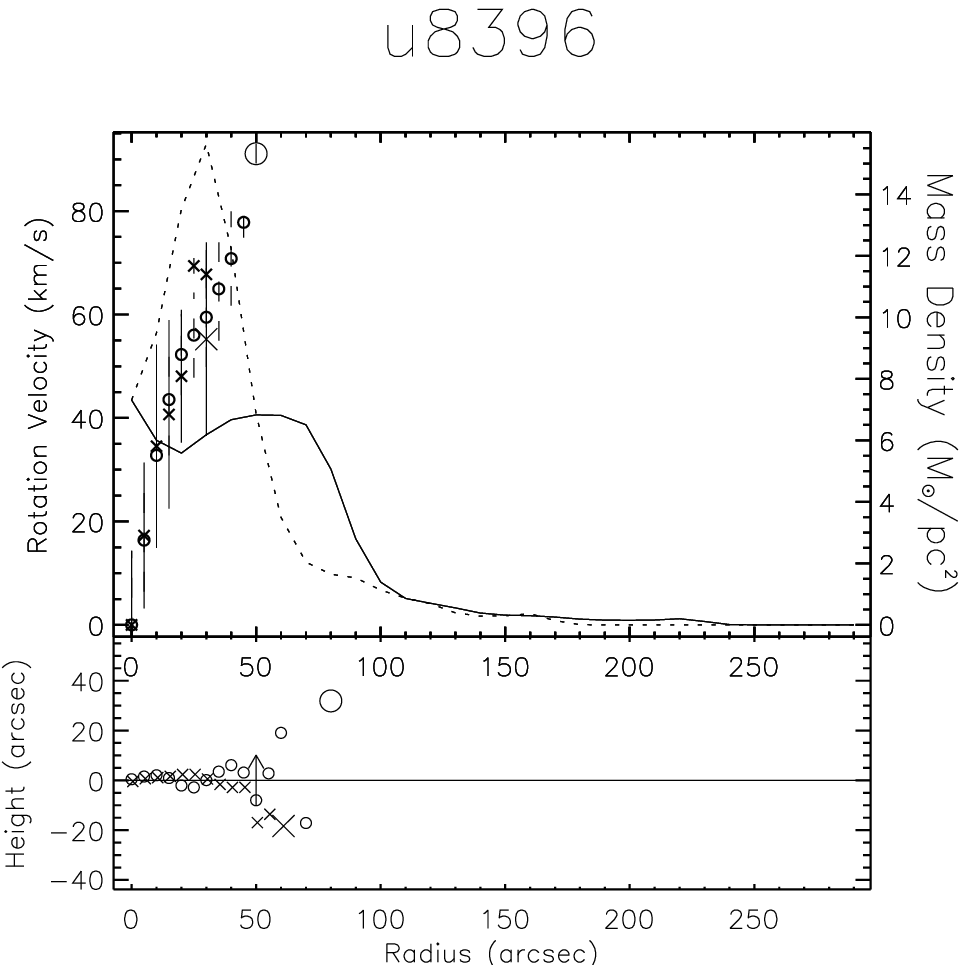}{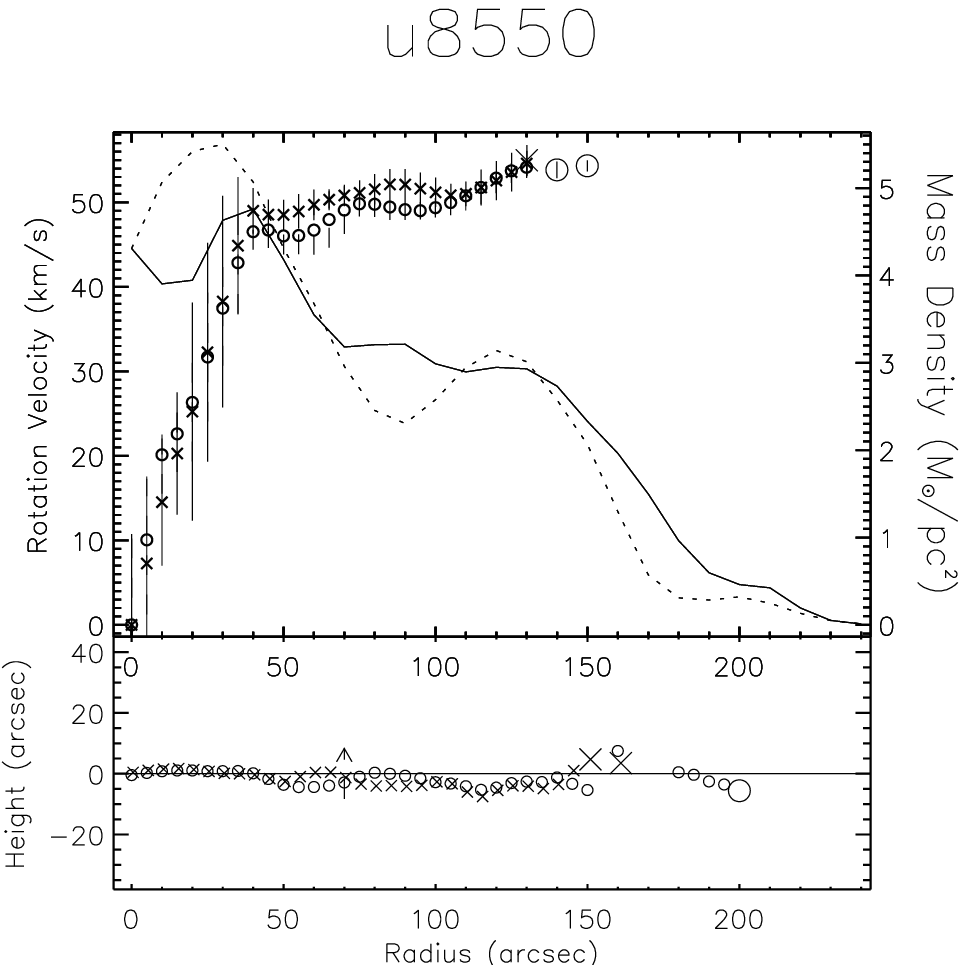}{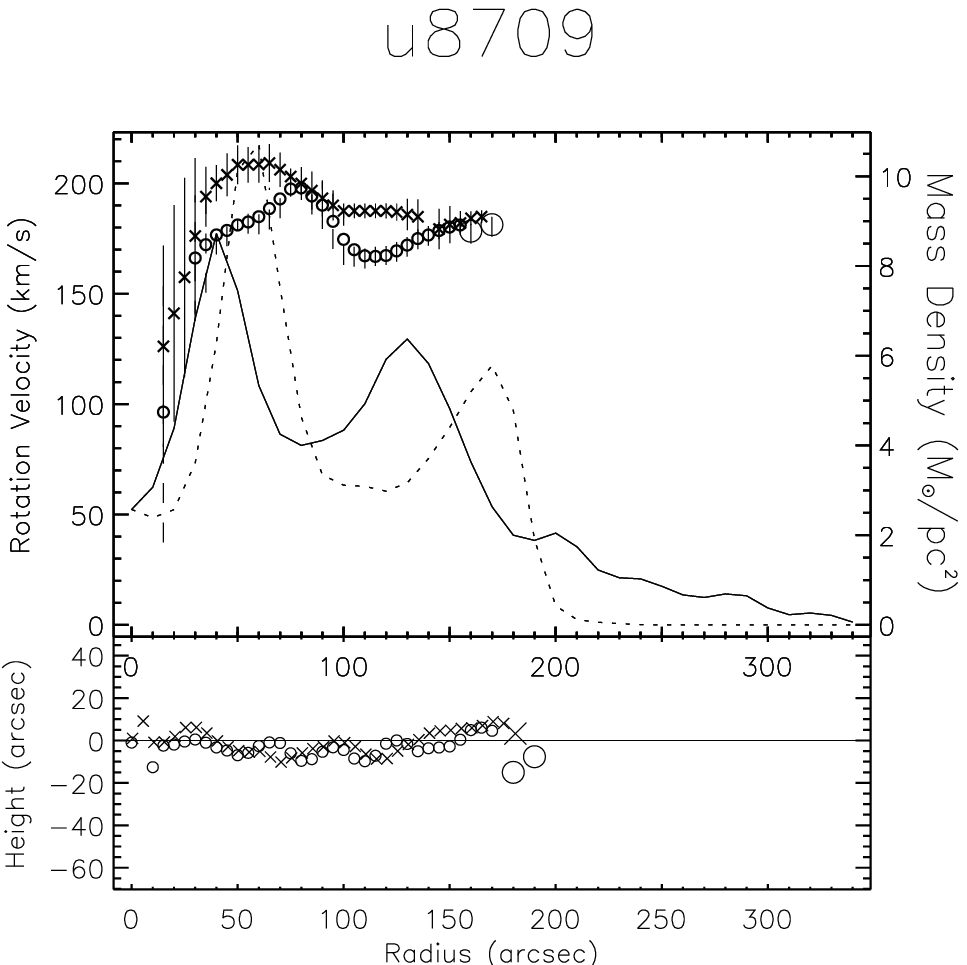}
\plotthree{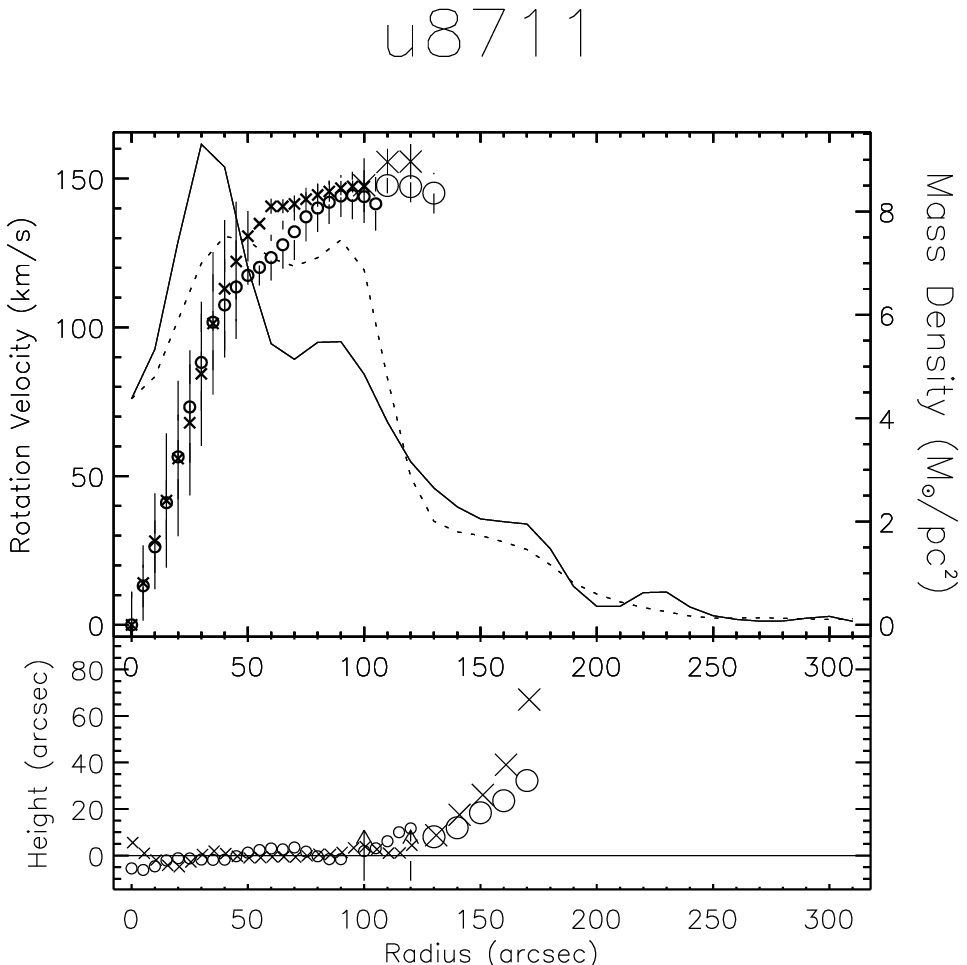}{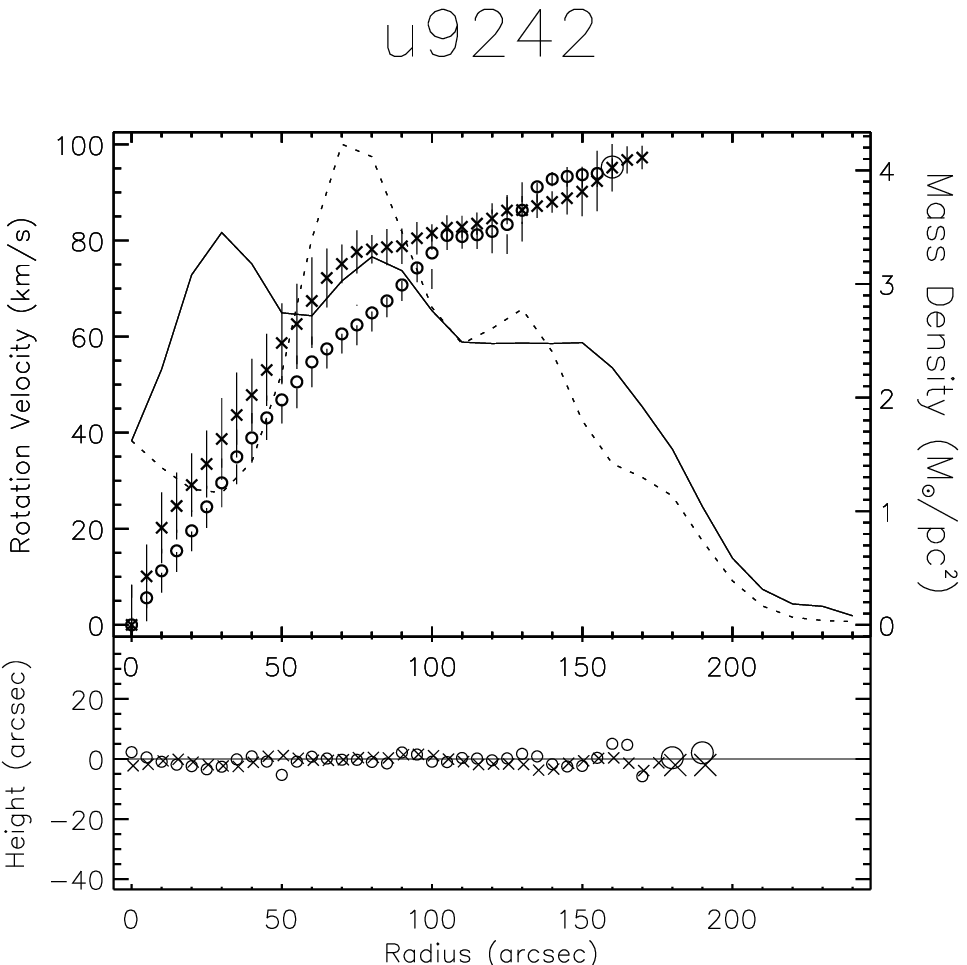}{}
\caption{-- {\it continued}} 

\end{figure*}

\subsection{Warps and environment}
\label{warpe}

\begin{figure*}
\plotfull{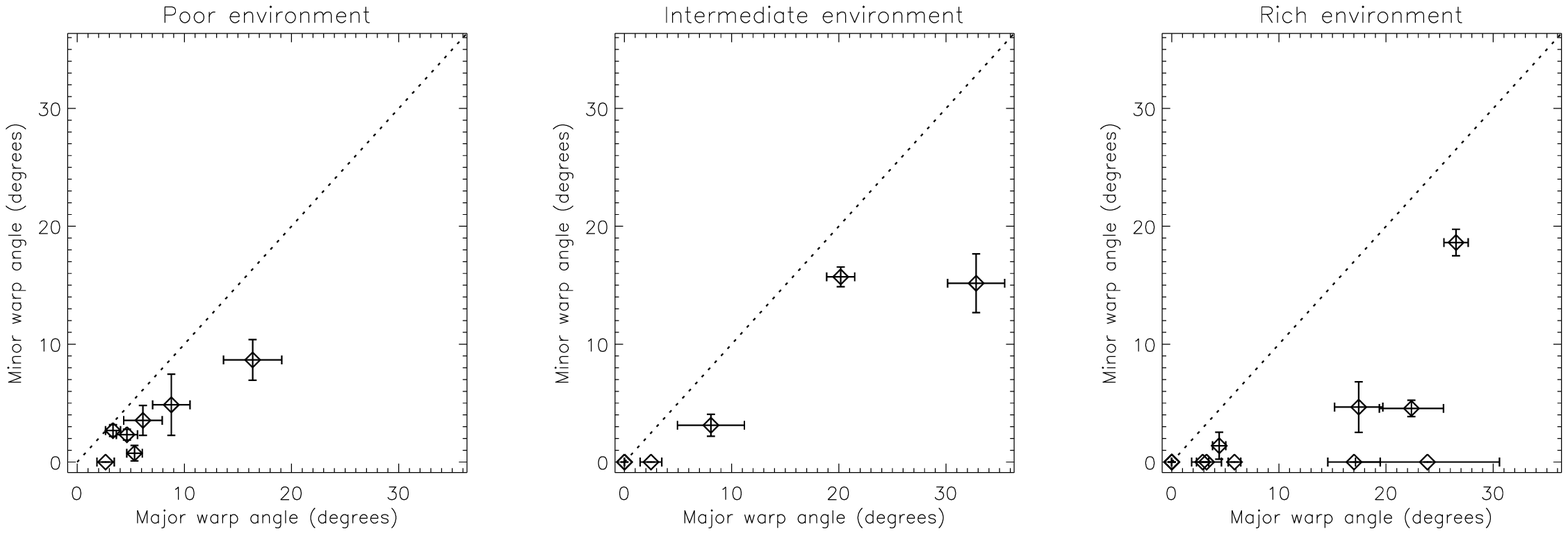}
\caption{Warp angles for each environment class (see definition in
  Sect.~\ref{environment}). The left panel shows the warps of galaxies in poor
  environments, the middle panel the galaxies in intermediate
  environments, and the right panel those in rich
  environments. For each galaxy, the larger of both side's warp angles
  has been plotted
  on the x axis, versus the smaller one on the y axis (so the upper left
  part of the plots is empty). The galaxies with a detected warp on
  only one side lie in the y=0 line, and the non warped galaxies are in
  the origin. The dotted line indicates equal warp
  angles on both sides of the galaxy.
  }
\label{fig:warpenv}
\end{figure*}

The lopsidedness seems to be correlated with the environment, in
the sense that in rich environments galaxies are more
lopsided. It has been claimed (Reshetnikov \& Combes 1998) that optical warps also
depend on the environment, in 
the sense that galaxies in rich environments are more 
commonly warped than isolated galaxies. These authors concluded
that this could indicate that tidal interactions are the cause of
warps, or at least that they reinforce them. In a study of the influence of
minor mergers on the warping of optical disks Schwarzkopf \& Dettmar
(2001) found warps in all their merging systems, but in
only half of their control sample. The amplitude of the warps in the
merging systems was larger as well. They concluded that tidal
interaction does generate warps but that this mechanism does not
account for all the observed ones.

We checked our sample to see if these
relations hold for the larger HI warps as well. Table~\ref{tab:warp}
shows that, if anything, the trend in frequency for the occurrence of HI warps in our sample seems
to go in the opposite sense:
galaxies in poor environments are more frequently warped than galaxies
in dense environments. However, all galaxies with an extended HI disk
with respect to the optical disk are warped, therefore the frequency
of warps depends on the extension of the HI disks. Thus, in our sample
there are more galaxies with no extended HI in rich environments than
in poor environments. The warp parameters (amplitude and asymmetry)
are illustrated as a function of environment in  Fig.~\ref{fig:warpenv}.

There seems to be a tendency for galaxies in
poor environments to be more prone 
to warping, but to have smaller warp angles than in rich
environments. Warps also seem to be more symmetric in poor than in rich
environments. The explanation for the asymmetry and higher amplitude
of the warps in rich environments could lie in the tidal
interaction with neighboring galaxies (e.g. UGC 5452, UGC
8396). The indication that galaxies in poor environments are more
frequently warped may mean that tidal interactions are not the only
mechanism that produces warps. 

\subsection{HI vs. optical warps}

As mentioned above, warps may show a different behavior with
respect to environment when observed in the optical or in
HI. Unfortunately, not much work has been done about the relationship
between HI and optical warps in the same galaxies, and none with high
quality HI and optical 
data. 

\begin{center}
\begin{table}
\begin{center}
\begin{tabular}{cccc}
\hline
\multicolumn{2}{c}{ Galaxy}   &     Optical  &  HI  \\
 (NGC) & (UGC) &              &                     \\
\hline
3432     & 5986      &  --- & U       \\
3510     & 6126      &  N   & N       \\
3600     & 6283      &  N   & S       \\
4010     & 6964      &  N   & S       \\
4144     & 7151      &  --- & ---     \\
5023     & 8286      &  --- & N       \\
5229     & 8550      &  S   & S?      \\
5297     & 8709      &  --- & ---     \\
5301     & 8711      &  N   & N       \\
\hline
\end{tabular}
\caption{Comparison of warps in the optical \cite{ss} and HI (this
  work). The first two columns show the NGC and UGC names, and the
  third and fourth columns show the sense of the detected warp, which
  can be either N (anti-clockwise), S (clockwise) or U (bowl
  shaped). NGC 5229 doesn't have a clear detected warp in HI. }
\label{tab:ss}
\end{center}
\end{table}
\end{center}

S\'anchez-Saavedra et al. (1990) studied optical warps using POSS
plates. They looked at all NGC galaxies in the northern hemisphere
that had log$R_{25}$ larger than 0.57, and determined
whether the warp was clockwise 
or anti-clockwise (S or N warps). Nine of their galaxies are in our
sample, so we decided to compare with our results and find out whether
the sense of the warp 
is the same or not. Table~\ref{tab:ss} shows the results. Four of the
galaxies in common do not appear to be warped in the S\'anchez-Saavedra et al.
(1990) study. Two of the warped galaxies have the same sense as
found in
this work, two the opposite, and the last one has a dubious warp in
HI. In the case of UGC 6964 the
difference may be due to dust and/or spiral arms. The optical image
shows patchy emission and it indeed looks bent with an 'N' shape, while
the HI bends in the opposite direction in quite a spectacular
way. In the other case, UGC 6283, the difference is probably due
to the shape of the warp curve itself. The disk bends first slightly in one
direction, turns back and then bends in the opposite
direction. The warp seen in the optical is probably tracing that first
small 'kink' of the warp shape. So in this case the warp is correctly
detected, but the general sense of the warp is not, because the
optical disk ends
before the HI warp changes direction. This galaxy is not completely
edge-on and it is possible that this 'kink' could be due to the
presence of spiral arms.  

Our conclusion from this comparison is that the stellar warps studied
thus far are not directly comparable to the HI warps. The latter occur
at larger galactocentric radius, and generally have a considerably
higher amplitude. At this point it is unclear whether (faint) stellar
disks extend to these radii or not.

\subsection{The radial HI profile and warping}
\label{radial}

In our search for a common property among warped galaxies (with respect to
the non warped ones), we have compared
the radial HI profiles of both types of galaxies. We have excluded galaxies that are 
interacting with nearby neighbors, because a tidal interaction probably
affects the radial profile of a galaxy. We also removed galaxies with very small
warps. Thus we have constructed two groups of galaxies: the
{\em bona fide} warped galaxies, UGC 1281, 2459, 3137,
3909, 5459, 6964, 7125, 7321, 7774, 8286 and 8711; and the {\em
  non}-warped galaxies UGC 7089, 7090, 7151, 7483, 9242.

We have scaled the radial profiles using R$_{\rm HI}$ as unit radius and
the radial densities with M$_{\rm HI}$/D$^2$, where D is the distance to
the galaxy. We have
tried other scaling factors (such as the radius where the density drops
to 0.5 the peak density; and the area below the profile) with no
effect on the conclusions presented here.

\begin{figure*}
\plotfull{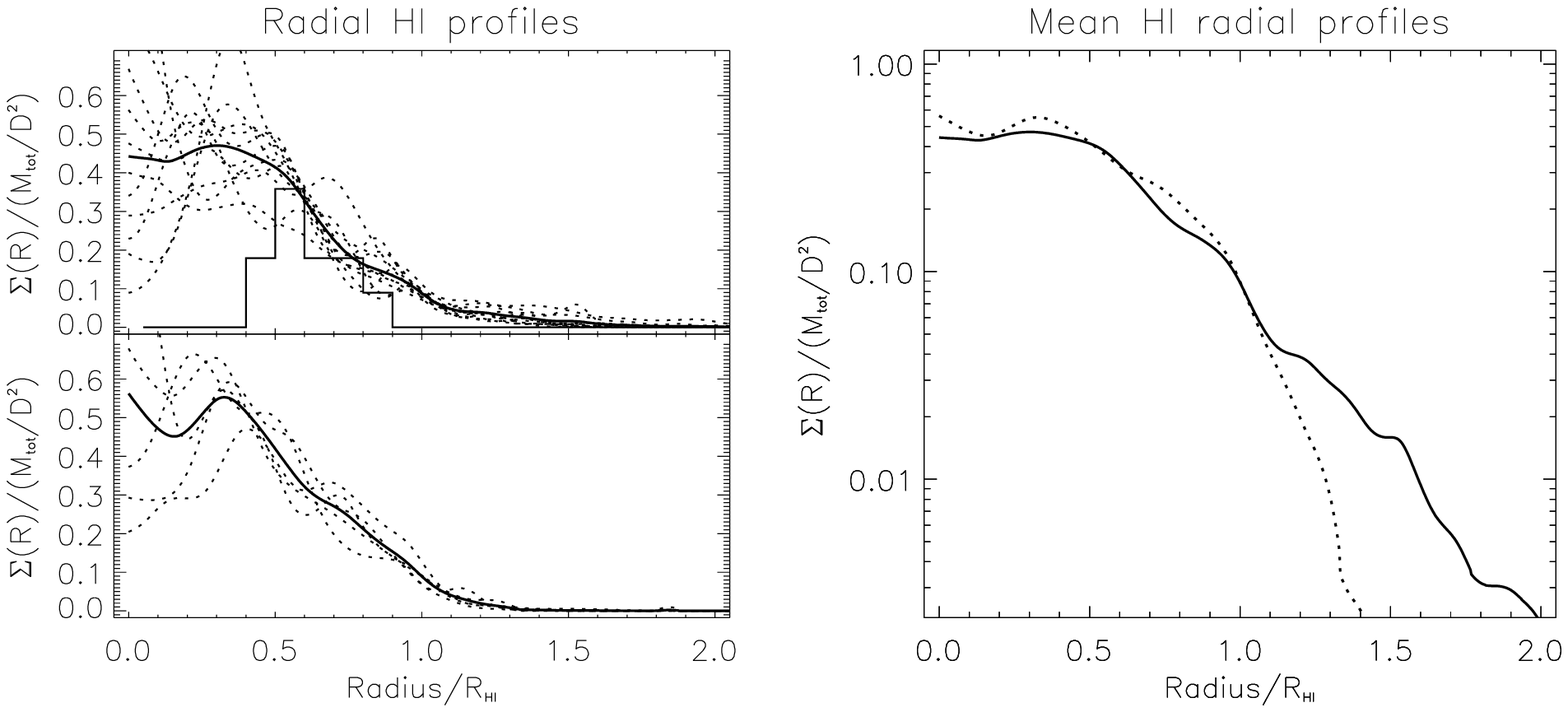}
\caption{Radial HI density profiles. The plot on the left shows the
  HI profiles for {\em bona fide} warped (upper panel) and {\em
    non}-warped (lower panel)
  galaxies (see Scet.~\ref{radial} for definition of both groups). The
  profiles of individual warped and non warped galaxies 
  are plotted in dotted line, and the thick line is the average
  profile for each group. In the upper panel a histogram of the
  radius where the warp begins is overplotted. We have treated both
  sides of the galaxy independently, to take into account that two of
  our {\em bona fine} warped galaxies have only one-sided warps. The
  plot on the right shows the mean profiles for both {\em bona fide} warped
  (solid line) and {\em non} warped galaxies (dotted line).}
\label{fig:hiprof}
\end{figure*}

Fig.~\ref{fig:hiprof} shows the individual profiles as well as the
mean profile for each group of galaxies. A histogram showing the radius at which
warps begin is also plotted. These plots show that warped galaxies
are more extended in HI than non warped galaxies: at around 1.1
R$_{\rm HI}$ the HI density of 
a galaxy with no warp falls much faster than that of a warped galaxy, which
still has some more gas extending up to 1.5 and even 2 R$_{\rm HI}$. This
extra gas we see in the warped galaxies is the same feature as the
``shoulder'' mentioned in Sancisi (1983).


Thus, the typical radial HI profile of a warped galaxy has the
following shape: in the inner parts it is roughly constant or slowly decreases,
around 0.5 R$_{\rm HI}$ it drops faster, and further out it extends at
low levels ending at 1.5-2 R$_{\rm HI}$. It is in the second part of
the profile (the steepest part, around the optical edge) where a warp
develops in most of the galaxies.  

\begin{figure*}
\plotfull{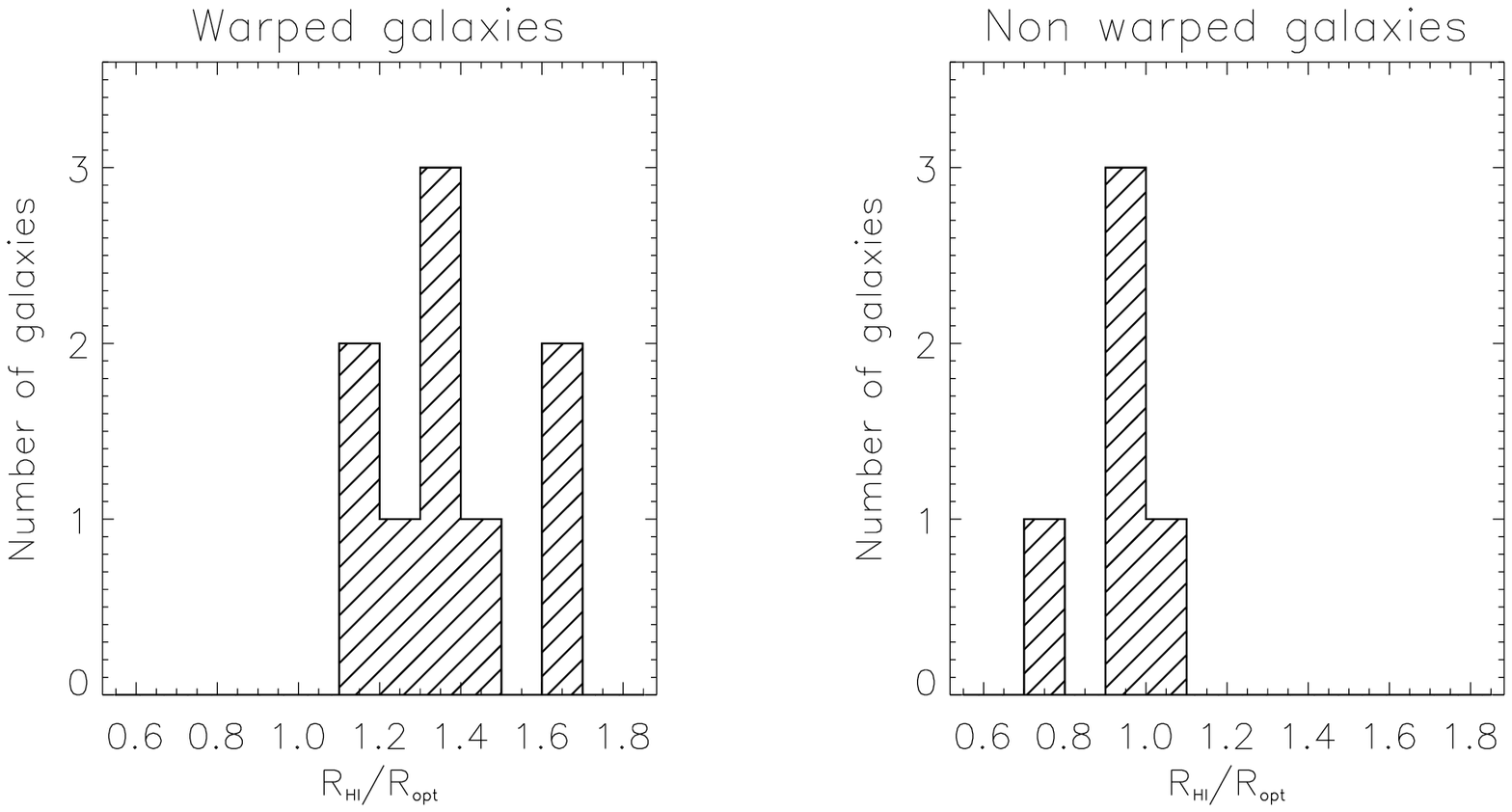}
\caption{Histograms of the ratio R$_{\rm HI}$ to R$_{\rm opt}$ for
  {\em bona fide} warped (left panel) and {\em non}-warped (right
  panel) galaxies (see Sect.~\ref{radial} for definition of both
  groups). The histograms show that warped galaxies are more extended
  in HI than the non warped ones.}
\label{fig:rhiopt}
\end{figure*}

We have also measured the extent of the HI with respect to the optical
in both non warped and warped galaxies. Fig.~\ref{fig:rhiopt} shows
the histograms for both groups, and indicates that warped
galaxies have HI layers more extended with respect to the optical than
galaxies with no warps. There is no galaxy in our sample with an
extended HI disk with respect to the optical disk and no warp. 

\subsection{Warp amplitudes}

\begin{figure*}
\plotfull{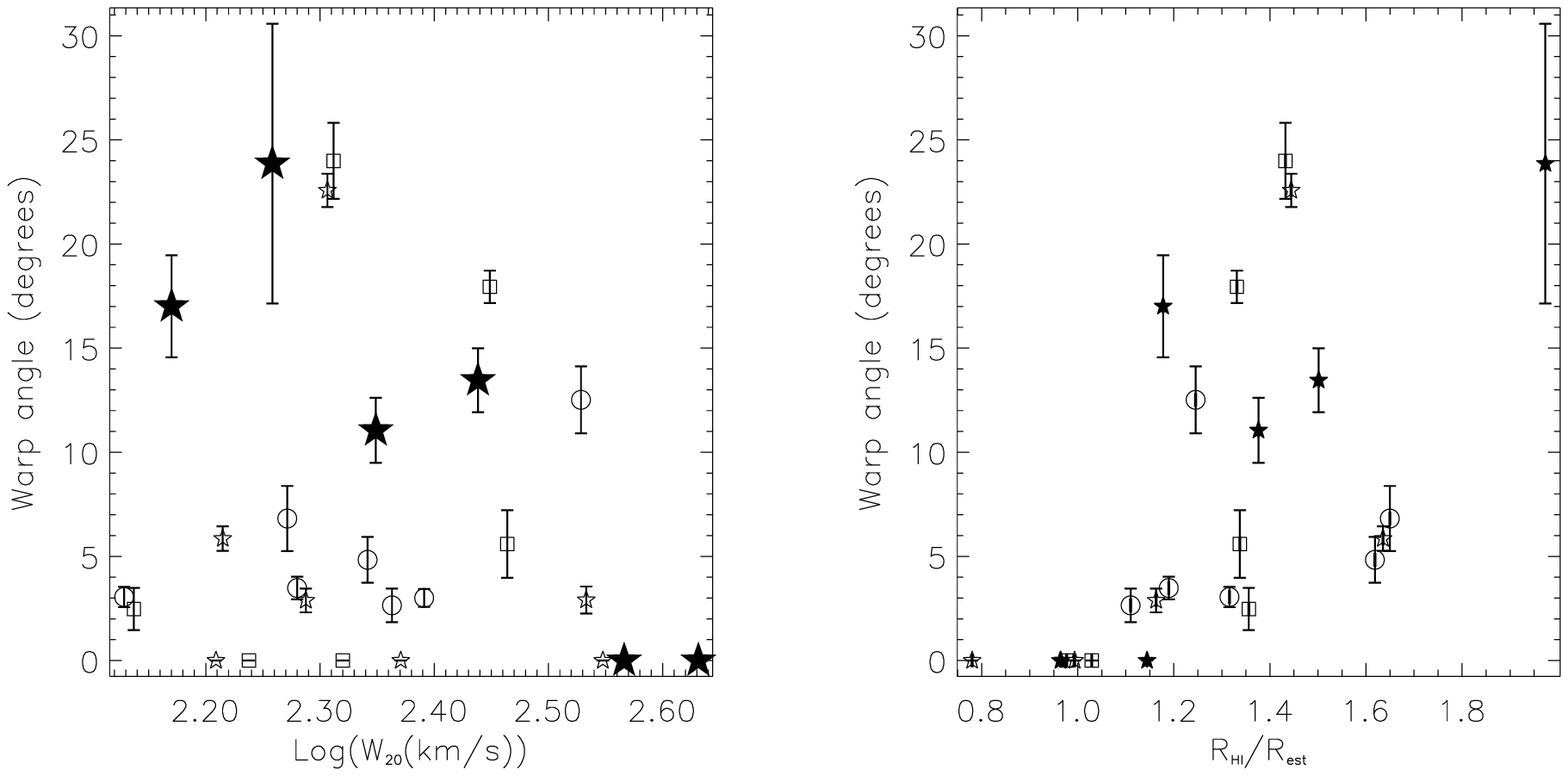}
\caption{Warp amplitude vs. the width of the global HI profile at
  20\% level (left) and vs. 
  R$_{\rm HI}$/R$_{\rm opt}$ (right). The symbols indicate different
  environmental conditions: isolated (open circles), intermediate
  (squares), rich environment 
  (empty stars) and interacting (filled large stars). For a detailed
  description of the environment classification, see
  Sect.~\ref{environment}}
\label{fig:warpamp}
\end{figure*}

We have plotted the warp amplitude (warp angle) vs. the width of the
global HI profile at 20\% level and
vs. the extent of the HI with respect to the optical in Fig.
\ref{fig:warpamp}. The objects with obvious
interaction with nearby companions are shown with large
filled symbols. For this class of objects there seems to be a relation
between the width of the profile and the warp amplitude in
the sense that galaxies with broader profiles have smaller warps. 

For the rest of the galaxies we have not found any relation with
either the visible mass of the galaxy or the width of the global HI
profile. If warps are caused by some force realigning  
the disk of the galaxy (as opposed to inflow of HI gas at a different
angle than that of the inner disk), this means that the force should scale (on
average) with the mass of the system. The amplitude of warps does not seem to be
related to either the density lopsidedness or the kinematical
lopsidedness. The same is true for the asymmetry of warps.

Fig. \ref{fig:warpamp} also shows the amplitude of the warps vs. the
HI extent with respect to the optical. We can see that non warped
galaxies are at low R$_{\rm HI}$/R$_{\rm opt}$, and that large warps only
occur in galaxies with large R$_{\rm HI}$/R$_{\rm opt}$.


\subsection{Mean surface density}

\begin{figure*}
\plotfull{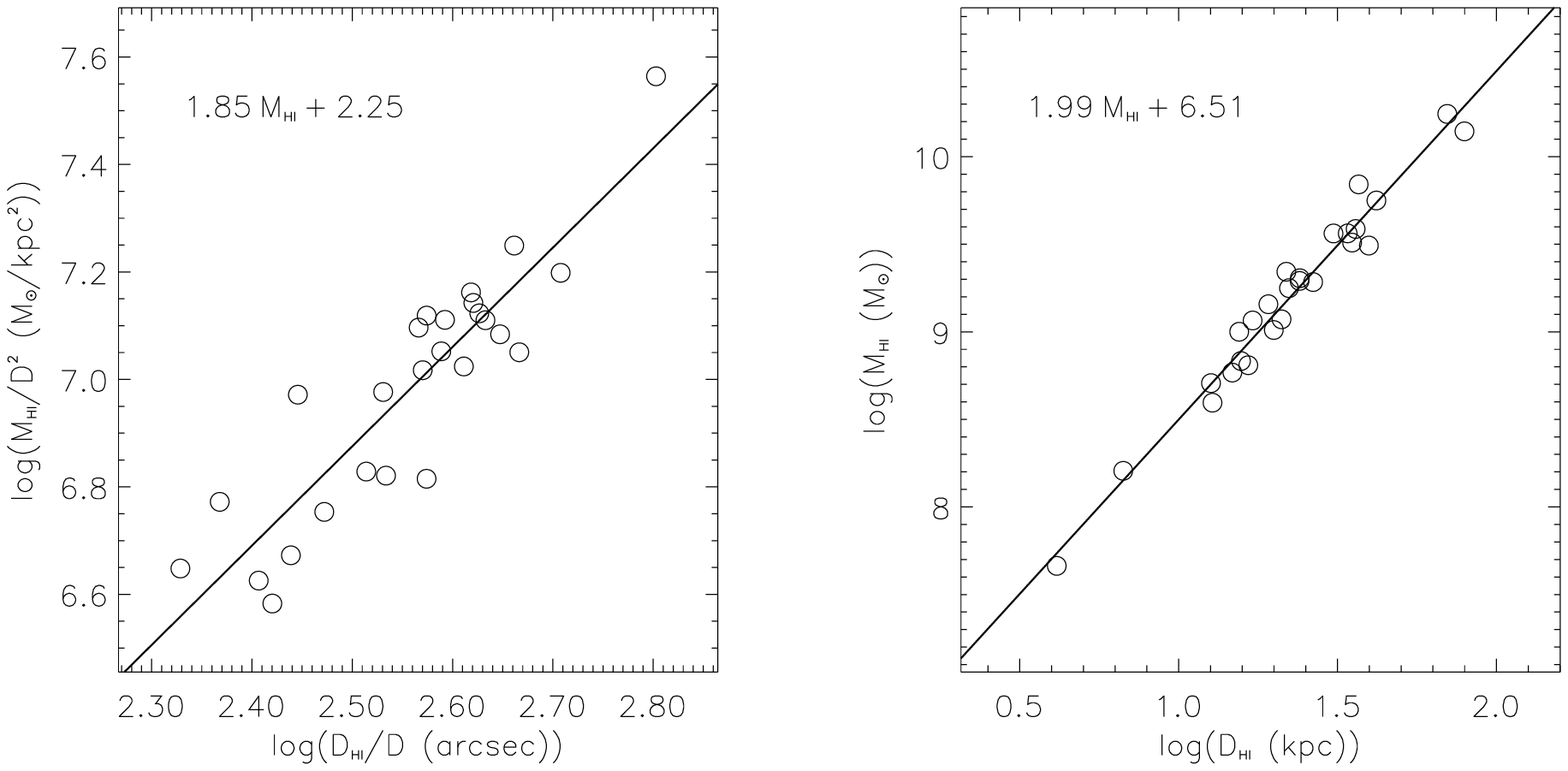}
\caption{HI mass vs. HI radius for all the galaxies in the sample. The
left panel shows the relation among measured quantities ($M_{\rm{HI}}$/D$^2$,
D$_{\rm{HI}}$/D, where D is the distance to the galaxies), while the right
panel shows the relation after the distance has been used to derive
linear diameters and masses of the galaxies.}
\label{fig:mhirhi}
\end{figure*}

It has been shown that there is a tight relation between
HI mass and HI radius \cite{broeils,verheyen}, with a slope close to 2 in a
log-log scale. Fig.~\ref{fig:mhirhi} shows both M$_{\rm{HI}}/D^2$ vs. D$_{\rm{HI}}/D$
relation (measured quantities) and M$_{\rm{HI}}$ vs. D$_{\rm{HI}}$ (after
multiplication by the distance to each galaxy). The slope of the relation changes
from 1.85 to 1.99 in these plots, which illustrates how uncertainty in
the distances to the galaxies can artificially 'enhance' relations and
alter the slopes.
But even a slope of 1.85 implies that the average HI surface
density ($\langle\sigma_{\rm{HI}}\rangle$) is more or less constant
from galaxy to galaxy. The average value is $4.0\pm1.0 M_\odot {\rm pc}^{-2}$,
consistent with the measurements by Broeils \& Rhee (1997).

%

There seems to be a trend of the mean surface density with respect to
the asymmetry of warps, in the sense that galaxies with larger mean
surface densities seem to possess more symmetric warps.  

\section{Conclusions}

We have studied the warps in the outer HI layers of a sample of 26 edge-on
galaxies with particular attention to warp amplitudes, asymmetry,
occurrence rates and their relationship with the environment. We have also studied
the density and kinematic lopsidedness of the HI disks. 

There are indications that environment may play a role in
warping. It is claimed that in rich environments optical warps seem to
be larger and more
frequent \cite{res}, and HI warps are larger and more 
asymmetric. However, the ubiquity of warps, even in low density regions and
the symmetry of some warps suggest that their origin is not simply
the result of interactions with the environment. The fact that all
galaxies with an HI layer more 
extended than the optical show warps of very different amplitudes
could indicate that warping is similar in a way to spiral
structure. Most perturbations of a disk galaxy, arising e.g. from
companions or secondary infall, will provoke a warped response.

The universality of this response makes it hard to pinpoint the
perturbation that preceded it. While the trend with environment
suggests that tidal effects play an important role, it seems likely
that there are other, perhaps more intrinsic, effects at work that
cause even quite isolated galaxies to warp. In this sense, the
situation may be similar to that for spiral structure, another
universal response whose origin is not clear in all details.

Late infall of HI gas might
be responsible for the warps we detect. This would explain the 'extra'
HI gas that warped galaxies have outside R$_{\rm HI}$ while the inner
profiles of warped and non warped galaxies are similar. This late
infall might cause star formation at the outskirts of the
optical disk that could be detected in H$\alpha$ imaging or by looking
at its color.

To summarize, we list the main conclusions:

\begin{itemize}

\item

We detect warps in 20 out of our 26 sample galaxies confirming that 
warping of the HI disks is a very common phenomenon in disk
galaxies. In fact all galaxies that have an HI disk more extended
than the optical are warped.

\item

The amplitude of warps varies considerably from galaxy to galaxy. Also
for a given galaxy there can be a large asymmetry in amplitude and shape
between the two sides. A large number of warps in our sample are
asymmetric. 
Most of the galaxies with both sides warped are antisymmetric (S
shape warps). We only have two cases of U-type warps, and both
galaxies are strongly interacting with nearby companions and are very
disturbed. 

\item

The warping of the disks usually starts near the edge of the optical
disk where the HI density drops down.

\item

The connection between HI and optical warps is not clear.
HI warps are found in general at larger radii than the
optical ones, and as a consequence they probe a different region of
the potential of the galaxy. A joint optical+HI study of warps could
give important insights on the formation mechanism(s) of warps.

\item

There seems to be a dependence of warps on environment in the sense that galaxies in
rich environments tend to have larger and more asymmetric warps than
galaxies in poor environments. 

\item

The presence of density lopsidedness (and in a weaker way that of
ki\-ne\-matical lopsidedness) seems to be related to the presence of
nearby companions. 

\end{itemize}

\section*{Acknowledgments}

IGR wishes to thank Jorge Jim\'enez-Vicente for many stimulating
discussions and useful suggestions. We are grateful to Martin Vogelaar
and Hans Terlow for help and assistance in the GIPSY software package.
The Westerbork Synthesis Radio Telescope is operated by the
Netherlands Foundation for Research in Astronomy with financial
support from the Netherlands Organization for Scientific Research
(NWO). This research has made use of the
NASA/IPAC Extragalactic Database (NED) which is operated by the Jet
Propulsion Laboratory, California Institute of Technology, under contract with
the National Aeronautics and Space Administration. 

\newcommand{\apj}{ApJ}
\newcommand{\mnras}{MNRAS}


\setcounter{dummyyoyo}{\thetable}
\begin{center}
\begin{table*}
\begin{center}
\begin{tabular}{|c|l|c|c|r|r|r|r}
\hline
Galaxy & \multicolumn{1}{c|}{Companion} & RA & DEC & \multicolumn{1}{c|}{dist} & \multicolumn{1}{c|}{V$_{NED}$} & \multicolumn{1}{c|}{V$_{HI}$} \\
\hline
\hline
1281  &                          & 01 46 38.1 & 32 20 15   &        &   157  &   157\\
\hline
2459  &                          & 02 57 08.4 & 48 50 43   &        &   2464 &   2469\\
      & HFLLZOA G144.00-08.53    & 02 58 57.7 & 48 43 03   &   19.6 &        &   2456\\
\hline
3137  &                          & 04 39 22.4 & 76 19 37   &        &   992  &   993\\
\hline
3909  &                          & 07 30 54.0 & 73 49 28   &        &   945  &   943\\
\hline
4278  &                          & 08 10 27.4 & 45 53 43   &        &   563  &   555\\
      & NGC 2537A                & 08 10 09.0 & 46 08 46   &   15.4 &   443  &   \\
      & NGC 2537                 & 08 09 42.6 & 46 08 32   &   16.7 &   447  &   445\\
\hline
4806  &                          & 09 06 30.2 & 33 19 38   &        &   1947 &   1948\\
      & KUG 0906+333A            & 09 06 16.0 & 33 19 31   &   3.0  &        &   $\simeq$1850\\
      & -------                  & 09 06 43.2 & 33 20 52   &   3.0  &        &   $\simeq$1970\\
      & -------                  & 09 06 58.3 & 33 09 14   &   11.9 &        &   $\simeq$2020\\
      & IC 2445                  & 09 10 10.9 & 32 00 53   &   91.4 &   1968 & \\
      & CG 0010                  & 09 09 50.3 & 32 53 15   &   96.1 &   1860 & \\
      & UGC 04777                & 09 03 34.8 & 32 49 14   &   96.7 &   2052 & \\
\hline
5452  &                          & 10 04 17.1 & 33 16 20   &        &   1342 &   1340\\
      & UGC 5446                 & 10 03 36.3 & 33 11 27   &   9.8  &   1383 &   1362\\
      & UGC 5482                 & 10 07 22.9 & 33 31 23   &   59.5 &   1466 & \\
      & KUG 1006+322             & 10 06 04.2 & 33 15 22   &   65.0 &   1390 & \\
      & UGC 5393                 & 09 58 46.3 & 33 22 41   &   69.4 &   1448 & \\
\hline
5459  &                          & 10 04 54.7 & 53 19 39   &        &   1112 &   1109\\
      & UGC 5460                 & 10 04 55.0 & 52 05 20   &   74.3 &   1093 & \\
      & UGC 5479                 & 10 06 51.9 & 54 44 53   &   87.0 &   1105 & \\
\hline
5986  &                          & 10 49 42.9 & 36 53 09   &        &   616  &   615\\
      & UGC 5983                 & 10 49 27.4 & 36 51 36   &   3.5  &   $^{(1)}$     & \\
      & SBS 1054+365             & 10 54 59.8 & 36 31 30   &   67.1 &   603  &   \\
      & VV 747                   & 10 55 00.2 & 36 31 43   &   67.1 &   634 & \\
      & CGCG 184-040 NED02       & 10 55 00.0 & 36 31 00   &   67.3 &   665 & \\
\hline
6126  &                          & 11 01 00.6 & 29 09 17   &        &   705  &   704\\
      & [M98k] 110110.0+285730   & 11 01 10.0 & 28 57 30   &   12.1 &        &   $\simeq$660\\
      & UGCA 225                 & 11 02 15.6 & 29 24 34   &   22.4 &   646  &   $\simeq$640\\
      & UGC 6102                 & 10 59 05.2 & 28 57 30   &   27.8 &   702  &   $\simeq$700\\
      & NGC 3486                 & 10 57 40.4 & 29 14 37   &   44.0 &   681 & \\
      & BTS 026                  & 10 57 57.0 & 29 58 18   &   63.2 &   675 & \\
\hline
6283  &                          & 11 13 06.5 & 41 51 50   &        &   719  &   713\\
\hline
6964  &                          & 11 56 02.9 & 47 32 20   &        &   907  &   902\\
      & NGC 3949                 & 11 51 05.2 & 48 08 13   &   61.5 &   807 & \\
      & NGC 3985                 & 11 54 06.8 & 48 36 44   &   67.3 &   950 & \\
      & UGCA 259                 & 11 56 18.8 & 46 00 46   &   91.6 &   1154 & \\
      & MRK 1460                 & 11 48 12.8 & 48 31 46   &   98.5 &   768 & \\
\hline
7089  &                          & 12 03 25.6 & 43 25 25   &        &   776  &   774\\
      & UGC 7094                 & 12 03 38.5 & 43 14 05   &   11.6 &   780  &   786\\
      & NGC 4111                 & 12 04 31.0 & 43 20 37   &   12.8 &   807 & \\
      & MAPS-NGP O\_217\_0067378   & 12 03 26.8 & 43 10 52   &   14.5 &   756 & \\
      & NGC 4117                 & 12 05 14.1 & 43 24 17   &   19.7 &   943 & \\
      & NGC 4118                 & 12 05 20.8 & 43 23 22   &   21.0 &   661 & \\
      & NGC 4138                 & 12 06 58.6 & 43 57 46   &   50.3 &   888 & \\
      & NGC 4143                 & 12 07 04.8 & 42 48 43   &   54.3 &   985 & \\
      & IC 0750                  & 11 56 17.6 & 43 00 92   &   82.0 &   701 & \\
      & IC 0749                  & 11 55 59.5 & 43 00 44   &   84.9 &   784 & \\
      & NGC 4183                 & 12 10 46.6 & 43 58 33   &   86.3 &   930 & \\
      & NGC 4051                 & 12 00 36.4 & 44 48 35   &   88.5 &   725 & \\
      & UGC 7129                 & 12 06 23.4 & 42 01 10   &   90.3 &   926 & \\
      & NGC 4013                 & 11 55 56.6 & 44 13 32   &   94.2 &   834  &   \\
\hline
\multicolumn{5}{l}{\scriptsize (1) No velocity available, but the optical image shows interaction with UGC 5986}
\end{tabular}
\caption{Companions to the galaxies in the sample within 100
  arcmin. We have only considered galaxies that have a heliocentric
  velocity that differs less than 150 km/s from that of the host
  galaxy, or have a clear pattern of interaction with it. We present
  the names of the host galaxy, the names of the companions,
  positions, distances of the companions to the host galaxy,
  heliocentric velocity from NED, and measured heliocentric
  velocities. }
\label{tab:comp}
\end{center}
\end{table*}
\end{center}

\setcounter{table}{\thedummyyoyo}

\begin{center}
\begin{table*}
\begin{center}
\begin{tabular}{|c|l|c|c|r|r|r|}
\hline
Galaxy & \multicolumn{1}{c|}{Companion} & RA & DEC & \multicolumn{1}{c|}{dist} & \multicolumn{1}{c|}{V$_{NED}$} & \multicolumn{1}{c|}{V$_{HI}$} \\
\hline\hline
7090  &                          & 12 03 28.5 & 47 45 20   &        &   566  &   581\\
      & PC 1200+4755\hspace{18mm} & 12 00 26.7 & 47 55 57   &   32.3 &   600  &   \\
      & NGC 4144                 & 12 07 27.5 & 46 44 09   &   73.4 &   265 & \\
\hline
7125  &                          & 12 06 10.2 & 37 04 51   &        &   1071 &   1072\\
      & UGC 7207                 & 12 09 47.8 & 37 17 30   &   45.2 &   1051 & \\
      & UGC 7257                 & 12 12 32.1 & 36 14 13   &   91.8 &   942 & \\
\hline
7151  &                          & 12 07 27.5 & 46 44 09   &        &   265  &   263\\
      & NGC 4096                 & 12 03 28.5 & 47 45 20   &   73.4 &   566 & \\
      & NGC 4242                 & 12 15 01.2 & 45 53 47   &   93.1 &   517 & \\
      & MRK 1471                 & 12 15 21.9 & 47 41 10   &   98.7 &   484 & \\
\hline
7321  &                          & 12 15 02.0 & 22 49 05   &        &   408  &   407\\
\hline
7483  &                          & 12 21 41.8 & 31 47 56   &        &   1253 &   1247\\
      & UGC 7428                 & 12 19 32.6 & 32 22 21   &   44.0 &   1137 &       \\
      & KUG 1218+310             & 12 18 05.6 & 31 04 36   &   63.3 &   1010 &       \\
      & NGC 4314                 & 12 20 01.8 & 30 10 21   &   99.9 &   963  &       \\
\hline
7774  &                          & 12 33 57.4 & 40 16 49   &   526  &   527  &       \\
      & UGC 07678                & 12 29 34.4 & 40 06 32   &   51.3 &   685  &       \\
      & UGC 7751                 & 12 32 46.6 & 41 20 10   &   64.8 &   605  &       \\
      & UGC 7719                 & 12 31 34.6 & 39 17 42   &   65.2 &   681  &       \\
      & UGCA 290                 & 12 34 56.5 & 39 01 08   &   76.5 &   445  &       \\
      & NGC 4618                 & 12 39 09.0 & 41 25 29   &   90.5 &   544  &       \\
      & NGC 4625                 & 12 39 29.0 & 41 32 50   &   98.5 &   609  &       \\
\hline
8246  &                          & 13 07 44.3 & 34 26 47   &        &   813  &   807\\
      & NGP9 F269-1340356        & 13 08 09.0 & 34 30 09   &   6.1  &        &   $\simeq$850 \\
      & UGC 8323                 & 13 12 29.3 & 35 08 42   &   72.0 &   856  &       \\
      & UGC 8261                 & 13 08 41.9 & 35 46 00   &   80.1 &   852  &       \\
      & IC 4213                  & 13 09 52.1 & 35 56 11   &   93.1 &   815  &       \\
      & UGC 8181                 & 13 03 02.9 & 33 10 02   &   96.5 &   886  &       \\
\hline
8286  &                          & 13 09 58.1 & 44 18 14   &        &   407  &   406\\
\hline
8396  &                          & 13 19 09.3 & 38 47 57   &        &   946  &   947 \\
      & NGC 5112                 & 13 19 41.4 & 38 59 44   &   13.3 &   965  &   971 \\
      & UGC 8315                 & 13 11 53.6 & 39 24 41   &   92.2 &   1164 &       \\
\hline
8550  &                          & 13 31 58.4 & 48 10 16   &        &   364  &   359\\
      & MESSIER 051b             & 13 27 52.4 & 47 31 32   &   56.6 &   465  &       \\
      & MESSIER 051a             & 13 27 46.0 & 47 27 22   &   60.3 &   463  &       \\
      & SBS 1331+493             & 13 31 19.5 & 49 21 28   &   71.5 &   599  &       \\
\hline
8709  &                          & 13 44 18.4 & 44 07 18   &        &   2407 &   2402\\
      & NGC 5296                 & 13 44 13.5 & 44 06 02   &   1.5  &   2224 &       \\
      & UGC 08733                & 13 46 33.8 & 43 39 37   &   36.9 &   2338 &       \\
      & UGC 08798                & 13 50 44.5 & 44 04 25   &   69.4 &   2275 &       \\
      & NGC 5336                 & 13 50 05.4 & 43 29 21   &   73.2 &   2338 &       \\
\hline
8711  &                          & 13 44 21.4 & 46 21 25   &        &   1503 &   1506\\
\hline
9242  &                          & 14 23 19.7 & 39 45 50   &        &   1440 &   1438\\
      & NGC 5582                 & 14 18 41.3 & 39 55 18   &   54.3 &   1435 &       \\
      & MRK 0676                 & 14 17 23.7 & 40 05 11   &   70.9 &   1734 &       \\
\hline
\end{tabular}
\caption{ -- {\it continued}}
\end{center}
\end{table*}
\end{center}

\setcounter{dummyfofo}{\thefigure}
\appendix
\section{Determination of the warp radius on an edge-on galaxy}
\label{rwarp}

In this appendix we explain how we have determined the warp radius for our
galaxies using Gaussian fits perpendicular to the
major axis of the galaxy. We take both the centroids of the Gaussian
fits (we refer to them as datapoints) and their uncertainties ($\sigma_i$). 
The basic idea is to determine the radius at which the
galactic disk begins to bend away from the plane defined by the inner
disk in a systematic way. 

We take the innermost R$_{25}$/2 part of the disk as the inner disk,
which we assume is flat. We fit a straight line to it and determine
the vertical offset and the dispersion ($\sigma_{\rm inner}$). The
offset will correct (to first order) possible errors in the
determination of the center of the galaxy, and we can use the
dispersion as an estimate of how noisy our disk is (influence of not
completely edge-on spiral arms and other irregularities).  We then
define the warp radius as the location where a significantly large
portion of the disk (at least R$_{25}$/4) departs coherently more than
$\sigma_{\rm inner}+\sigma_i$ from the plane defined by the inner
disk. $\sigma_i$ are the uncertainties of the individual datapoints in
the warped portion of the disk.

If the uncertainties in the datapoints are large we could miss some
warps with this procedure. Even with large errors, a coherent smooth
warp could be detected if we smooth the warp curve and apply the
procedure again. Thus we smooth the warp curve to different
resolutions and determine a
preliminary warp radius at each resolution. The final warp radius is
the minimum value of all the preliminary values. The outcome of 
this method is shown in the Atlas in Fig. \ref{atlas}, where  warp
curves and warp radii are shown on the same panel. 


\onecolumn

\setcounter{figure}{\thedummyfofo}
\begin{figure*}
\plotsmall{vacio2.ps}
\caption{
In the following pages we present an overview of the HI and optical
data for all the galaxies in the sample. For each galaxy we have
created a six panel figure, in which we have displayed the relevant
information.  We have used the 30$''$ resolution data for the HI
except for the XV diagram where we used the full resolution data.
These panels are available in electronic form in the WHISP pages ({\em
  http://www.astro.rug.nl/$\sim$whisp}). For each galaxy we have
tabulated the rotation curve, the HI radial surface density profile
and the warp curve. There is extra material as well that supplies
extra velocity information.{\em Top left}:
R-band optical image, rotated so that the major axis of the galaxy is
on the horizontal axis. Vertical dashed lines have been plotted at
R$_{25}$. 
{\em Top right}:
Warp curve derived from the total intensity HI map using the method
described in Sect.\ref{warpcurve}. Note that the scale is
not the same in the horizontal and vertical axes, and that this ratio
differs from galaxy to galaxy. This allows the reader to follow the
warp curve more precisely. A plot with the same scale on both axes is
shown in Fig.~\ref{fig:plotwl2}. The vertical dot-dashed lines mark
the warp radius.
{\em Middle left}:
Total HI map. The thick contour is the locus of the points that have
three sigma rms noise as defined in Sect.~\ref{totalhi}. Note that this
does not mean that the line profiles have emission at three sigma level, as
low-level emission may add-up to more than three sigma. The other
contours levels are at 1, 2, 3, 4, 8, 16,... times the lowest contour,
whose level is indicated in the caption of each galaxy). The beam size
is indicated by the symbol at the lower right, R$_{\rm HI}$ (radius
at 1 $M_{\odot}/{\rm pc}^2$) by vertical dashed lines and the warp radii by
vertical dash-dotted lines.
{\em Middle right}:
Radial HI surface density profile. The right (left) side is plotted in
solid (dotted) lines and the mean profile with a thick line. Section
\ref{hiprof} explains how the 
observed integrated emission is deconvolved to derive the radial
profile. The solid (dotted) arrow indicates the warp radius for the
right (left) side of the galaxy. In the cases where the warp radius is
the same for both sides a double-head arrow is plotted.
{\em Bottom left}:
Position-Velocity (heliocentric) diagram along the major axis of the galaxy. 
Full
resolution data are plotted here. For a more extended rotation curve
determined combining high resolution 
and 30 arcsec resolution data, see Fig.~\ref{fig:plotwl2}.
Contours have been drawn at -3$\sigma$, -1.5$\sigma$, 1.5$\sigma$, 3$\sigma$,
4.5$\sigma$, 6$\sigma$, 9$\sigma$, 12$\sigma$,... 
(the value of 3$\sigma$ is indicated in the caption of each galaxy). 
Negative contours are dashed. The derived rotation
curve (see section \ref{rc}) is shown as black dots with error
bars. Note that these maps give the HI distribution only along the
major axis and, therefore, do not contain the warped HI.
{\em Bottom right}:
Global profile, with the systemic velocity indicated by the dotted
horizontal line. 
}
\label{atlas}
\end{figure*}
\setcounter{figure}{\thedummyfofo}


\begin{figure}
\plotatlas{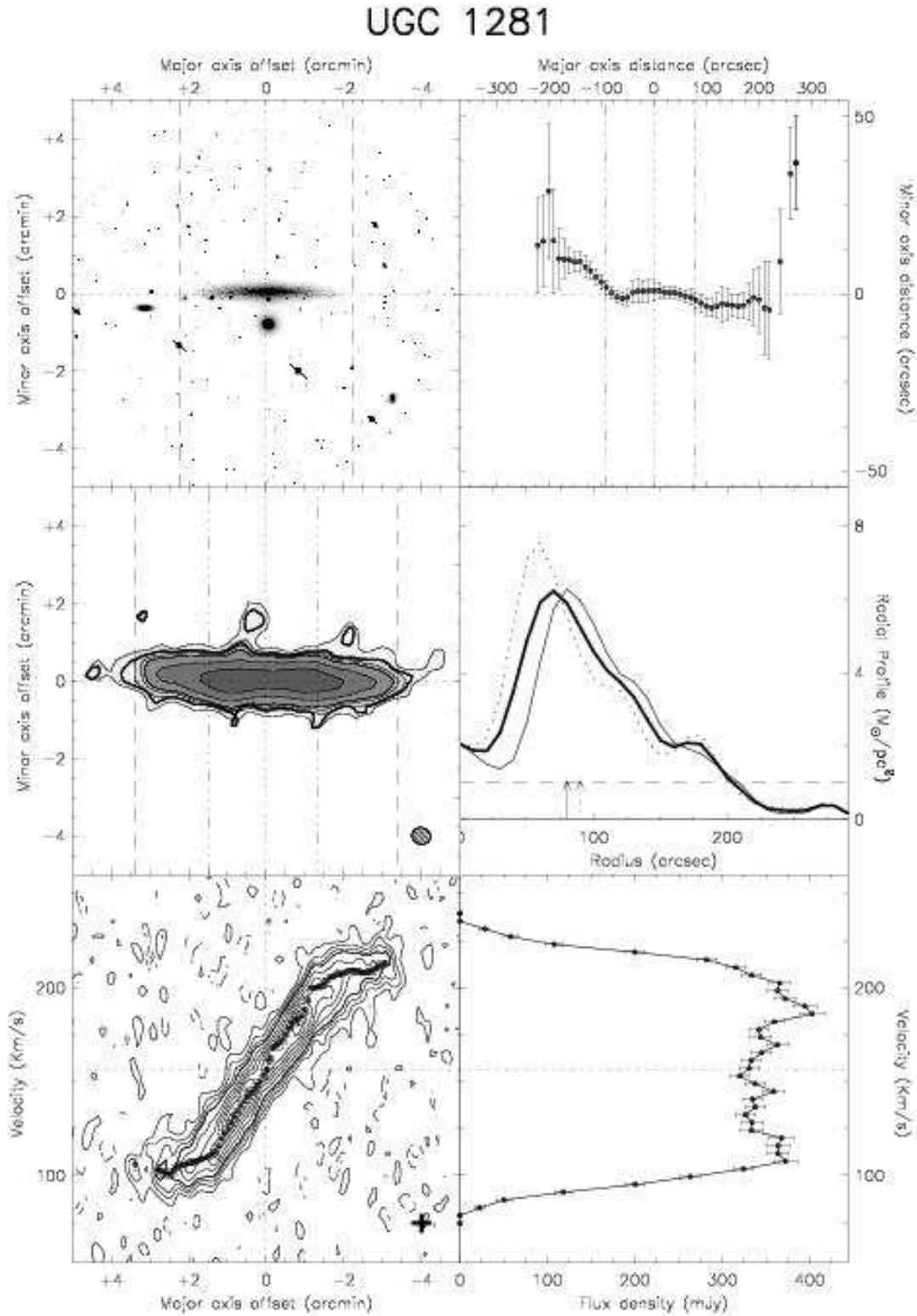}
\caption{-- {\it continued}.
\noindent {\bf UGC 1281:} The warp is visible in the extreme velocity
channels on both sides of the disk.
For this galaxy the lowest contour in the total HI map is $1.5\,10^{20}$ HI atoms/cm$^2$, and in the XV diagram $3\sigma = 13.4$ K.
}
\end{figure}
\setcounter{figure}{\thedummyfofo}
\clearpage

\begin{figure}
\plotatlas{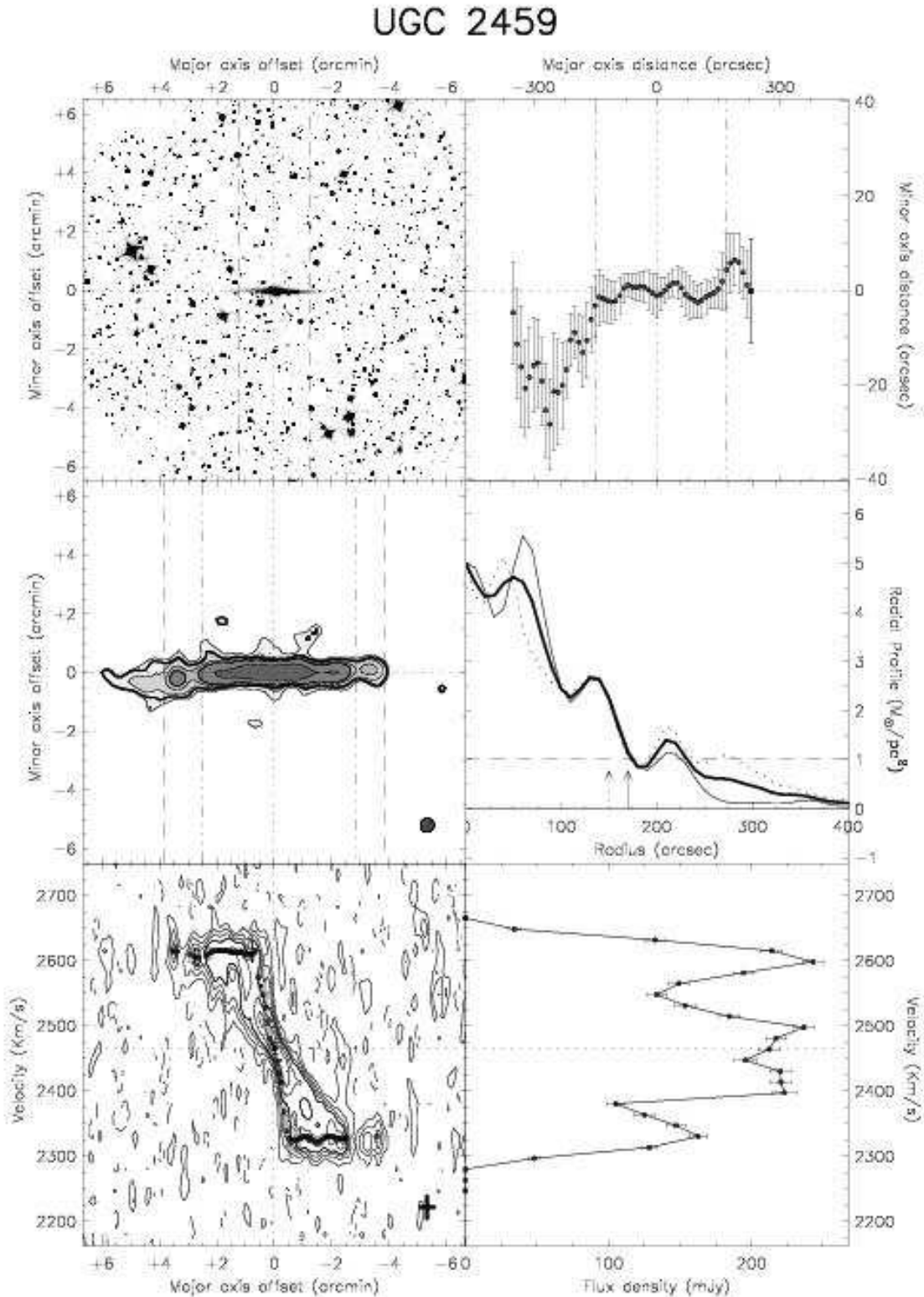}
\caption{-- {\it continued}.
\noindent {\bf UGC 2459:} This galaxy is located at very low galactic
latitude (-8.6\degr) and the optical image is highly affected by
interstellar extinction. The total HI map shows evidence for spiral
arms around R$_{\rm HI}$ and just before that the disk warps on both
sides. The distribution of HI is asymmetric: the right side warp
is very small, while on the left side the disk bends downwards and
then turns back towards the plane of the inner disk as in the Galaxy.
For this galaxy the lowest contour in the total HI map is $3\,10^{20}$ HI atoms/cm$^2$, and in the XV diagram $3\sigma = 8.53$ K.
}
\end{figure}
\setcounter{figure}{\thedummyfofo}
\clearpage

\begin{figure}
\plotatlas{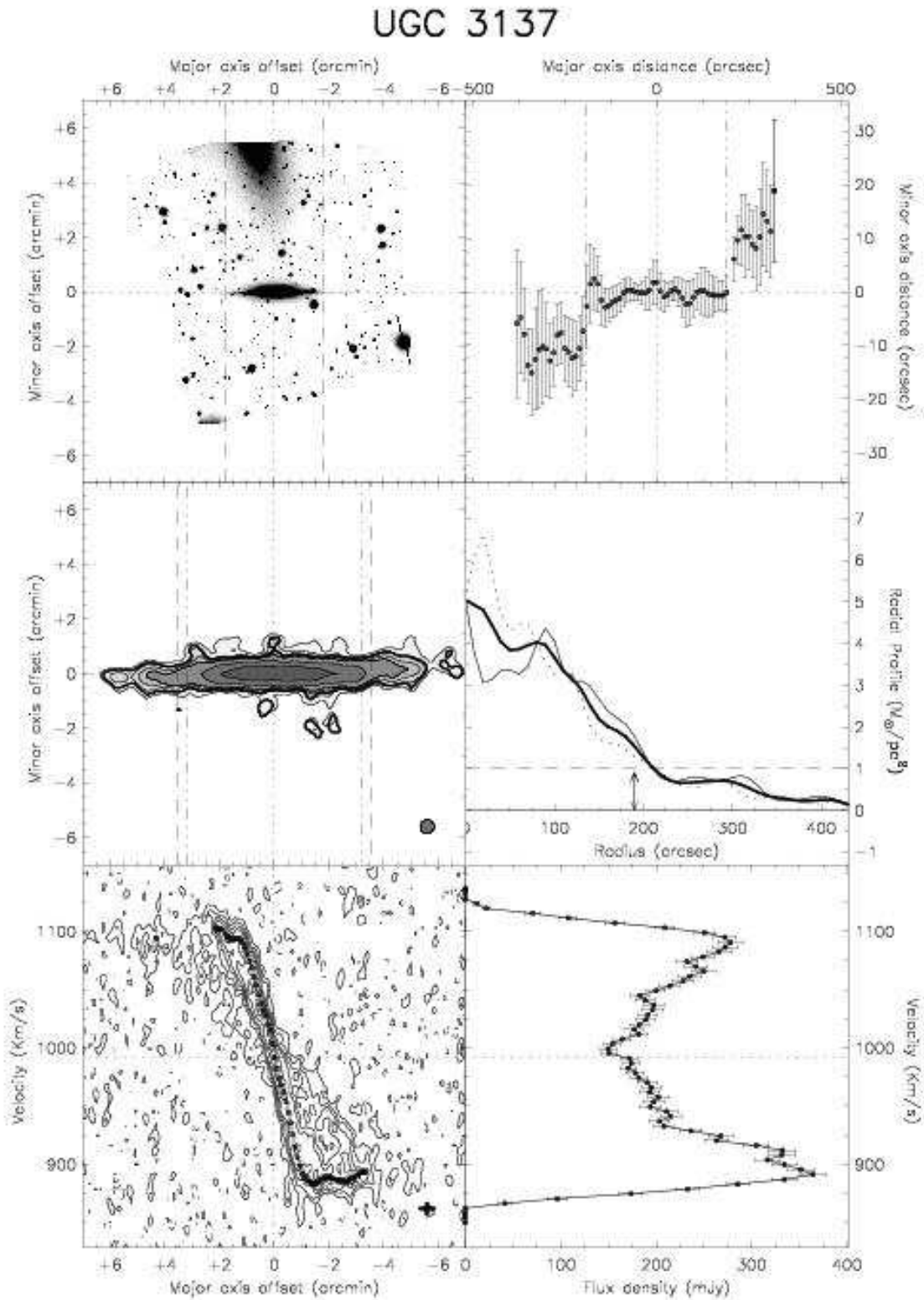}
\caption{-- {\it continued}.
\noindent {\bf UGC 3137:} This galaxy has a low level, very
symmetric warp that bends suddenly and then stays at a fixed height,
resembling a step. 
For this galaxy the lowest contour in the total HI map is $1.6\,10^{20}$ HI atoms/cm$^2$, and in the XV diagram $3\sigma = 18.9$ K.
}
\end{figure}
\setcounter{figure}{\thedummyfofo}
\clearpage

\begin{figure}
\plotatlas{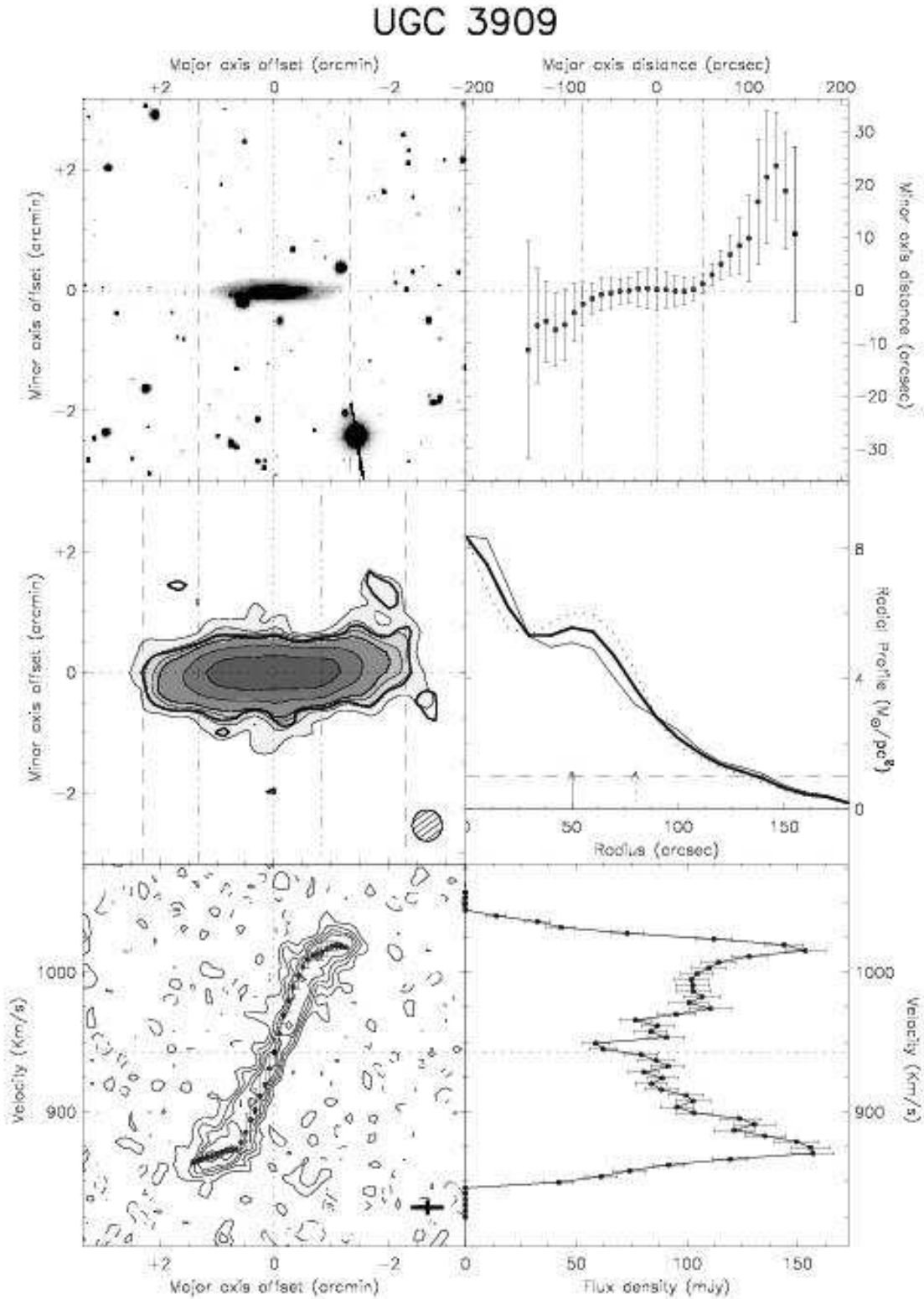}
\caption{-- {\it continued}.
\noindent {\bf UGC 3909:} The optical disk seems to warp in the same
sense as the HI.
For this galaxy the lowest contour in the total HI map is $1.2\,10^{20}$ HI atoms/cm$^2$, and in the XV diagram $3\sigma = 21.7$ K.
}
\end{figure}
\setcounter{figure}{\thedummyfofo}
\clearpage

\begin{figure}
\plotatlas{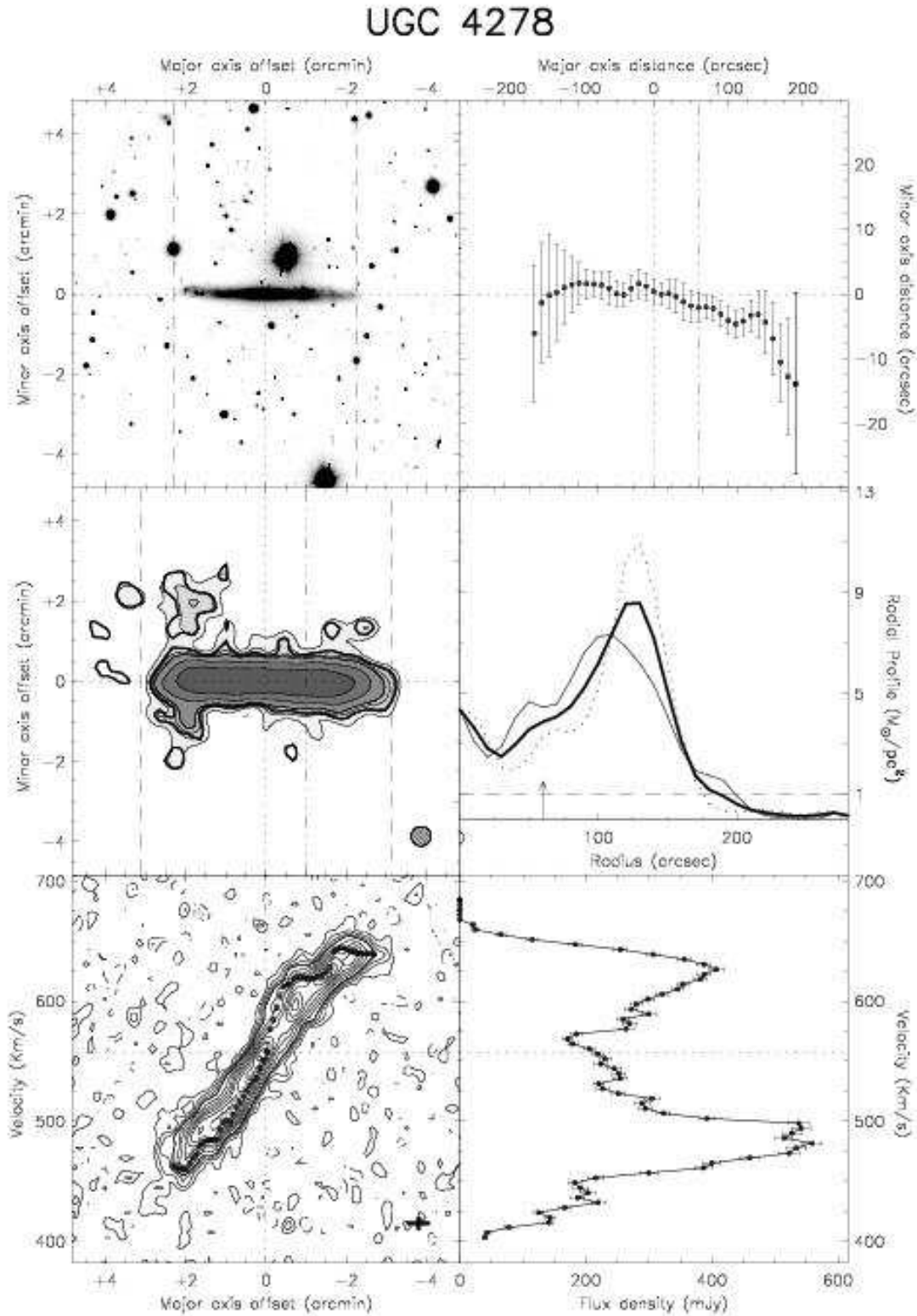}
\caption{-- {\it continued}.
\noindent {\bf UGC 4278:} The main disk of this galaxy is surrounded
by some faint clouds on the left side. The galaxy is also
kinematically lopsided.
For this galaxy the lowest contour in the total HI map is $1.6\,10^{20}$ HI atoms/cm$^2$, and in the XV diagram $3\sigma = 12.8$ K.
}
\end{figure}
\setcounter{figure}{\thedummyfofo}
\clearpage

\begin{figure}
\plotatlas{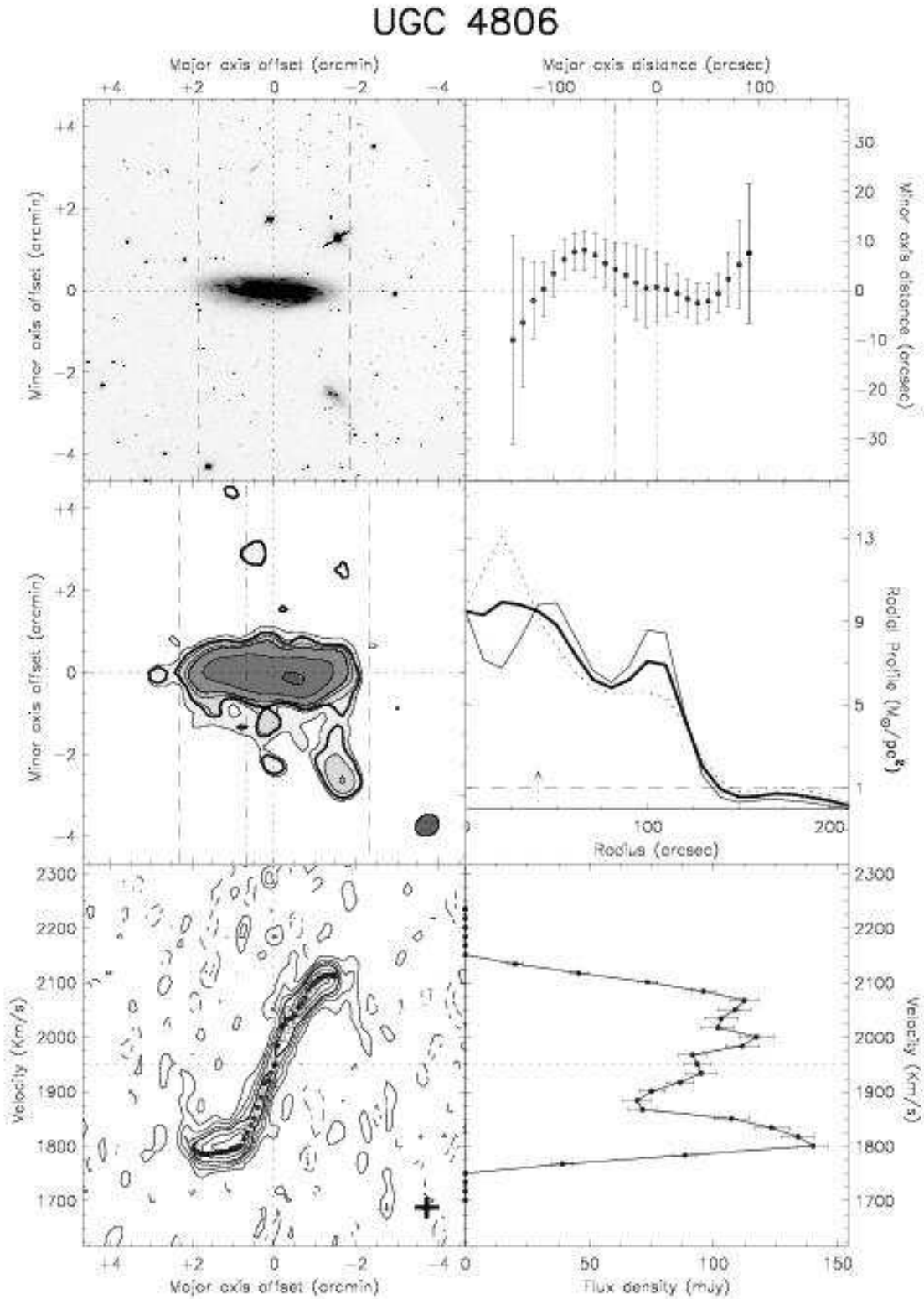}
\caption{-- {\it continued}.
\noindent {\bf UGC 4806:} We detected three galaxies in HI within a
radius of 12$'$ around this galaxy. The HI disk is only slightly
larger than the optical and shows no warp.
For this galaxy the lowest contour in the total HI map is $2.2\,10^{20}$ HI atoms/cm$^2$, and in the XV diagram $3\sigma = 6.42$ K.
}
\end{figure}
\setcounter{figure}{\thedummyfofo}
\clearpage

\begin{figure}
\plotatlas{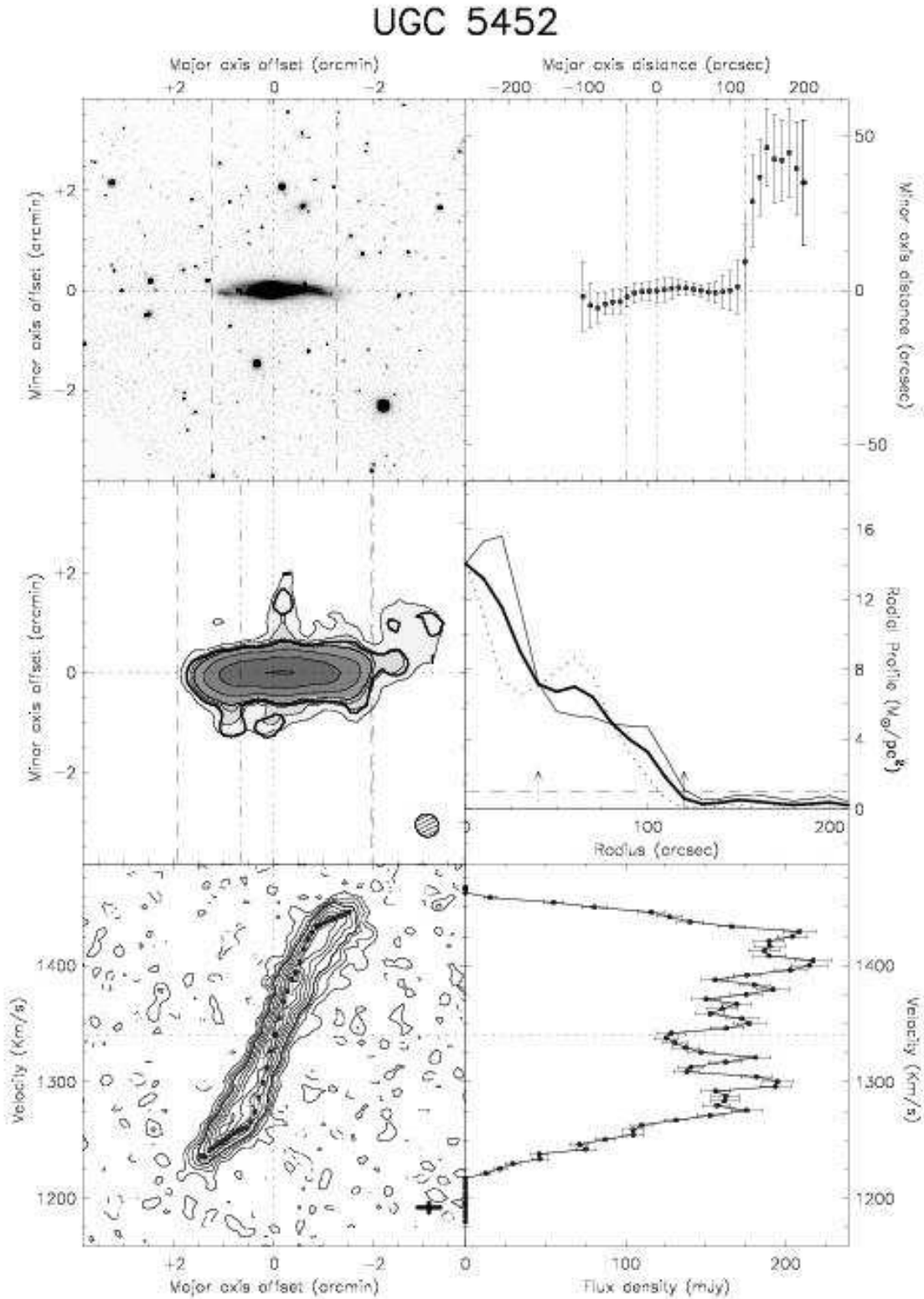}
\caption{-- {\it continued}.
\noindent {\bf UGC 5452:} The emission in the region of the warp (right side, 
see total HI map) is very faint and barely detected.
The one-sided warp of this galaxy pointing
towards the galaxy UGC 5446 suggests a tidal origin for it. The
optical image seems to bend in the opposite direction to that of the HI warp
in the right side.
For this galaxy the lowest contour in the total HI map is $1.4\,10^{20}$ HI atoms/cm$^2$, and in the XV diagram $3\sigma = 9.29$ K.
}
\end{figure}
\setcounter{figure}{\thedummyfofo}
\clearpage

\begin{figure}
\plotatlas{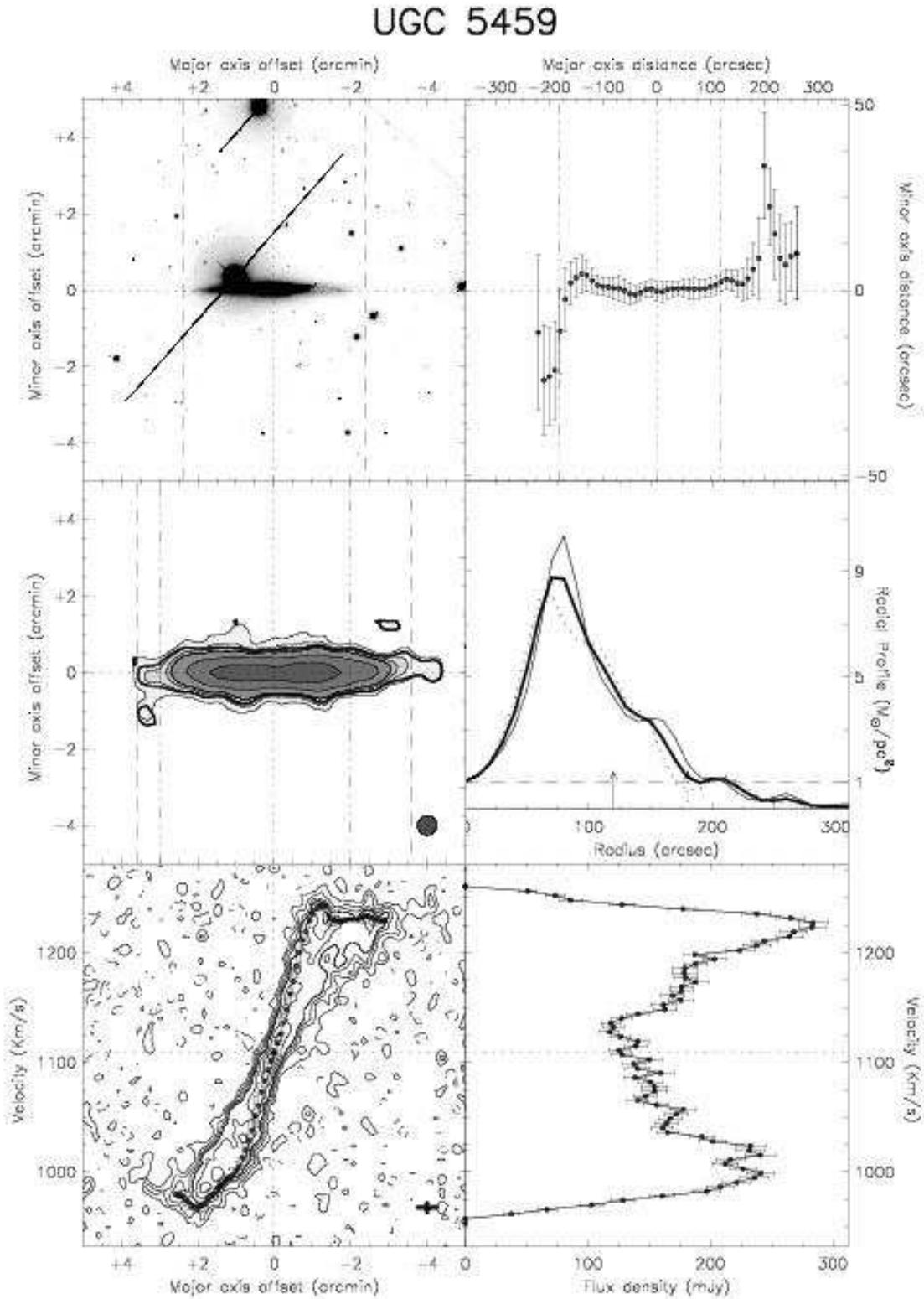}
\caption{-- {\it continued}.
\noindent {\bf UGC 5459:} This is a prototype of a kinematically
lopsided galaxy. The rotation curve rises fast and remains
approximately flat on
the receding side, whereas it keeps smoothly rising on the approaching side.
For this galaxy the lowest contour in the total HI map is $2.3\,10^{20}$ HI atoms/cm$^2$, and in the XV diagram $3\sigma = 15.7$ K.
}
\end{figure}
\setcounter{figure}{\thedummyfofo}
\clearpage

\begin{figure}
\plotatlas{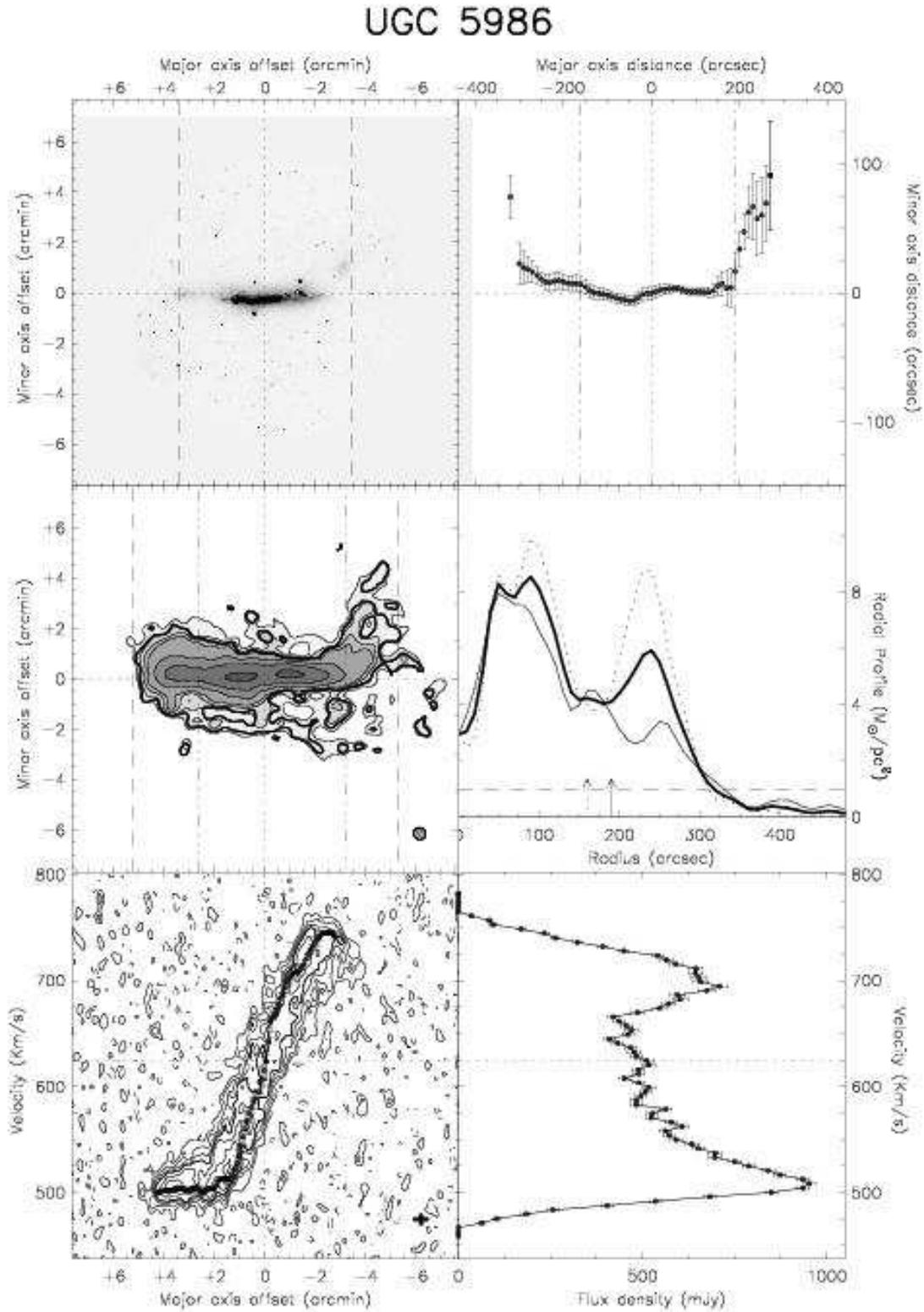}
\caption{-- {\it continued}.
\noindent {\bf UGC 5986:} This galaxy shows major distortions both
morphologically and kinematically. The galaxy UGC 5983 (only 3.5$'$
away) is likely to be responsible for some of these distortions. The
central part of the galaxy shows non-circular motions of the gas. 
For this galaxy the lowest contour in the total HI map is $2\,10^{20}$ HI atoms/cm$^2$, and in the XV diagram $3\sigma = 15.7$ K.
}
\end{figure}
\setcounter{figure}{\thedummyfofo}
\clearpage

\begin{figure}
\plotatlas{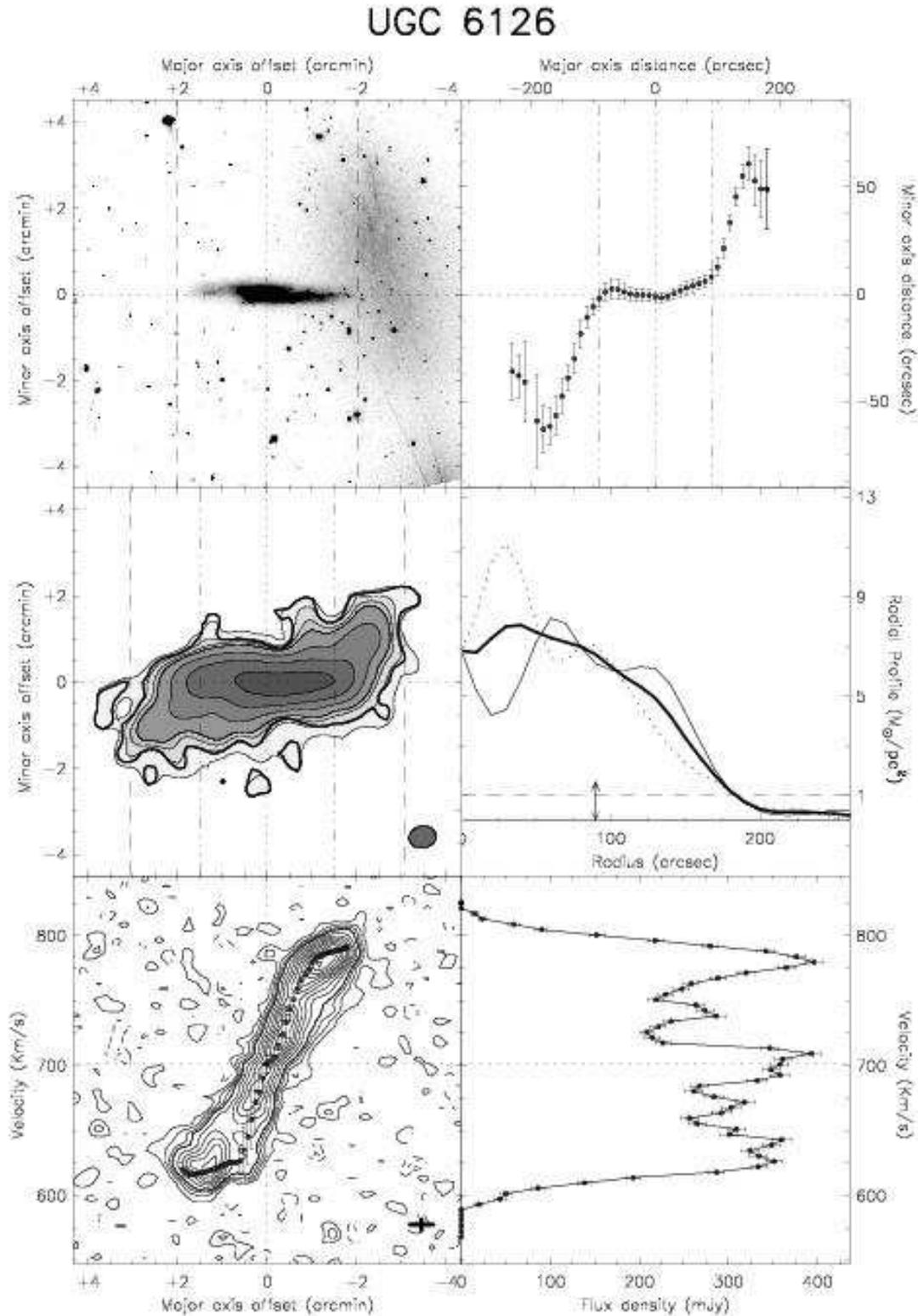}
\caption{-- {\it continued}.
\noindent {\bf UGC 6126:} This galaxy has one of the largest warps of
the sample. It is surrounded by companion galaxies, three of which are
detect in HI. Taylor et al. (1994) observed this
galaxy with a higher resolution with the VLA and detected low level HI
emission above the end of the disk on the right side. Careful inspection of our
datacube smoothed to a 60$''$ beam and velocity resolution of 25 km/s
resulted in a marginal detection of that HI cloud. The warp has more
structure in density than usual at full resolution and it has 
a velocity slightly higher than the maximum of the rotation curve
(which in general is not true for the warps in other galaxies).
For this galaxy the lowest contour in the total HI map is $1.1\,10^{20}$ HI atoms/cm$^2$, and in the XV diagram $3\sigma = 7.68$ K.
}
\end{figure}
\setcounter{figure}{\thedummyfofo}
\clearpage

\begin{figure}
\plotatlas{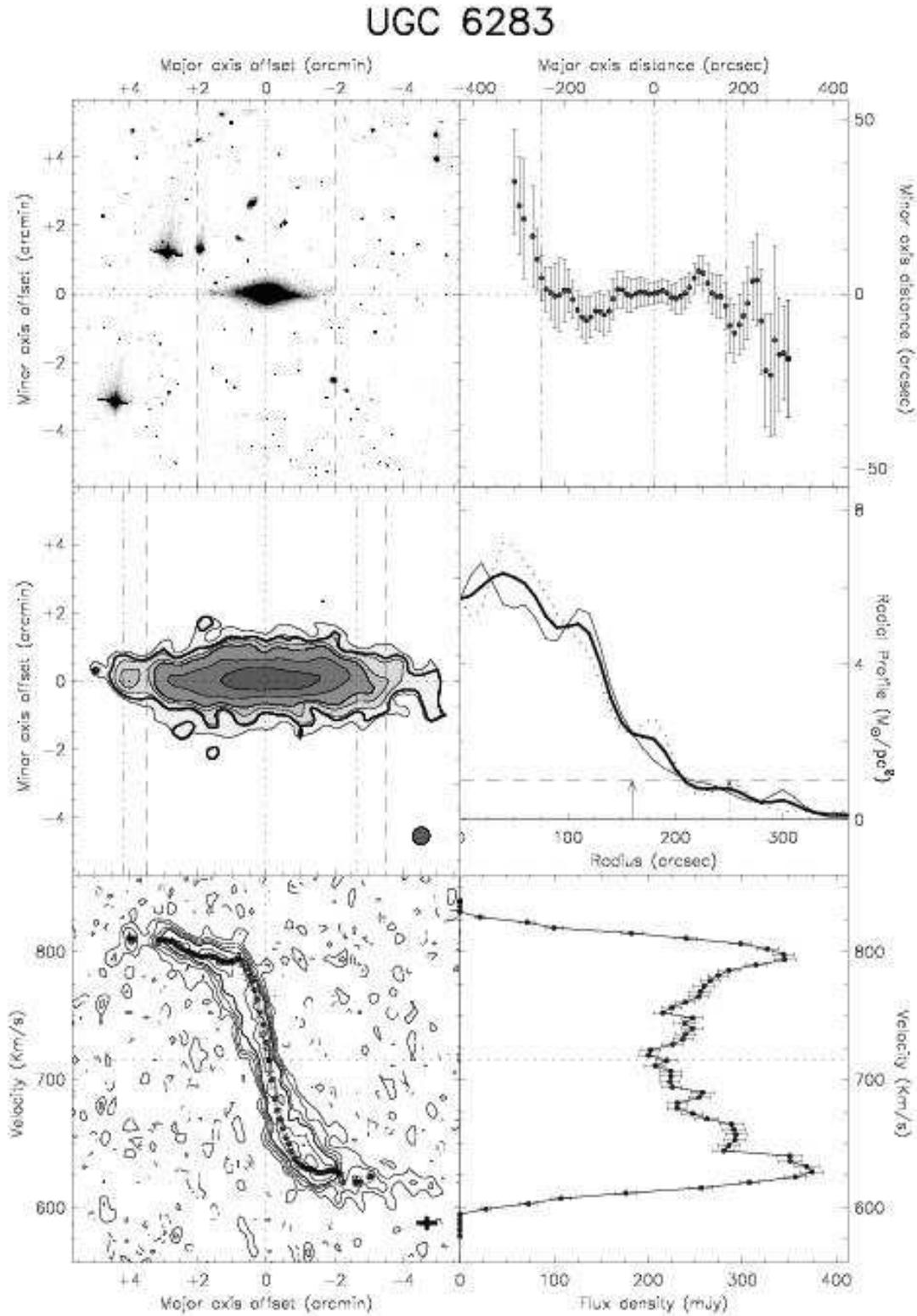}
\caption{-- {\it continued}.
\noindent {\bf UGC 6283:} This galaxy is the earliest type galaxy
(Sab) of our sample and has a very prominent bulge. This inner bulge is inclined
with respect to the plane defined by the HI disk by $\simeq$ 3\degr. The optical image also shows a faint disk component aligned
with the HI disk (probably the warp detected in S\'anchez-Saavedra
et al. (1990)).
For this galaxy the lowest contour in the total HI map is $1.6\,10^{20}$ HI atoms/cm$^2$, and in the XV diagram $3\sigma = 13.1$ K.
}
\end{figure}
\setcounter{figure}{\thedummyfofo}
\clearpage

\begin{figure}
\plotatlas{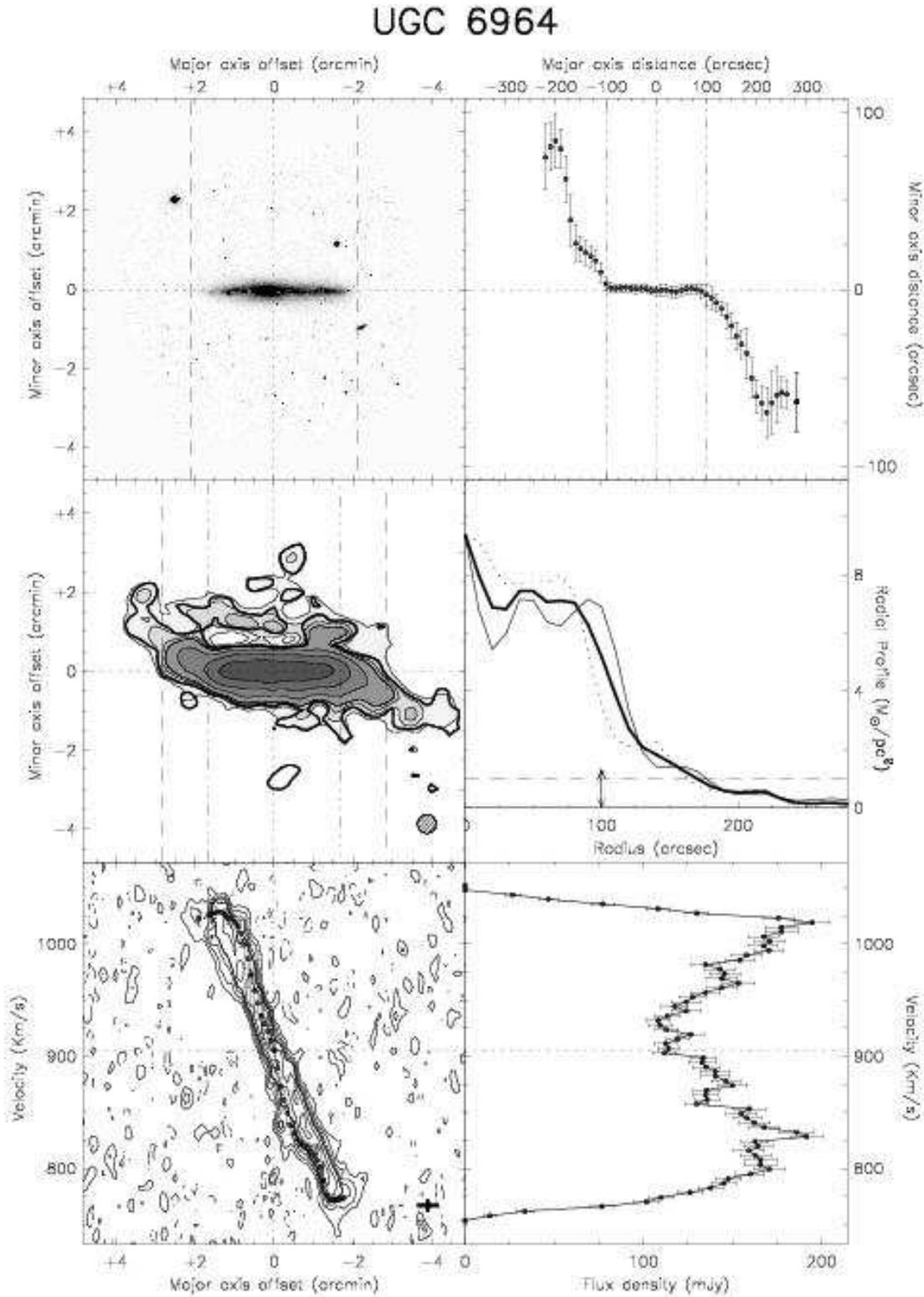}
\caption{-- {\it continued}.
\noindent {\bf UGC 6964:} This galaxy shows quite a large and
symmetric warp. Note the lopsided optical image. The XV diagram
shows a ``bump'' very close to the center and a closer inspection of the
datacube suggests that this could be the signature of infalling (or
outflowing) gas onto the disk. That feature is kinematically connected
to the extra HI gas at the top-right part of the disk.
For this galaxy the lowest contour in the total HI map is $9.2\,10^{19}$ HI atoms/cm$^2$, and in the XV diagram $3\sigma = 19.7$ K.
}
\end{figure}
\setcounter{figure}{\thedummyfofo}
\clearpage

\begin{figure}
\plotatlas{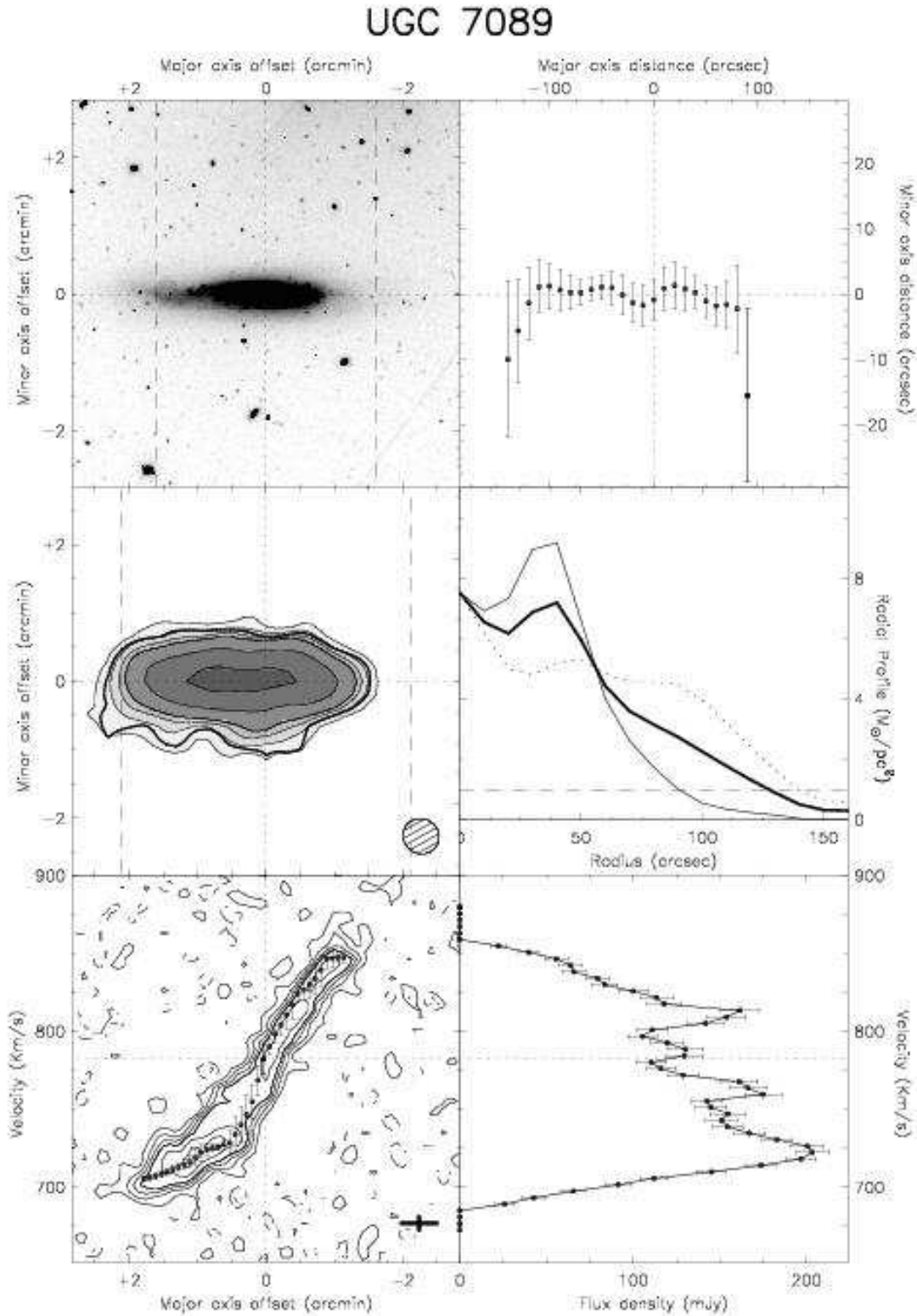}
\caption{-- {\it continued}.
\noindent {\bf UGC 7089:} This galaxy is very lopsided both in
the optical and the HI distribution. The HI does not extend
further than the optical.
For this galaxy the lowest contour in the total HI map is $1.4\,10^{20}$ HI atoms/cm$^2$, and in the XV diagram $3\sigma = 14$ K.
}
\end{figure}
\setcounter{figure}{\thedummyfofo}
\clearpage

\begin{figure}
\plotatlas{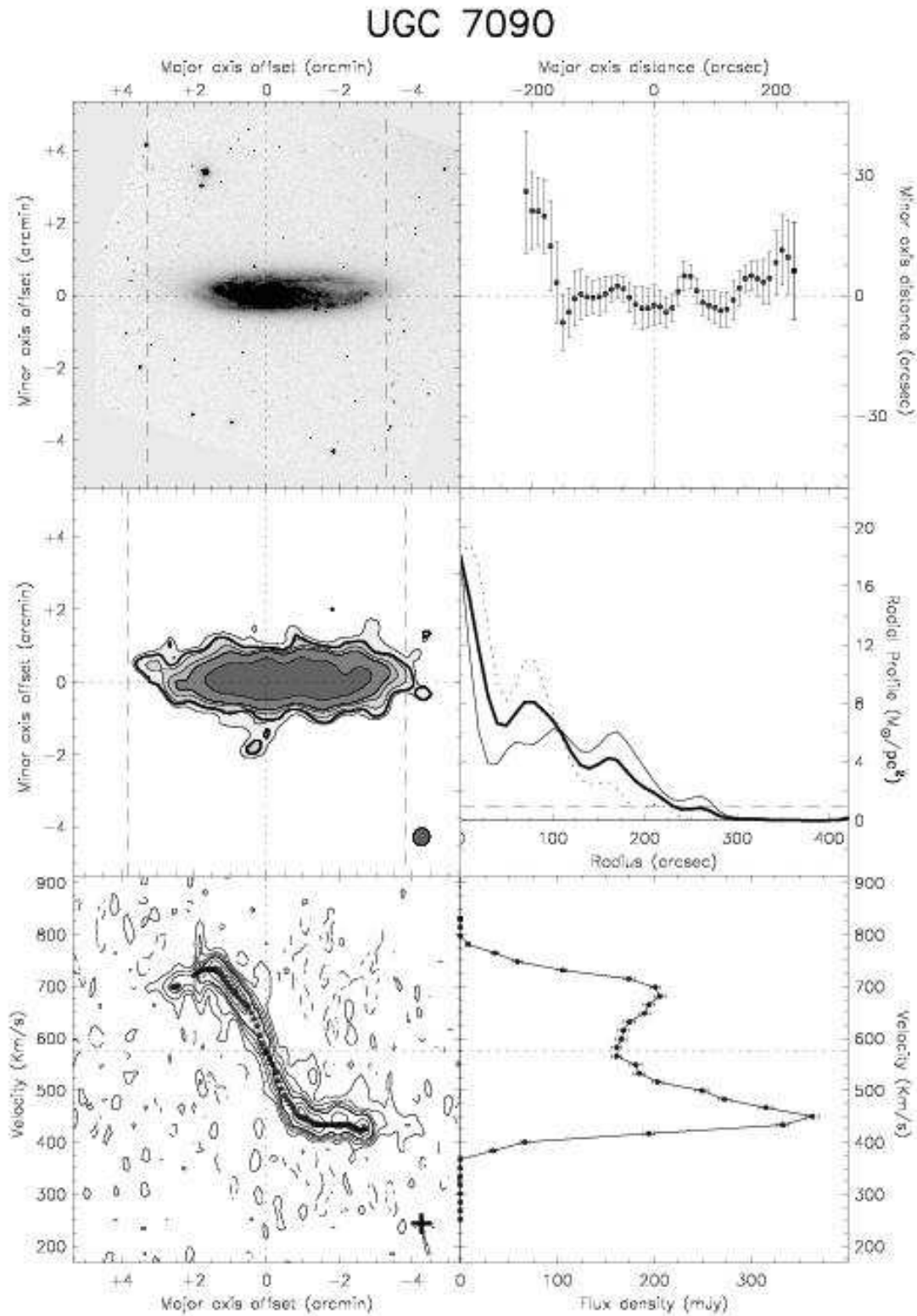}
\caption{-- {\it continued}.
\noindent {\bf UGC 7090:} This galaxy has very strong kinematical and
density lopsidedness indices and a very asymmetric global profile.
For this galaxy the lowest contour in the total HI map is $3\,10^{20}$ HI atoms/cm$^2$, and in the XV diagram $3\sigma = 7.97$ K.
}
\end{figure}
\setcounter{figure}{\thedummyfofo}
\clearpage

\begin{figure}
\plotatlas{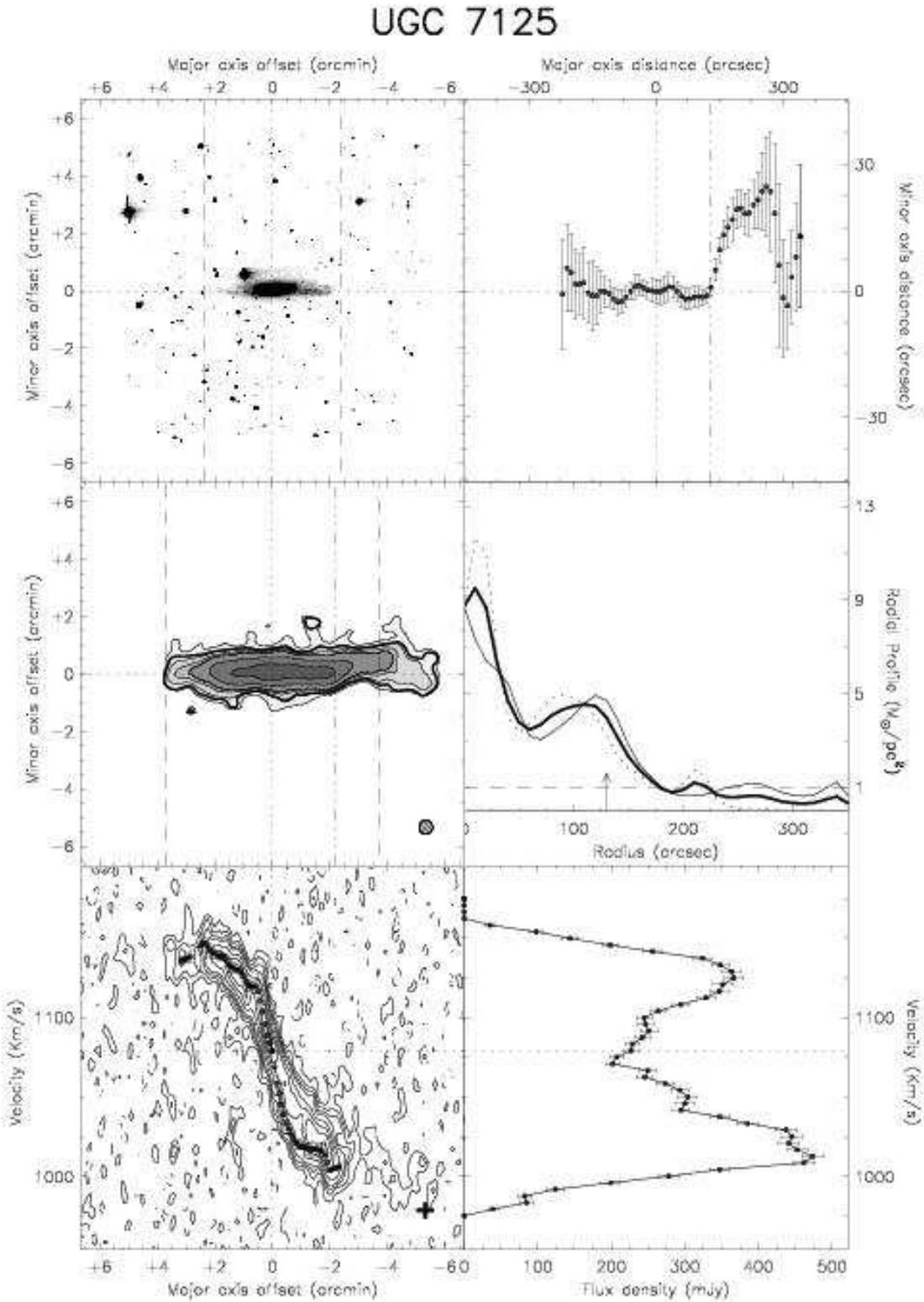}
\caption{-- {\it continued}.
\noindent {\bf UGC 7125:} The warp of this galaxy on the right part of
the disk resembles that of the Milky Way: it rises and then turns
back towards the plane, crossing it.
For this galaxy the lowest contour in the total HI map is $1.7\,10^{20}$ HI atoms/cm$^2$, and in the XV diagram $3\sigma = 11.6$ K.
}
\end{figure}
\setcounter{figure}{\thedummyfofo}
\clearpage

\begin{figure}
\plotatlas{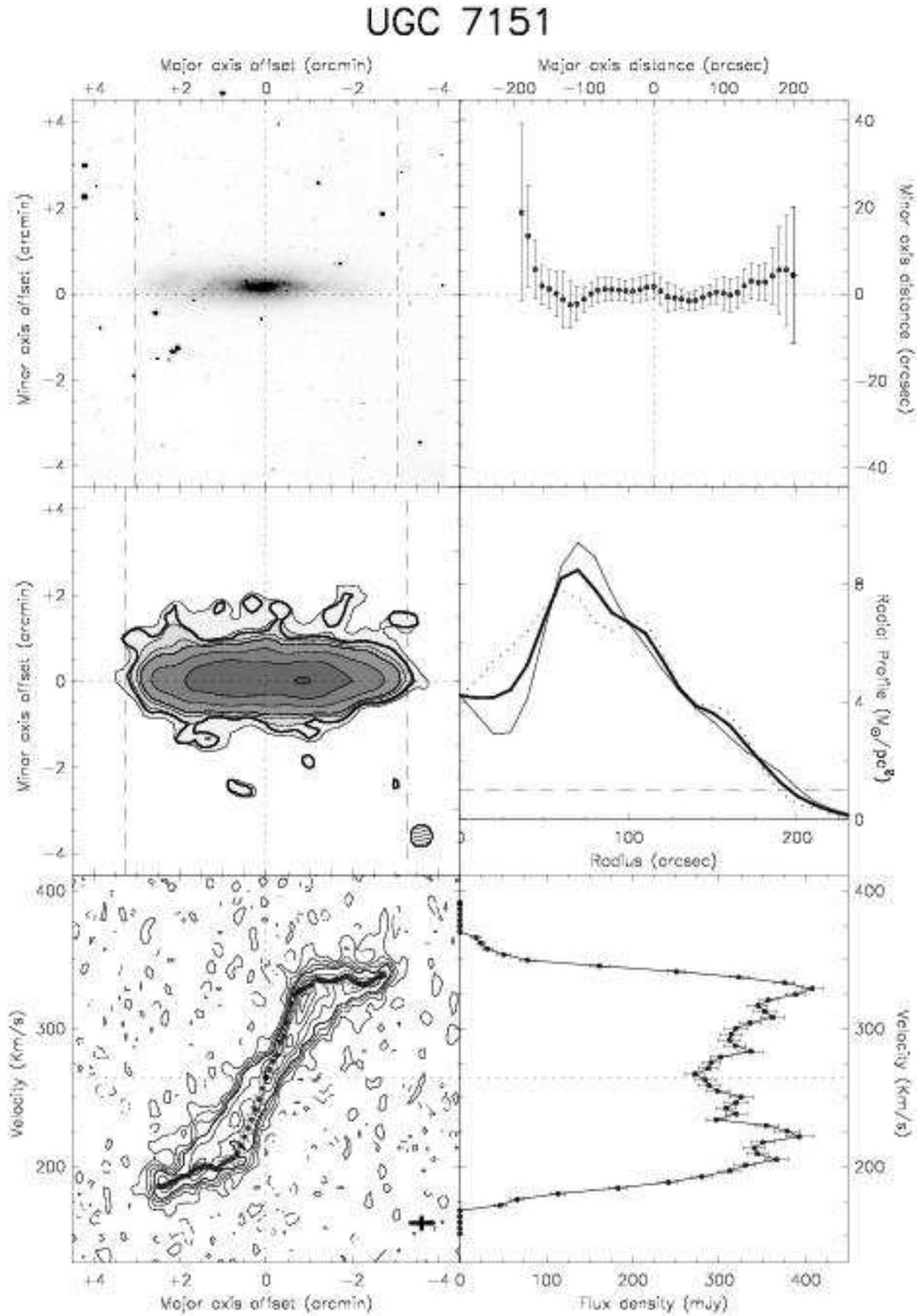}
\caption{-- {\it continued}.
\noindent {\bf UGC 7151:} The HI is confined to the extent
of the optical disk in this galaxy. The rotation curve
reaches a flat part on the receding side but rises continuously 
until the last measured point on the approaching side.
For this galaxy the lowest contour in the total HI map is $1.4\,10^{20}$ HI atoms/cm$^2$, and in the XV diagram $3\sigma = 14.5$ K.
}
\end{figure}
\setcounter{figure}{\thedummyfofo}
\clearpage

\begin{figure}
\plotatlas{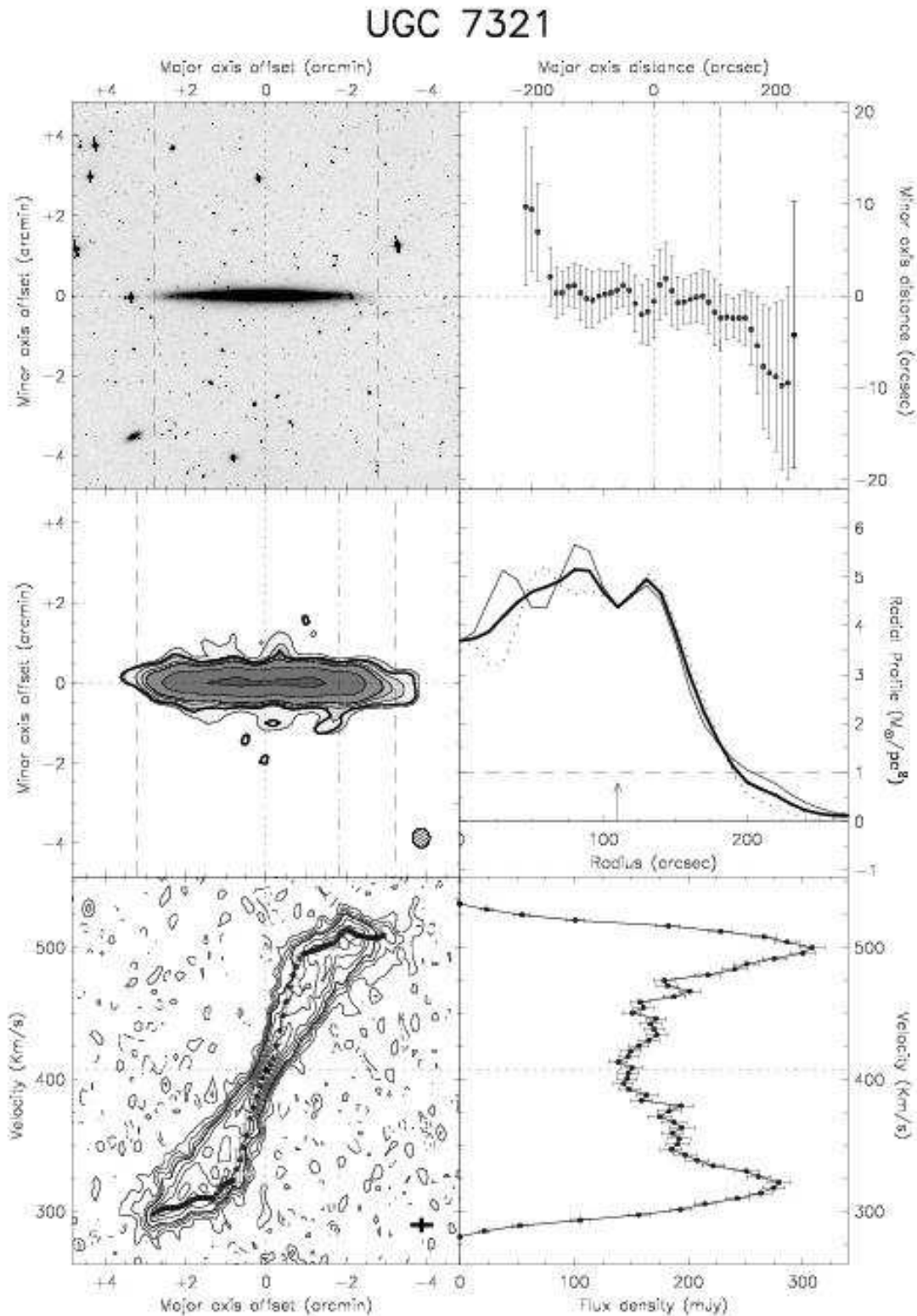}
\caption{-- {\it continued}.
\noindent {\bf UGC 7321:} This galaxy ultrathin \cite{goad}. There is
a mild warp visible on the  
receding side. This places some constrains on the
possible formation scenarios for the warp of this galaxy because some
of the processes that generate warps do also heat up the disk.
For this galaxy the lowest contour in the total HI map is $2.4\,10^{20}$ HI atoms/cm$^2$, and in the XV diagram $3\sigma = 8.58$ K.
}
\end{figure}
\setcounter{figure}{\thedummyfofo}
\clearpage

\begin{figure}
\plotatlas{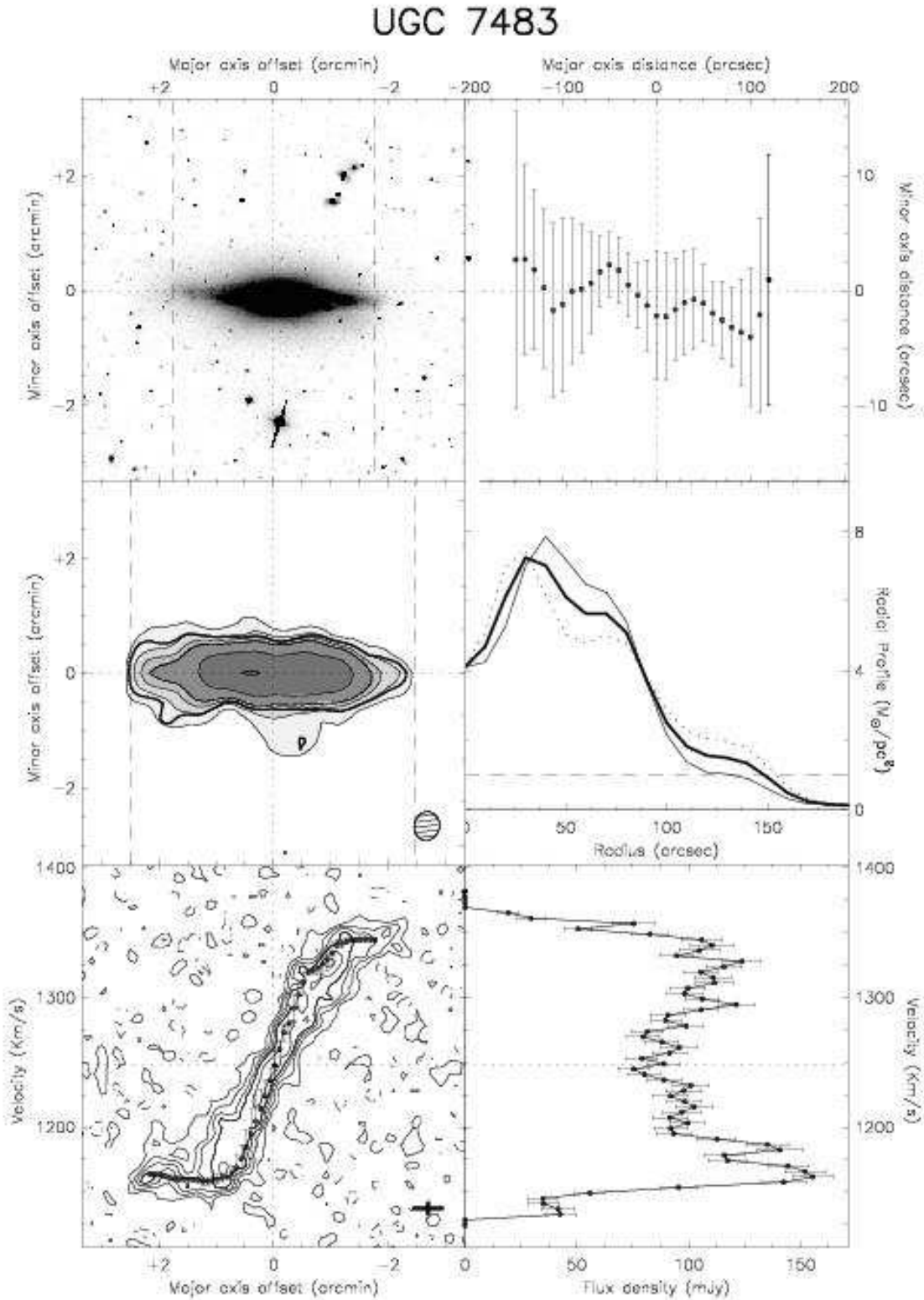}
\caption{-- {\it continued}.
\noindent {\bf UGC 7483:} We found this galaxy to be classified as
SBc. The optical picture shows a very prominent bulge and a small disk
(even in the HI) so it is likely that it is an earlier type. As with
other galaxies with bulges, we find that the optical disk is
misaligned with the HI disk by few degrees.
For this galaxy the lowest contour in the total HI map is $2\,10^{20}$ HI atoms/cm$^2$, and in the XV diagram $3\sigma = 11.4$ K.
}
\end{figure}
\setcounter{figure}{\thedummyfofo}
\clearpage

\begin{figure}
\plotatlas{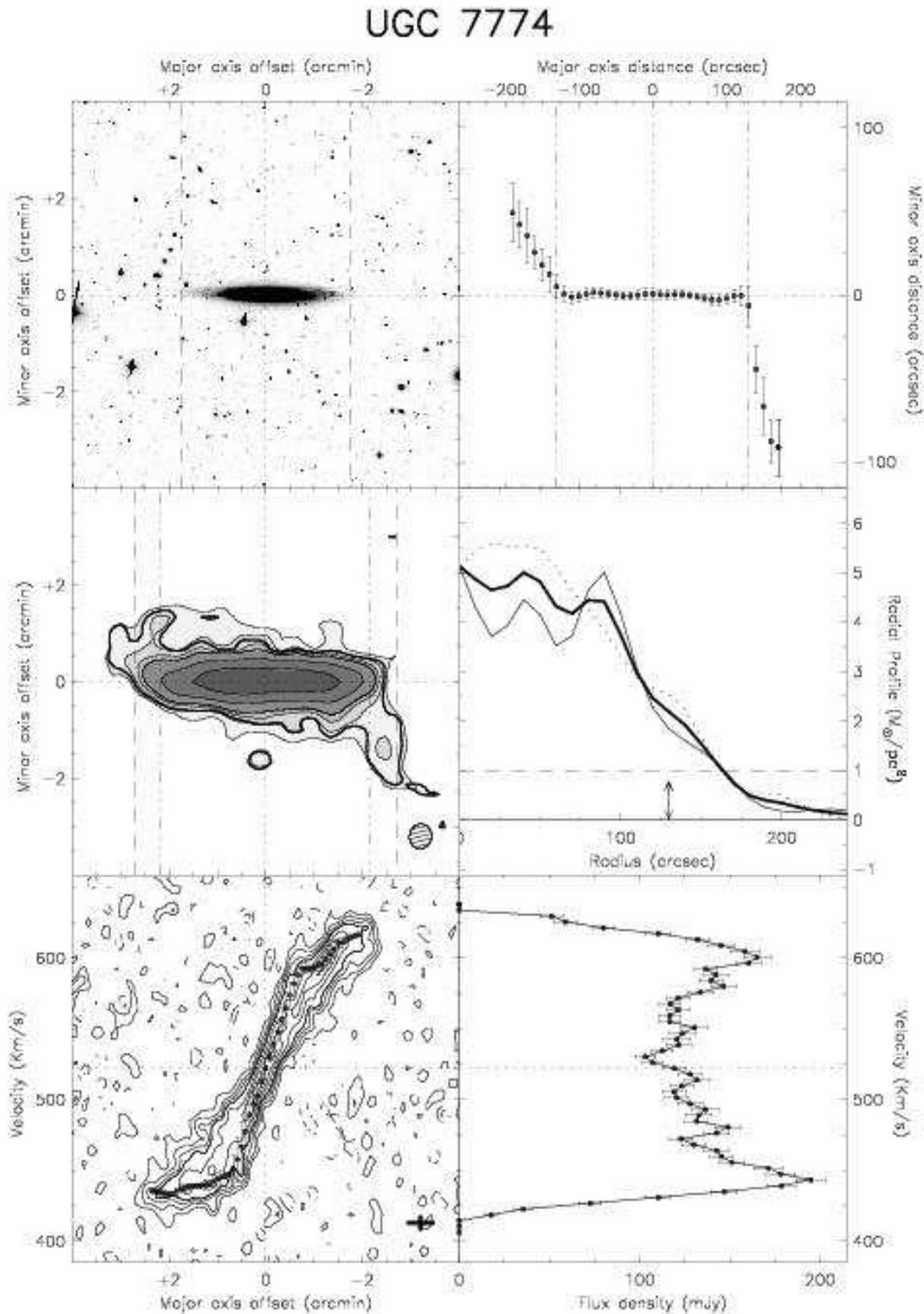}
\caption{-- {\it continued}.
\noindent {\bf UGC 7774:} This is the most remarkable galaxy in our
sample. The right side of the disk is warped at a very high angle and the
line of nodes (LON) is very close to the line of sight. However, the LON of
the SE warp is not close to the line of sight, creating a very
asymmetric warp.
Standard tilted-ring models would be unable to fit the
warp of this galaxy due to the difference in the LON for the receding
and approaching sides. The closest known galaxy is at
300 kpc (projected separation), so tidal interaction is unlikely to be
the cause of this extremely asymmetric warp.
For this galaxy the lowest contour in the total HI map is $1.3\,10^{20}$ HI atoms/cm$^2$, and in the XV diagram $3\sigma = 12.5$ K.
}
\end{figure}
\setcounter{figure}{\thedummyfofo}
\clearpage

\begin{figure}
\plotatlas{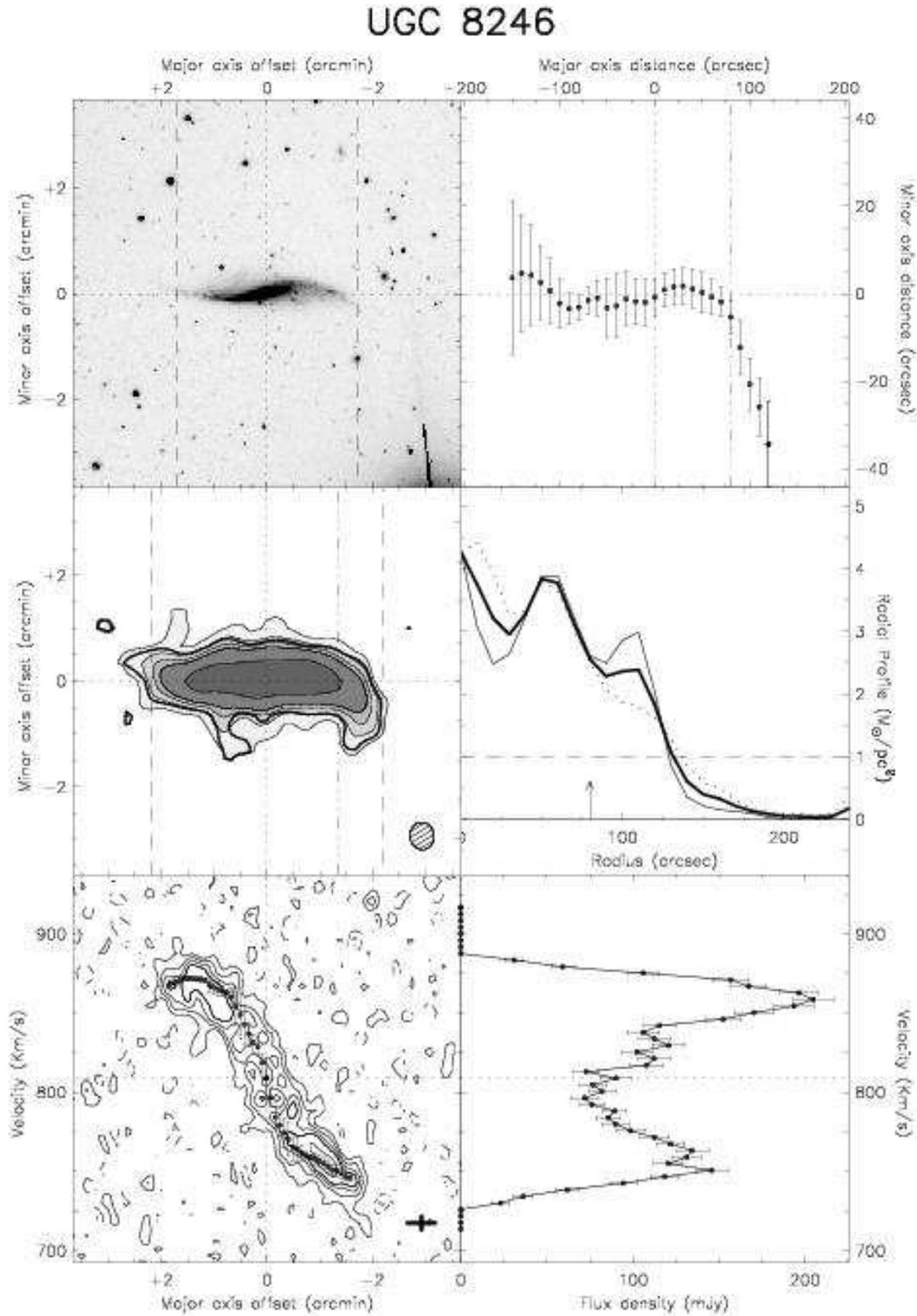}
\caption{-- {\it continued}.
\noindent {\bf UGC 8246:} This galaxy is very close (6 arcmin) to NGP9
F269-1340356, detected at similar velocities in our 21 cm data. The
central parts of the galaxy show evidence of large non-circular motions, and
it is possible that the warped appearance is due to either spiral arms or
tidal features.
For this galaxy the lowest contour in the total HI map is $1.3\,10^{20}$ HI atoms/cm$^2$, and in the XV diagram $3\sigma = 10.6$ K.
}
\end{figure}
\setcounter{figure}{\thedummyfofo}
\clearpage

\begin{figure}
\plotatlas{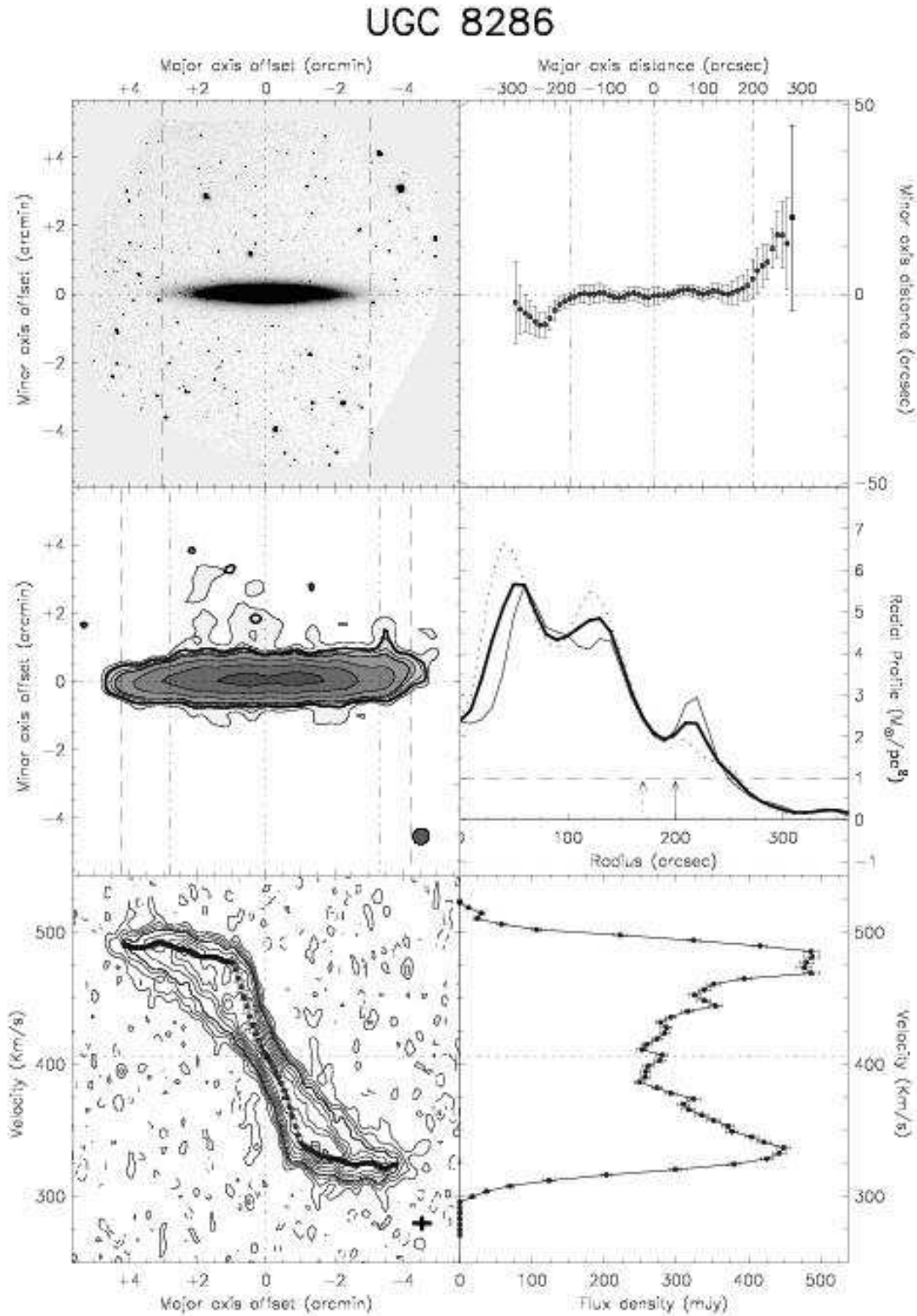}
\caption{-- {\it continued}.
\noindent {\bf UGC 8286:} This galaxy has a mild warp on 
both sides of the galaxy, starting around the edge of 
the optical disk.
For this galaxy the lowest contour in the total HI map is $1.2\,10^{20}$ HI atoms/cm$^2$, and in the XV diagram $3\sigma = 9.87$ K.
}
\end{figure}
\setcounter{figure}{\thedummyfofo}
\clearpage

\begin{figure}
\plotatlas{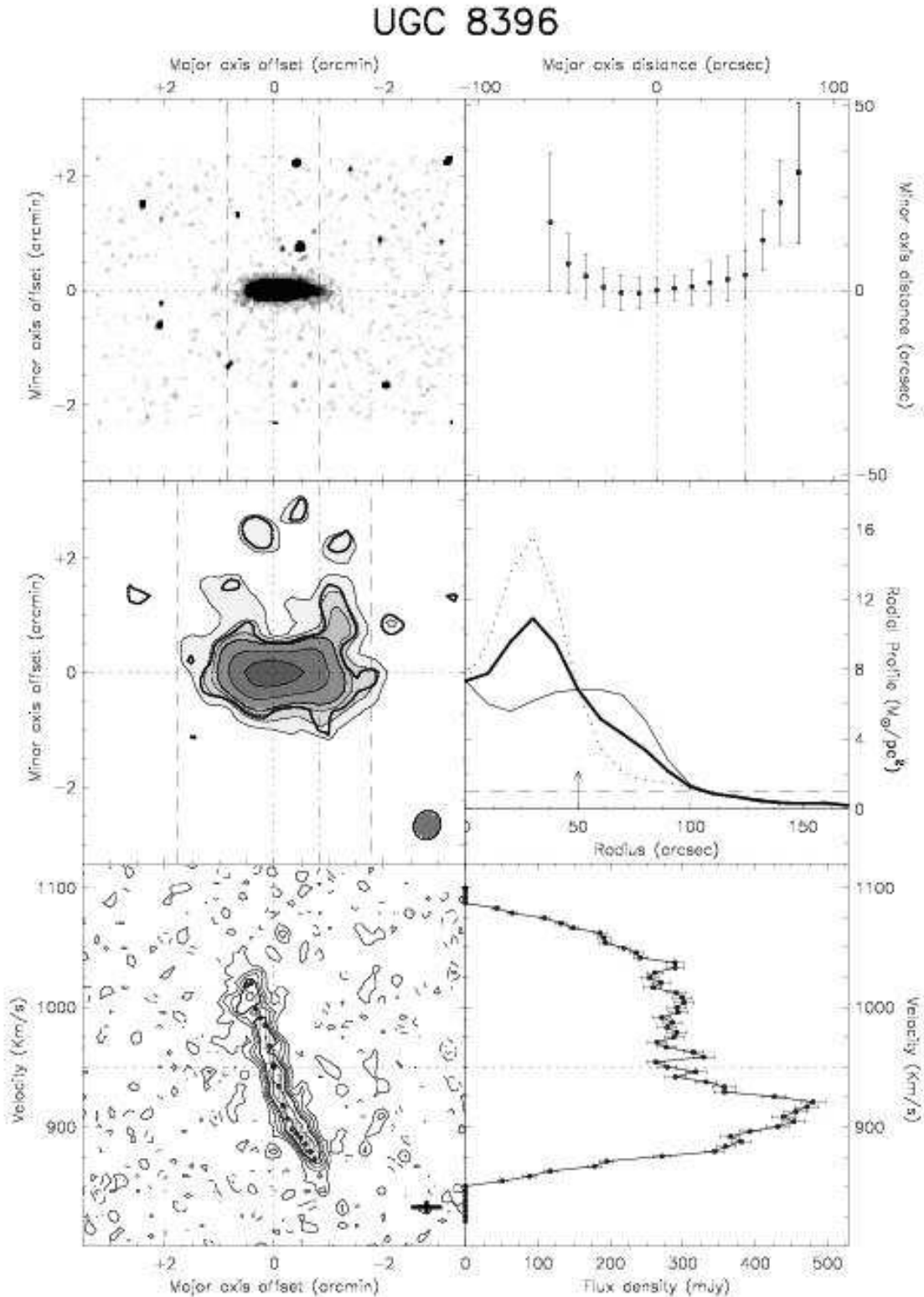}
\caption{-- {\it continued}.
\noindent {\bf UGC 8396:} This galaxy is very close to NGC 5112 (13
arcmin) and seems to be perturbed by it. The U shaped warp seems to
'point' towards NGC 5112. We also detected some ``loose'' HI 
gas between UGC 8396 and NGC 5112. 
For this galaxy the lowest contour in the total HI map is $1.5\,10^{20}$ HI atoms/cm$^2$, and in the XV diagram $3\sigma = 14.5$ K.
}
\end{figure}
\setcounter{figure}{\thedummyfofo}
\clearpage

\begin{figure}
\plotatlas{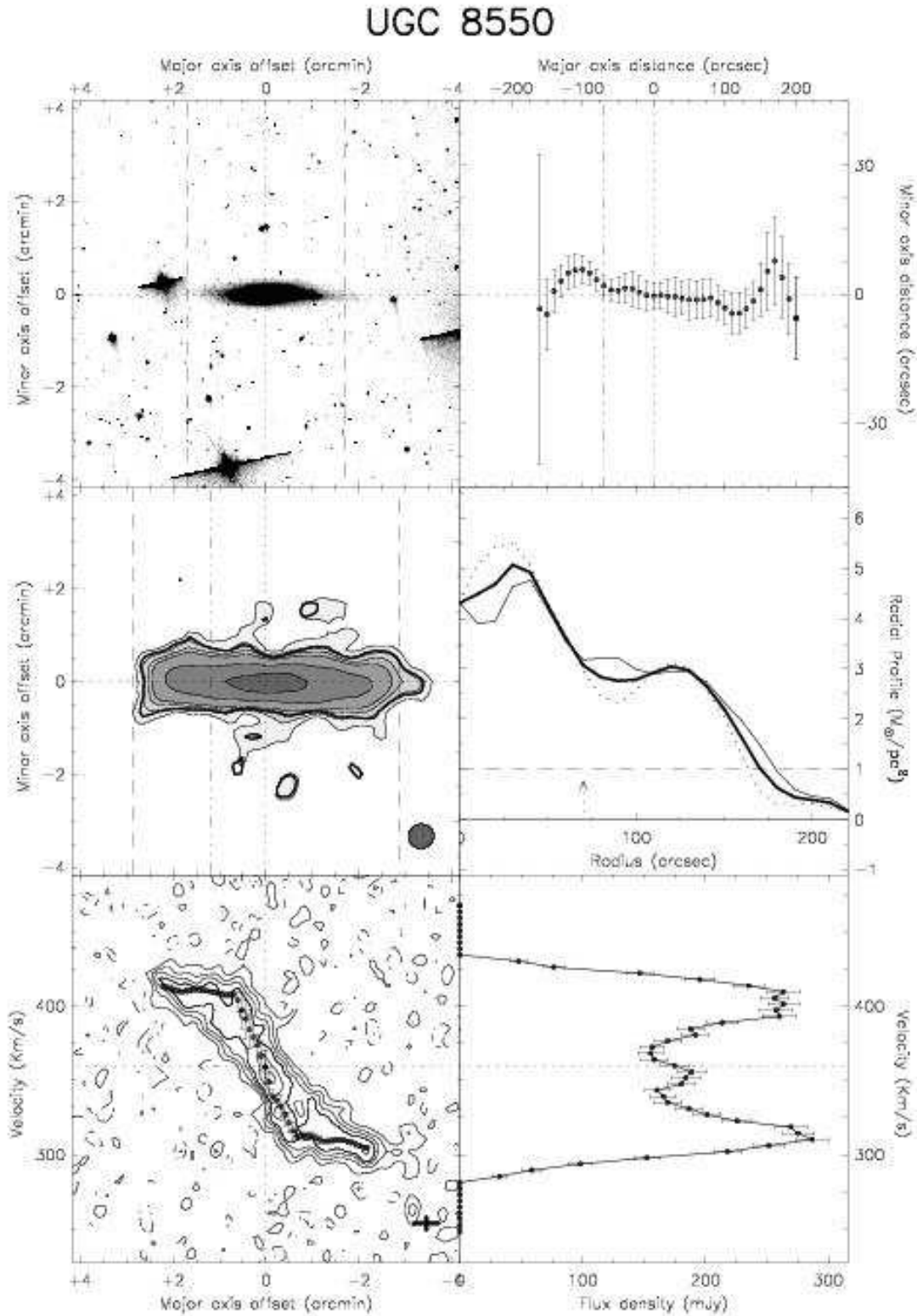}
\caption{-- {\it continued}.
\noindent {\bf UGC 8550:} The disk of this galaxy looks somewhat peculiar both
in the optical and the HI. The optical image shows a faint extension
of the disk misaligned with the inner isophotes, probably due to non
edge-on spiral arms.
For this galaxy the lowest contour in the total HI map is $1.4\,10^{20}$ HI atoms/cm$^2$, and in the XV diagram $3\sigma = 16.8$ K.
}
\end{figure}
\setcounter{figure}{\thedummyfofo}
\clearpage

\begin{figure}
\plotatlas{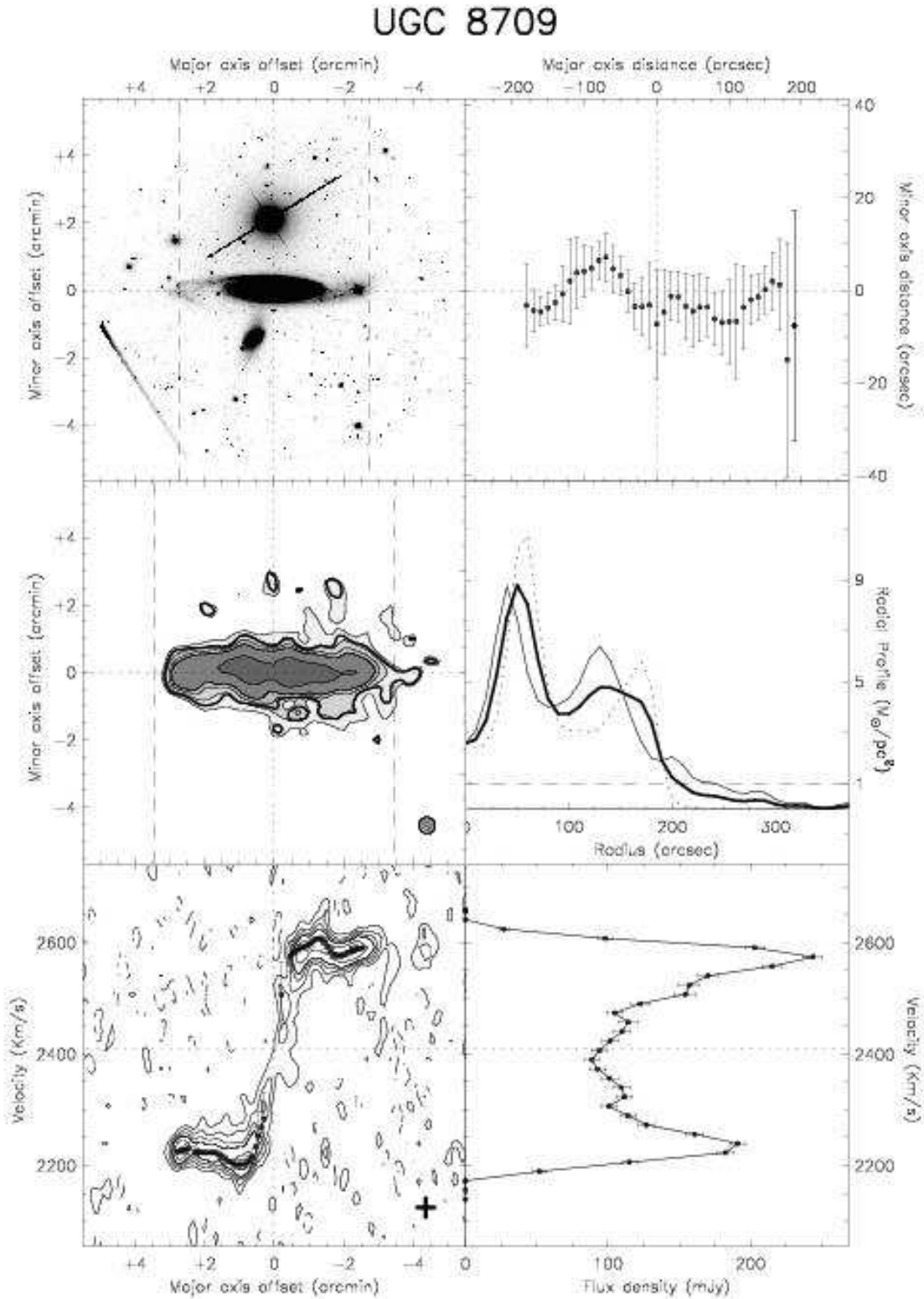}
\caption{-- {\it continued}.
\noindent {\bf UGC 8709:} This galaxy is not completely edge-on and two
spiral arms are clearly visible on the optical picture. Even though it has
a very close companion galaxy (NGC 5296), it has no warp.
For this galaxy the lowest contour in the total HI map is $2.8\,10^{20}$ HI atoms/cm$^2$, and in the XV diagram $3\sigma = 7.67$ K.
}
\end{figure}
\setcounter{figure}{\thedummyfofo}
\clearpage

\begin{figure}
\plotatlas{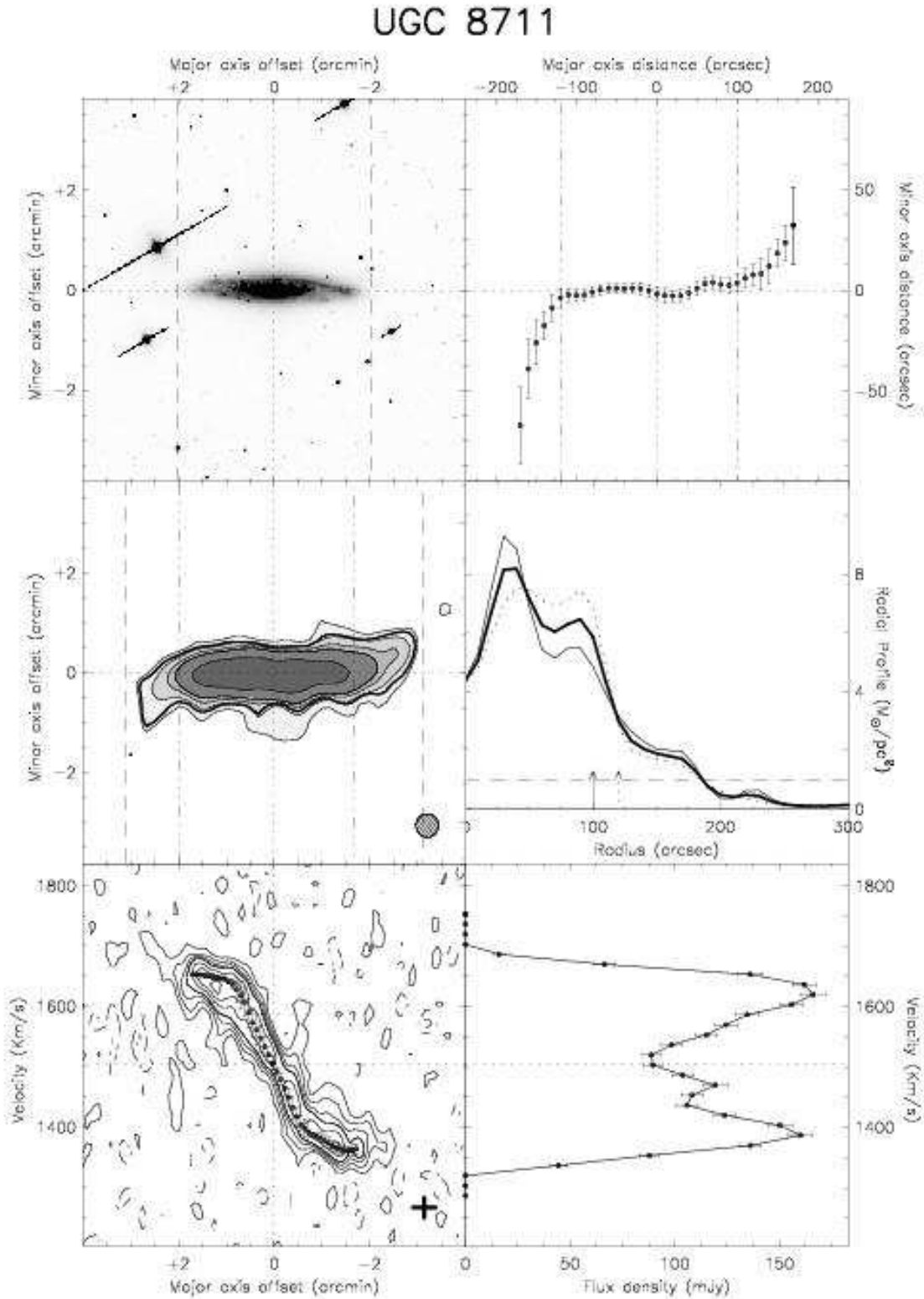}
\caption{-- {\it continued}.
\noindent {\bf UGC 8711:} This galaxy is isolated and exhibits
quite a large warp. The warp can not be mistaken for inclined spiral
arms because these run in the opposite directions. 
For this galaxy the lowest contour in the total HI map is $3.2\,10^{20}$ HI atoms/cm$^2$, and in the XV diagram $3\sigma = 9.59$ K.
}
\end{figure}
\setcounter{figure}{\thedummyfofo}
\clearpage

\begin{figure}
\plotatlas{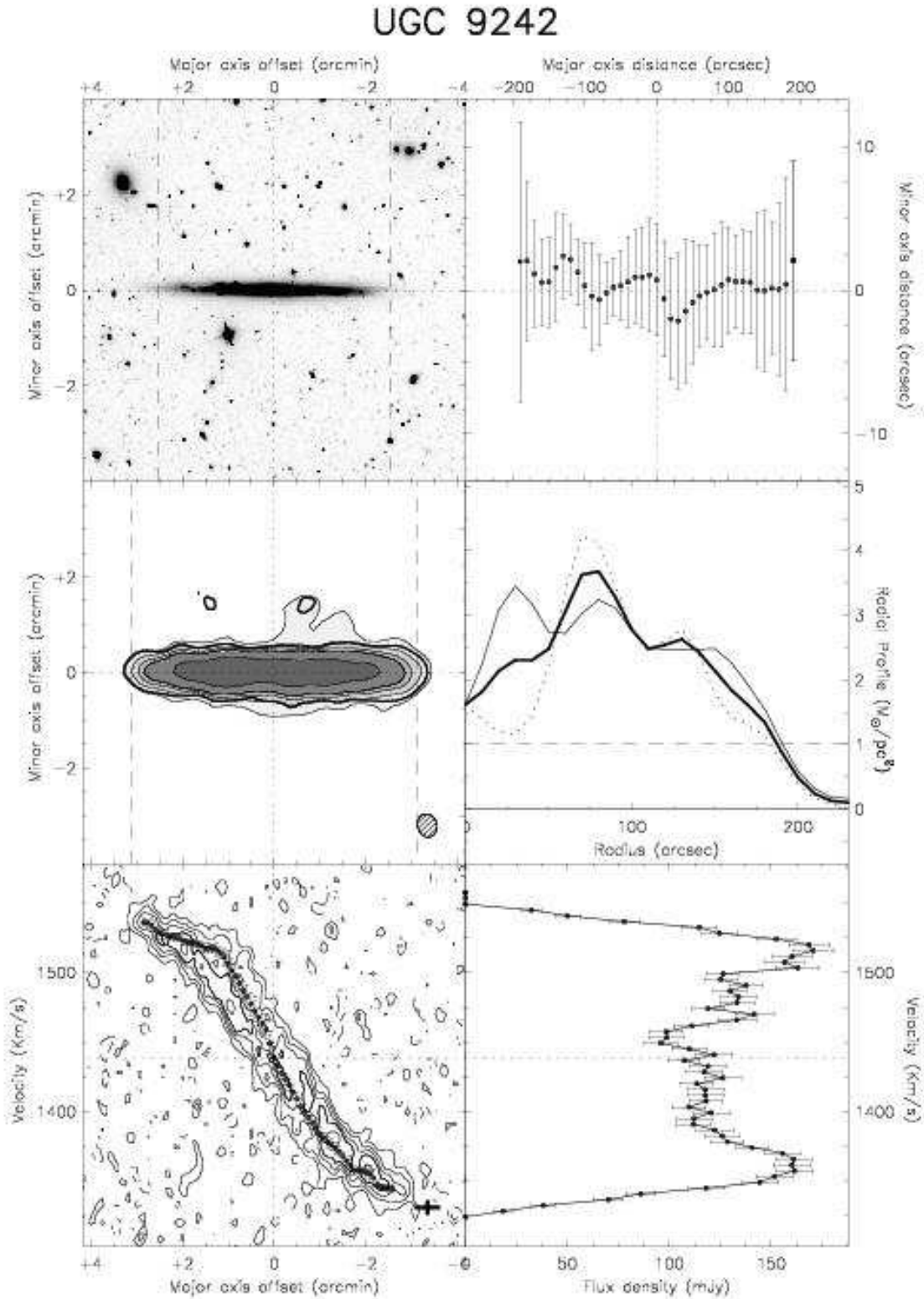}
\caption{-- {\it continued}.
\noindent {\bf UGC 9242:} This is another ultrathin galaxy. It is the flattest of
the galaxies in our sample, with deviations smaller than 2.5$''$ over
the 400$''$ diameter of the galaxy. There is evidence of non-circular motions
in the central regions. 
For this galaxy the lowest contour in the total HI map is $1.9\,10^{20}$ HI atoms/cm$^2$, and in the XV diagram $3\sigma = 20$ K.
}
\end{figure}
\setcounter{figure}{\thedummyfofo}
\clearpage


\end{document}